\documentclass[11pt, letterpaper, final]{amsart}
\usepackage[foot]{amsaddr}

\usepackage{setspace,fullpage}

\usepackage{amsmath,amssymb,latexsym, amscd, esint}
\usepackage{exscale,cite,graphics,xcolor}
\usepackage[dvips]{epsfig}
\usepackage{scalerel}

\newcommand\reallywidehat[1]{\arraycolsep=0pt\relax%
\begin{array}{c}
\stretchto{
  \scaleto{
    \scalerel*[\widthof{\ensuremath{#1}}]{\kern-.5pt\bigwedge\kern-.5pt}
    {\rule[-\textheight/2]{1ex}{\textheight}} 
  }{\textheight} %
}{0.5ex}\\           
#1\\                 
\rule{-1ex}{0ex}
\end{array}
}

\usepackage[colorlinks]{hyperref}
\usepackage[notref, notcite]{showkeys}

\renewcommand{\geq}{\geqslant}
\renewcommand{\leq}{\leqslant}

\title[The capillary-gravity Whitham equation]{Numerical bifurcation and stability for \\ the capillary-gravity Whitham equation}

\author[Charalampidis]{Efstathios~G.~Charalampidis}
\address{Mathematics Department, California Polytechnic State University San Luis Obispo, CA 93407-0403, USA}
\email{echarala@calpoly.edu}

\author[Hur]{Vera~Mikyoung~Hur}
\address{Department of Mathematics, University of Illinois at Urbana-Champaign Urbana, IL 61801, USA}
\email{verahur@math.uiuc.edu}
\thanks{VMH is supported by the NSF through award DMS-2009981.}

\date{\today}

\begin{document}

\begin{abstract}
We adopt a robust numerical continuation scheme to examine the global bifurcation of periodic traveling waves of the capillary-gravity Whitham equation, which combines the dispersion in the linear theory of capillary-gravity waves and a shallow water nonlinearity. 
We employ a highly accurate numerical method for space discretization and time stepping, to address orbital stability and instability for a rich variety of the solutions. Our findings can help classify capillary-gravity waves and understand their long-term dynamics. 
\end{abstract}

\maketitle

\section{Introduction}\label{sec:intro}

Whitham in his 1967 paper \cite{Whitham1967} (see also \cite{Whitham;book}) put forward
\begin{equation}\label{eqn:Whitham0}
u_t+c_\text{ww}(|\partial_x|)u_x+\frac32\sqrt{\frac{g}{h}} uu_x=0
\end{equation}
to address, qualitatively, breaking and peaking of waves in shallow water. Here and throughout, $u(x,t)$ is related to the displacement of the fluid surface from the rest state at position~$x$ and time~$t$, 
and $c_{\rm{\tiny ww}}(|\partial_x|)$ is a Fourier multiplier operator, defined as 
\begin{equation}\label{def:c0}
\reallywidehat{c_\text{ww}(|\partial_x|)f}(k)=\sqrt{\frac{g\tanh(kh)}{k}}\widehat{f}(k),
\end{equation}
where $g$ is the gravitational constant and $h$ the undisturbed fluid depth. Notice that $c_\text{ww}(k)$ is the phase speed of $2\pi/k$ periodic waves in the linear theory of water waves \cite{Whitham;book}. 

For relatively shallow water or, equivalently, relatively long waves for which $kh\ll1$, one can expand the right hand side of \eqref{def:c0} to obtain
\[
c_{\rm{\tiny ww}}(k)=\sqrt{gh}\Big(1-\frac16k^2h^2\Big)+O(k^4h^4),
\]
whereby arriving at the celebrated Korteweg--de Vries (KdV) equation:
\begin{equation}\label{eqn:KdV}
u_t+\sqrt{gh}\Big(u_x+\frac16h^2 u_{xxx}\Big)+\frac32\sqrt{\frac{g}{h}}uu_x=0.
\end{equation}
Therefore \eqref{eqn:Whitham0} and \eqref{def:c0} can be regarded as augmenting \eqref{eqn:KdV} to include the {\em full} dispersion in the linear theory of water waves and, thus, improving over \eqref{eqn:KdV} for short and intermediately long waves. Indeed, numerical studies (see \cite{MKD;comp}, for instance) reveal that the Whitham equation performs on par with or better than the KdV equation and other shallow water models in a wide range of amplitude and wavelength parameters. 

The KdV equation admits solitary and cnoidal waves but no traveling waves can `peak'. By contrast, the so-called extreme Stokes wave possesses a $120^\circ$ peaking at the crest. Also no solutions of \eqref{eqn:KdV} can `break'. That means, the solution remains bounded but its slope becomes unbounded. See \cite[Section~13.14]{Whitham;book} for more discussion. 
This is perhaps not surprising because the dispersion\footnote{The phase speed is $\sqrt{gh}(1-\frac16k^2h^2)$, poorly approximating $c_\text{ww}(k)$ when $kh\gg1$.} of the KdV equation is inadequate for explaining high frequency phenomena of water waves. Whitham \cite{Whitham1967,Whitham;book} conjectured breaking and peaking for \eqref{eqn:Whitham0} and \eqref{def:c0}. Recently, one of the authors \cite{Hur;breaking} proved breaking, and Ehrnstr\"om and Wahl\'en  \cite{EW;peaking} proved peaking. 
Johnson and one of the authors \cite{HJ;whitham} proved that a small amplitude and periodic traveling wave of \eqref{eqn:Whitham0} and \eqref{def:c0} is modulationally unstable, provided that $kh>1.145\dots$, comparing with the well-known Benjamin--Feir instability of a Stokes wave. 
By contrast, all cnoidal waves of \eqref{eqn:KdV} are modulationally stable. 

When the effects of surface tension are included, Johnson and one of the authors \cite{HJ;capillary} proposed to replace \eqref{def:c0} by 
\begin{equation}\label{def:cT0}
\reallywidehat{c_\text{ww}(|\partial_x|;T)f}(k)
=\sqrt{(g+Tk^2)\frac{\tanh(kh)}{k}}\widehat{f}(k),
\end{equation}
where $T$ is the surface tension coefficient. Notice that $c_\text{ww}(k;T)$ is the phase speed in the linear theory of capillary-gravity waves \cite{Whitham;book}. When $T=0$, \eqref{def:cT0} becomes \eqref{def:c0}. Since
\[
c_\text{ww}(k;T)=\sqrt{gh}\Big(1-\frac12\Big(\frac13-\frac{T}{gh^2}\Big)k^2h^2\Big)+O(k^4h^4)
\quad\text{as $kh\to0$},
\]
one arrives at the KdV equation for capillary-gravity waves:
\begin{equation}\label{eqn:KdVT}
u_t+\sqrt{gh}\Big(u_x+\frac12\Big(\frac13-\frac{T}{gh^2}\Big)h^2u_{xxx}\Big)+\frac32\sqrt{\frac{g}{h}}uu_x=0
\end{equation}
in the long wavelength limit unless $T/gh^2=1/3$. Notice that \eqref{eqn:KdV} and \eqref{eqn:KdVT} behave alike, qualitatively, possibly after a sign change. By contrast, whenever $T>0$, $c_\text{ww}(k;T)\to\sqrt{T|k|}$ as $kh\to\infty$, so that \eqref{eqn:Whitham0} and \eqref{def:cT0} become 
\begin{equation}\label{eqn:fKdV1/2}
u_t+\sqrt{T}|\partial_x|^{1/2}u_x+\frac32\sqrt{\frac{g}{h}}uu_x=0
\end{equation}
to leading order whereas when $T=0$, $u_t+\sqrt{g}|\partial_x|^{-1/2}u_x+\frac32\sqrt{\frac{g}{h}}uu_x=0$.

In recent years, the {\em capillary-gravity Whitham equation} has been a subject of active research \cite{HJ;capillary,Kalisch;physD,Kalisch;cWhitham,carter;fluids,EJ;WW} (see also \cite{HP;BW,HP;FDCH,Hur;vWhitham,Pandey;comp,KP;LWP}). Particularly, Remonato and Kalisch \cite{Kalisch;physD} combined a spectral collocation method and a numerical continuation approach to discover a rich variety of local bifurcation of periodic traveling waves of \eqref{eqn:Whitham0} and \eqref{def:cT0}, notably, crossing and connecting solution branches. Here we adopt a robust numerical continuation scheme to corroborate the results 
and produce convincing results for global bifurcation. Our findings support local bifurcation theorems (see \cite{EJ;WW}, for instance) and help classify {\em all} periodic traveling waves. 

We employ an efficient numerical method for solving stiff nonlinear PDEs implemented with fourth-order time differencing, to experiment with (nonlinear) orbital stability and instability for a plethora of periodic traveling waves of \eqref{eqn:Whitham0} and \eqref{def:cT0}. (Spectral) modulational stability and instability were investigated numerically \cite{carter;fluids} and also analytically for small amplitude \cite{HJ;capillary}. To the best of the authors' knowledge, however, nonlinear stability and instability have not been addressed. Our novel findings include, among many others, orbital stability for the $k=1$ branch whenever $T>0$ versus instability for $k\geq2,\in\mathbb{N}$ branches for great wave height when $0<T/gh^2\leq1/3$. 

The methodology here is potentially useful for tackling the capillary-gravity wave problem and other nonlinear dispersive equations. 

\section{Preliminaries}\label{sec:prelim}

We 
rewrite \eqref{eqn:Whitham0} and \eqref{def:cT0} succinctly as 
\begin{equation}\label{eqn:Whitham}
u_t+c_\text{ww}(|\partial_x|;T)u_x+(u^2)_x=0,
\quad\text{where}\quad
\reallywidehat{c_\text{ww}(|\partial_x|;T)f}(k)
=\sqrt{(1+Tk^2)\frac{\tanh(k)}{k}}\widehat{f}(k).
\end{equation}
Seeking a periodic traveling wave of \eqref{eqn:Whitham}, let 
\[
u(x,t)=\phi(z), \quad z=k(x-ct),
\]
where $c\in\mathbb{R}$ ($\neq 0$) is the wave speed, $k>0$ the wave number, and $\phi$ satisfies, by quadrature and Galilean invariance,
\begin{equation}\label{eqn:phi}
-c\phi+c_\text{ww}(k|\partial_z|;T)\phi+\phi^2=0. 
\end{equation}
We assume that $\phi$ is $2\pi$ periodic and even in the $z$ variable, so that $2\pi/k$ periodic in the $x$ variable. 
Notice
\begin{equation}\label{def:cT}
c_\text{ww}(k|\partial_z|;T)
\left\{\begin{matrix}\cos \\ \sin \end{matrix}\right\}(nz)
=c_\text{ww}(kn;T)\left\{\begin{matrix}\cos \\ \sin \end{matrix}\right\}(nz),\quad n\in\mathbb{Z}.
\end{equation}


For any $T\geq0$, $k>0$ and $c\in\mathbb{R}$, clearly, $\phi(z)=0$ solves \eqref{eqn:phi} and \eqref{def:cT}. Suppose that $T$ and $k$ are fixed. A necessary condition for nontrivial solutions to bifurcate from such trivial solution at some $c$ is that 
\[
\text{$c_\text{ww}(k|\partial_z|;T)\phi-c\phi=0$ admits a nontrivial solution},
\]
if and only if $c=c_\text{ww}(kn;T)$ for some $n\in\mathbb{N}$, by symmetry, whence 
\[
\text{$\cos(nz)\in\ker(c_\text{ww}(k|\partial_z|;T)-c_\text{ww}(kn;T))$ in some space of even functions}.
\] 

When $T=0$, $c_\text{ww}(\cdot\,;0)$ monotonically decreases to zero over $(0,\infty)$, whence for any $k>0$, $c_\text{ww}(kn_1;T)>c_\text{ww}(kn_2;T)$ whenever $n_1,n_2\in\mathbb{N}$ and $n_1>n_2$. Therefore, for any $k>0$ for any $n\in\mathbb{N}$, $\ker(c_\text{ww}(k|\partial_z|;0)-c_\text{ww}(kn;0))=\text{span}\{\cos(nz)\}$ in spaces of even functions. 

When $T\geq1/3$, $c_\text{ww}(\cdot\,;T)$ monotonically increases over $(0,\infty)$ and unbounded from above, whereby for any $k>0$ for any $n\in\mathbb{N}$, $\ker(c_\text{ww}(k|\partial_z|;T)-c_\text{ww}(kn;T))=\text{span}\{\cos(nz)\}$.

When $T<1/3$, to the contrary, $c_\text{ww}(kn_1;T)=c_\text{ww}(kn_2;T)$ for some $k>0$ for some $n_1,n_2\in\mathbb{N}$ and $n_1\neq n_2$, provided that $T=T(kn_1,kn_2)$, where
\begin{equation}\label{def:T(k1,k2)}
T(kn_1,kn_2)=\frac{1}{kn_1}\frac{1}{kn_2}
\frac{n_1\tanh(kn_2)-n_2\tanh(kn_1)}{n_1\tanh(kn_1)-n_2\tanh(kn_2)},
\end{equation}
and $\ker(c_\text{ww}(k|\partial_z|;T)-c)=\text{span}\{\cos(n_1z),\cos(n_2z)\}$, where $c=c_\text{ww}(kn_1;T)=c_\text{ww}(kn_2;T)$; otherwise, $\ker(c_\text{ww}(k|\partial_z|;T)-c_\text{ww}(kn;T))=\text{span}\{\cos(nz)\}$. 

Suppose that $T\geq0$, $k>0$, and $T\neq T(kn,kn')$ for any $n,n'\in\mathbb{N}$ and $n\neq n'$, particularly, either $T=0$ or $T\geq1/3$. We assume without loss of generality $n=1$. There exists a one-parameter curve of nontrivial, $2\pi$ periodic and even solutions of \eqref{eqn:phi} and \eqref{def:cT}, denoted by
\begin{equation}\label{def:loc;bifur}
(\phi(z;s),c(s)),\quad |s|\ll1,
\end{equation}
in some function space (see \cite{EW;peaking,EJ;WW}, for instance, for details), and 
\begin{equation}\label{eqn:|s|<<1}
\begin{aligned}
&\begin{aligned}
\phi(z;s)=s\cos(z)&+\frac{s^2}{2}\left(\frac{1}{c_\text{ww}(k;T)-c_\text{ww}(0;T)}
-\frac{1}{c_\text{ww}(k;T)-c_\text{ww}(2k;T)}\cos(2z)\right)\\
&+\frac{s^3}{2}\frac{1}{(c_\text{ww}(k;T)-c_\text{ww}(2k;T))(c_\text{ww}(k;T)-c_\text{ww}(3k;T))}\cos(3z)
+O(s^4), \end{aligned}\\
&c(s)=c_\text{ww}(k;T)+s^2\left(\frac{1}{c_\text{ww}(k;T)-c_\text{ww}(0;T)}
-\frac12\frac{1}{c_\text{ww}(k;T)-c_\text{ww}(2k;T)}\right)+O(s^4)
\end{aligned}
\end{equation}
as $|s|\to0$. Moreover, subject to a `bifurcation condition' (see \cite{EW;peaking,EJ;WW}, for instance, for details), \eqref{def:loc;bifur} extends to all $s\in[0,\infty)$. See \cite{EW;peaking,EJ;WW} and references therein for a rigorous proof. 

When $T=0$ and, without loss of generality, $k=1$, $(\phi(z;s_j),c(s_j))\to (\phi(z),c)$ as $j\to\infty$ for some $s_j\to \infty$ as $j\to\infty$ for some $\phi\in C^{1/2}([-\pi,\pi])\bigcap C^\infty([-\pi,0)\bigcup(0,\pi])$ and $0<c<\infty$ such~that 
\begin{equation}\label{def:stopping0}
\phi(0)=\frac{c}{2}.
\end{equation}
See \cite{EW;peaking} and references therein for a rigorous proof. Indeed, such a limiting solution enjoys $\phi(z)\sim\frac{c}{2}-\sqrt{\frac \pi 8}|z|^{1/2}$ as $|z|\to0$. 

When $T\geq 4/\pi^2$, so that the Fourier transform of $c_\text{ww}(k|\partial_z|;T)^{-1}$ is `completely monotone' \cite{EJ;WW}, and $k=1$, on the other hand,
\begin{enumerate}
\item[]$\|(\phi(\cdot;s), c(s))\|\to\infty$ as $s\to\infty$ in some\footnote{a H\"older-Zygmund space, for instance \cite{EJ;WW}} function space; or
\item[]$(\phi(z;s),c(s))$ is periodic in the $s$ variable.
\end{enumerate}
See \cite{EJ;WW}, for instance, for a rigorous proof. Our numerical findings suggest $\min_{z\in[-\pi,\pi]}\phi(z;s)\to -\infty$ 
as $s\to\infty$. But such a limiting scenario would be physically unrealistic, for the capillary-gravity Whitham equation models water waves in the finite depth. We say that a $2\pi$ periodic solution of \eqref{eqn:phi} and \eqref{def:cT} is admissible if 
\[
\phi(z)-\frac{1}{2\pi}\int^\pi_{-\pi}\phi(z)~dz>-1\quad\text{for all $z\in[-\pi,\pi]$},
\]
so that the fluid surface would not intersect the impermeable bed, and we stop numerical continuation once we reach a {\em limiting admissible solution}, for which 
\begin{equation}\label{def:stopping}
\min_{z\in[-\pi,\pi]}\phi(z)-\frac{1}{2\pi}\int^\pi_{-\pi} \phi(z)~dz=-1.
\end{equation} 
Section~\ref{sec:result} provides examples.

Suppose, to the contrary, that $T=T(kn_1,kn_2)$ (see \eqref{def:T(k1,k2)}) for some $k>0$ for some $n_1,n_2\in\mathbb{N}$ and $n_1<n_2$, so that $c_\text{ww}(kn_1;T)=c_\text{ww}(kn_2;T)=:c$. Suppose that $n_1$ does not divide $n_2$. There exists a two-parameter sheet of nontrivial, periodic and even solutions of \eqref{eqn:phi} and \eqref{def:cT}, and
\begin{equation}\label{eqn:s12}
\begin{aligned}
&\phi(z;s_1,s_2)=s_1\cos(n_1z)+s_2\cos(n_2z)+O((|s_1|+|s_2|)^2),\\
&c(s_1,s_2)=c+O((|s_1|+|s_2|)^2), \\
&k(s_1,s_2)=k+O((|s_1|+|s_2|)^2)
\end{aligned}
\end{equation}
for $|s_1|,|s_2|\ll1$. We emphasize that $k$ is a bifurcation parameter. Otherwise, $n_1$ divides $n_2$, and \eqref{eqn:s12} holds for $|s_1|,|s_2|\ll1$ and $s_2>s_0>0$ for some $s_0$. In other words, there cannot exist $2\pi/n_1$ periodic and `unimodal' waves, whose profile monotonically decreases from its single crest to the trough over the period. See \cite{EJ;WW}, for instance, for a rigorous proof.
The global continuation of \eqref{eqn:s12} and limiting configurations have not been well understood analytically, though, and here we address numerically. 

Turning the attention to the (nonlinear) orbital stability and instability of a periodic traveling wave of \eqref{eqn:Whitham}, notice that \eqref{eqn:Whitham} possesses three conservation laws:
\begin{equation}\label{def:EPM}
\begin{aligned}
E(u)=&\int \left(\frac12 uc_\text{ww}(|\partial_x|;T)u+\frac13u^3\right)~dx \qquad\text{(energy)},\\
P(u)=&\int \frac12 u^2~dx\qquad\text{(momentum)}, \\
M(u)=&\int u~dx \qquad\text{(mass)},
\end{aligned}
\end{equation} 
and so does the KdV equation with fractional dispersion: 
\begin{equation}\label{eqn:fKdV}
u_t+|\partial_x|^\alpha u_x+(u^2)_x=0,\quad 0<\alpha\leq2,
\quad\text{where}\quad \widehat{|\partial_x|^\alpha f}(k)=|k|^\alpha\widehat{f}(k),
\end{equation}
for which $E(u)=\int (\frac12u|\partial_x|^\alpha u+\frac13u^3)~dx$ rather than the first equation of \eqref{def:EPM}. We pause to remark that \eqref{eqn:Whitham} becomes \eqref{eqn:fKdV}, $\alpha=1/2$, for high frequency (see \eqref{eqn:fKdV1/2}), and \eqref{eqn:KdV} and \eqref{eqn:KdVT} compare with $\alpha=2$. A solitary wave of \eqref{eqn:fKdV} is a constrained energy minimizer and orbitally stable, provided that $\alpha>1/2$, if and only if $P_c>0$. See \cite{LPS;orbital} and references therein for a rigorous proof. See also \cite{HJ;orbital} for an analogous result for periodic traveling waves. It seems not unreasonable to expect that the orbital stability and instability of a periodic traveling wave of \eqref{eqn:Whitham} change likewise at a critical point of $P$ as a function of $c$ although, to the best of the authors' knowledge, there is no rigorous analysis of constrained energy minimization. Indeed, numerical evidence \cite{Kalisch;specVV} supports the conjecture when $T=0$ and $k=1$. Here we take matters further to $T\geq0$ and $k\geq1,\in\mathbb{N}$. Also we numerically elucidate the instability scenario when $T=0$ and $k=1$.

\section{Methodology}\label{sec:numerical}

We begin by numerically approximating $2\pi$ periodic and even solutions of \eqref{eqn:phi} and \eqref{def:cT} by means of a spectral collocation method \cite{JBoyd,GO;spec}. See \cite{Kalisch;specVV}, among others, for nonlinear dispersive equations of the form \eqref{eqn:Whitham} and \eqref{eqn:fKdV}. We define the collocation projection as a discrete cosine transform as
\begin{equation}\label{def:phiN}
\phi_N(z)=\sum_{n=0}^{N-1}\omega(n)\widehat{\phi}(n)\cos(nz), 
\end{equation}
where the discrete cosine coefficients are
\begin{equation}\label{def:phi(n)}
\widehat{\phi}(n)=\omega(n)\sum_{m=1}^N \phi_N(z_m)\cos(nz_m)
\end{equation}
and 
\[
\omega(n)=\begin{cases}
\sqrt{1/N} &\text{for $n=0$},\\
\sqrt{2/N} &\text{otherwise};
\end{cases}
\]
the collocation points are
\[
z_m=\pi\frac{2m-1}{2N},\quad m=1,2,\dots,N.
\]
Therefore
\[
\phi(z)\approx\phi_N(z)=\sum_{m=1}^N\sum_{n=0}^{N-1}\omega^2(n)\cos(nz_m)\cos(nz)\phi(z_m).
\]
We can compute \eqref{def:phiN} and \eqref{def:phi(n)} efficiently using a fast Fourier transform (FFT) \cite{GO;spec}. 
For $T\geq0$ and $k>0$, likewise, 
\begin{align*}
c_\text{ww}(k|\partial_z|;T)\phi(z)\approx (c_\text{ww}(k|\partial_z|;T)\phi)_N(z)
:=\sum_{\ell=1}^N\sum_{n=0}^{N-1}\omega^2(n)c_\text{ww}(kn;T)\cos(nz)\cos(nz_\ell)\phi_N(z_\ell), 
\end{align*}
and we can compute $(c_\text{ww}(k|\partial_z|;T)\phi)_N(z_m)$, $m=1,2,\dots,N$, via an FFT. 

Suppose that $T\geq0$ and $k>0$ are fixed and we take $N=1024$. We numerically solve 
\begin{equation}\label{eqn:F=0}
-c\phi_N(z_m)+(c_\text{ww}(k|\partial_z|;T)\phi)_N(z_m)+\phi_N(z_m)^2=0,\quad m=1,2,\dots,N,
\end{equation}
by means of Newton's method. We parametrically continue the numerical solution over $c$ by means of a pseudo-arclength continuation method \cite{Doedel} (see \cite{Kalisch;specVV}, among others, for nonlinear dispersive equations of the form \eqref{eqn:Whitham} and \eqref{eqn:fKdV}), for which the (pseudo-)arclength of a solution branch is the continuation parameter, whereas a parameter continuation method would use $c$ as the continuation parameter and vary it sequentially. The pseudo-arclength continuation method can successfully bypass a turning point of $c$, at which a parameter continuation method fails because the Jacobian of \eqref{eqn:F=0} becomes singular, 
whence Newton's method diverges. See \cite{Kalisch;physD} for more discussion. 

We say that Newton's method converges if the residual
\[
\sqrt{\sum_{m=1}^N \big(-c\phi_N(z_m)+(c_\text{ww}(k|\partial_z|;T)\phi)_N(z_m)+\phi_N(z_m)^2\big)^2}\leq 10^{-10}
\]
(see \eqref{eqn:F=0}), and this is achieved provided that an initial guess is sufficiently close to a true solution of \eqref{eqn:phi} and \eqref{def:cT}. To this end, we take a small amplitude cosine function and $c\approx c_\text{ww}(k;T)$ as an initial guess for the local bifurcation at $\phi=0$ and $c=c_\text{ww}(k;T)$, or \eqref{eqn:|s|<<1} so long as it makes sense. We have also solved \eqref{eqn:F=0} using a (Jacobian-free) Newton--Krylov method \cite{Kelley_nsoli} with absolute and relative tolerance of $10^{-10}$. The results correctly match those obtained from Newton's method, corroborating our numerical scheme. 

We have run a pseudo-arclength continuation code (see \cite{Jennie_KM}, for instance, for the applicability of the code used herein) with a fixed arclength stepsize, to numerically locate and trace solution branches. We have additionally run AUTO \cite{AUTO}, a robust numerical continuation and bifurcation software, to corroborate the results. All the results in Section~\ref{sec:result} have been obtained by employing AUTO with strict tolerances (\verb|EPSL = 1e-12| and \verb|EPSU=1e-12| in the AUTO's constants file). AUTO has the advantage, among many others, of making variable arclength stepsize adaptation (using the option \verb|IADS=1|) and of detecting branch points (using the option \verb|ISP=2| for \textit{all} special points). For the former, provided with minimum and maximum allowed absolute values of the pseudo-arclength stepsize, AUTO adjusts the arclength stepsize to be used during the continuation process.

Let $\phi$ and $c$ numerically approximate a $2\pi/k$ periodic traveling wave of \eqref{eqn:Whitham}, and we turn to numerically experimenting with its (nonlinear) orbital stability and instability. After a $2048$-point (full) Fourier spectral discretization in $x\in[-\pi,\pi]$, \eqref{eqn:Whitham} leads to
\begin{equation}\label{eqn:u}
u_t=\mathbf{L}u+\mathbf{N}(u),\quad\text{where}\quad
\mathbf{L}=-c_\text{ww}(|\partial_x|;T)\partial_x\quad\text{and}\quad\mathbf{N}(u)=-(u^2)_x,
\end{equation}
and we numerically solve \eqref{eqn:u}, subject to
\begin{equation}\label{eqn:IC}
u(x,0)=\phi(kx)+10^{-3}\max_{x\in[-\pi,\pi]}|\phi(kx)|\,U(-1,1),
\end{equation}
by means of an integrating factor (IF) method (see \cite{KassamTrefethen}, for instance, and references therein), where $U(-1,1)$ is a uniform random distribution. In other words, at the initial time, we perturb $\phi$ by uniformly distributed random noise of small amplitude, depending on the amplitude\footnote{We treat $\|\phi\|_{L^\infty}$ as the amplitude of $\phi$ although $\phi$ is not of mean zero, because the difference is insignificant.} of $\phi$. We pause to remark that the well-posedness for the Cauchy problem of \eqref{eqn:Whitham} can be rigorously established at least for short time in the usual manner via a compactness argument. 

Following the IF method, let $v=e^{-\mathbf{L}t}u$, so that \eqref{eqn:u} becomes
\begin{equation}\label{eqn:v}
v_{t}=e^{-\mathbf{L}t}\mathbf{N}(e^{\mathbf{L}t}v).
\end{equation}
This ameliorates the stiffness of \eqref{eqn:u}. Notice that \eqref{eqn:v} is of diagonal form, rather than matrix, in the Fourier space. We advance \eqref{eqn:v} forward in time by means of a fourth-order four-stage Runge--Kutta (RK4) method \cite{HW1} with a fixed time stepsize $\Delta t$:  
\[ v_{n+1}=v_n+\frac{a+2b+2c+d}{6},\]
where $v_n=v(t_n)$ and $t_n=n\Delta t$, $n=0,1,2,\dots$, 
\begin{align*}
a&=f(v_n,t_n)\Delta t,
&b&=f(v_n+a/2,t_n+\Delta t/2)\Delta t,\\
c&=f(v_n+b/2,t_n+\Delta t/2)\Delta t,
&d&=f(v_n+c,t_n+\Delta t)\Delta t
\end{align*}
and $f(v,t)=e^{-\mathbf{L}t}\mathbf{N}(e^{\mathbf{L}t}v)$ (see \eqref{eqn:v}). 
We take $\Delta t=10^{-4}\times2\pi/c$, where $c$ is the wave speed of $\phi$. While such value of $\Delta t$ ensures numerical stability, some computations, depending on $c$, require smaller\footnote{In most cases, $10^{-4}\times 2\pi/c =O(10^{-4})$ but in Figure~\ref{fig6}(b), for instance, $c=0.36$, whence $10^{-4}\times2\pi/c \approx 0.00174533$, whereas $10^{-5}\times 2\pi/c =O(10^{-4})$.} values, for instance, $\Delta t=10^{-4}\times \pi/c$ or $10^{-5}\times2\pi/c$. At each time step, we remove aliasing errors by applying the so-called $3/2$-rule so that the Fourier coefficients well decay for high frequencies (see \cite{JBoyd}, for instance). 

We assess the fidelity of our numerical scheme by monitoring \eqref{def:EPM} for numerical solutions. For unperturbed solutions, $E(t)$, $P(t), M(t)$ have all been conserved to machine precision, whereas for (randomly) perturbed ones, $E(t)$ and $P(t)$ to the order of $10^{-7}$ and $M(t)$ to machine precision. We have also corroborated our numerical results using two other methods: a higher-order Runge--Kutta method, such as the Runge--Kutta--Fehlberg (RKF) method \cite{HW1} with time stepsize adaptation and a two-stage (and, thus, fourth-order) Gauss--Legendre implicit\footnote{An implicit method, by construction, enjoys wider regions of absolute stability. This allows us to assess all our results, while avoiding numerical instability that might be observed in explicit methods such as RK4 and RKF methods. However, the IRK4 method requires a fixed point iteration (in the form of Newton's method) at each time step.} Runge--Kutta (IRK4) method \cite{HW2}. 
For stable solutions, the results obtained from the RK4 method correctly match those from the RKF and IRK4 methods. However, and for unstable solutions, the instability manifests at slightly different times for the same initial conditions. This is somewhat expected because the local truncation error (LTE) of each method and the convergence error of the IRK4 method act as perturbations. Recall that the RK4 and IRK4 methods have an LTE of the order of $(\Delta t)^{4}$ whereas the RKF method has $(\Delta t)^{6}$.

\section{Results}\label{sec:result}

\begin{figure}[htp]
\begin{center}
\includegraphics[height=.21\textheight, angle =0]{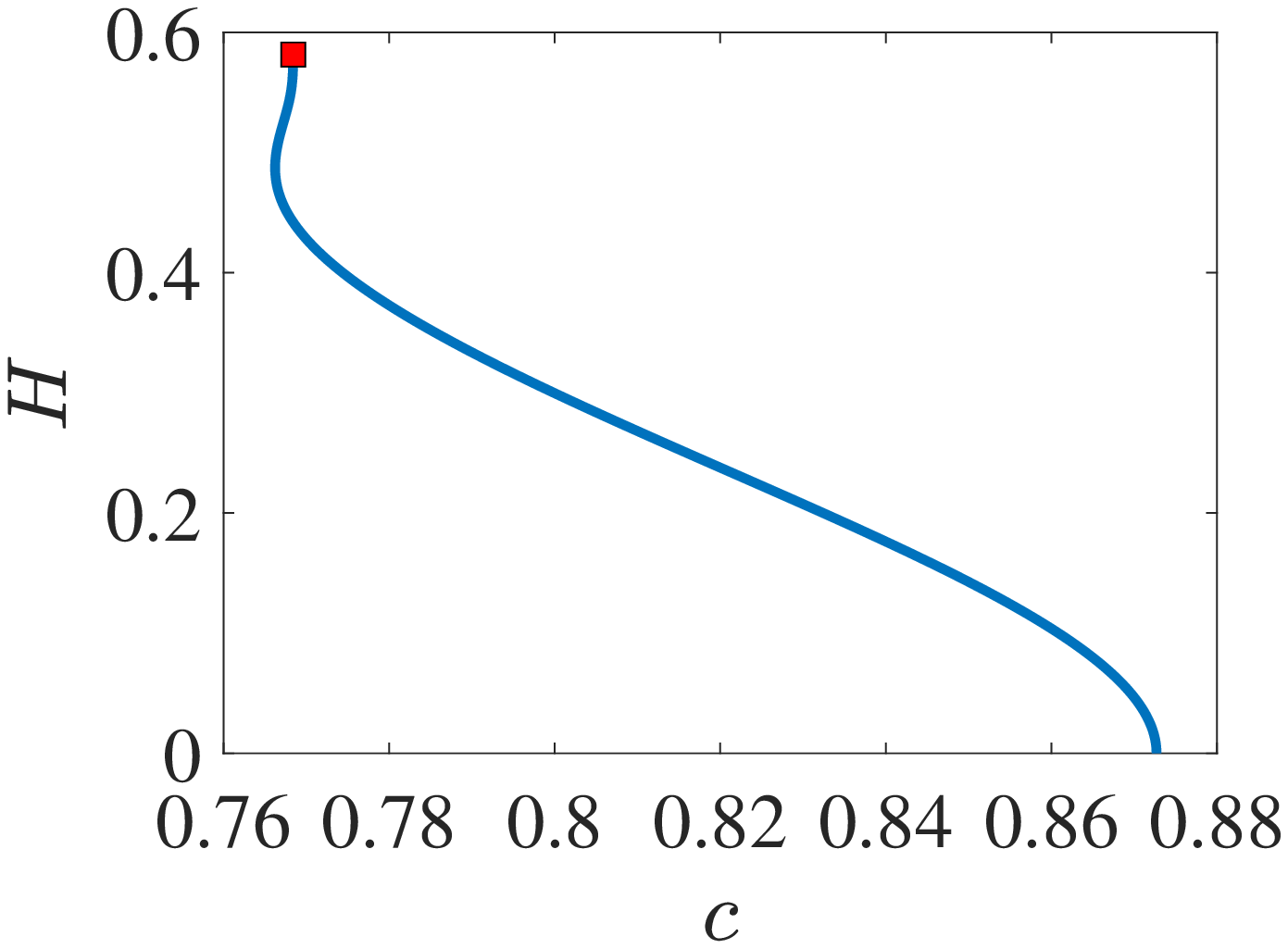}
\includegraphics[height=.21\textheight, angle =0]{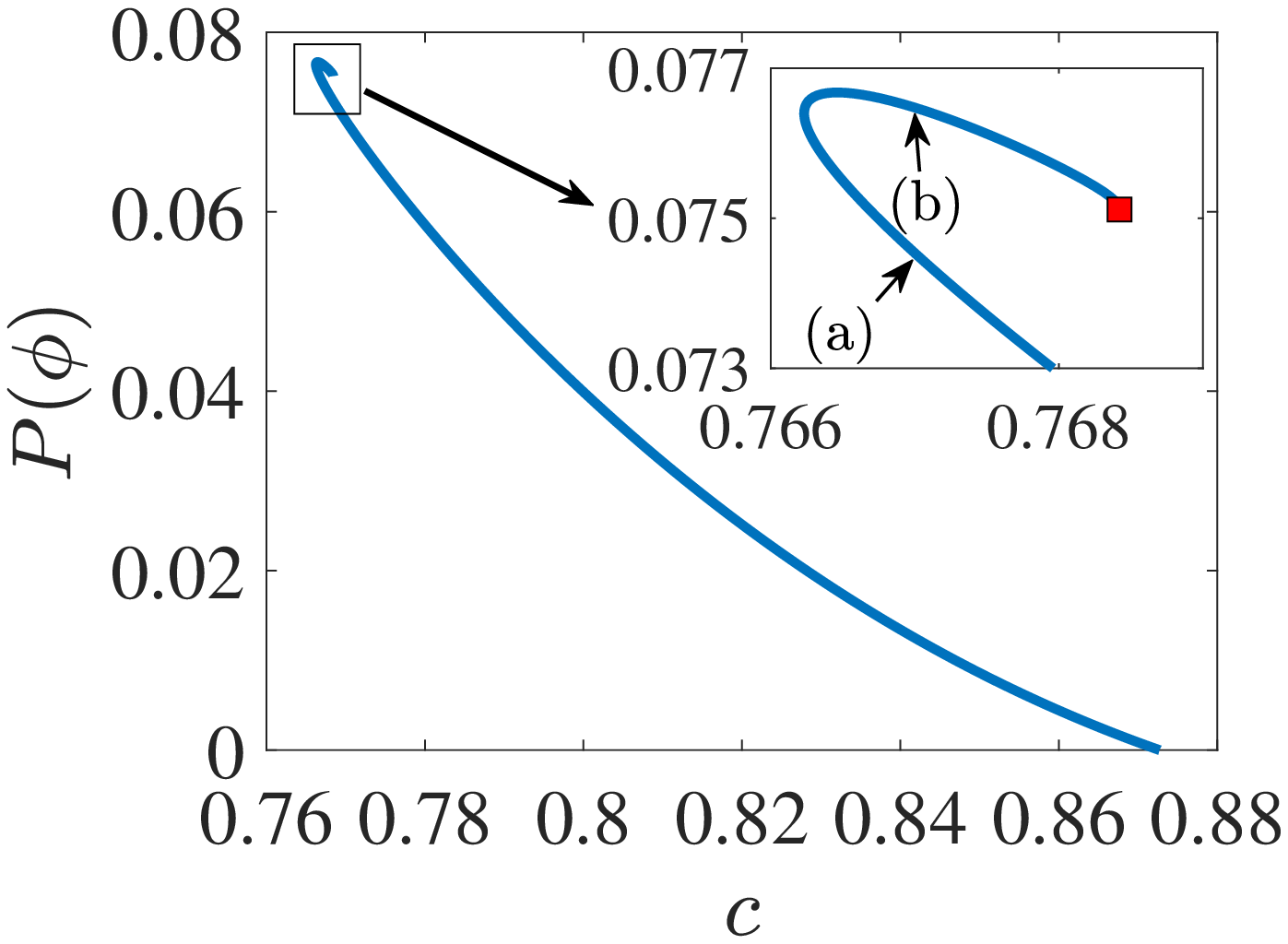}
\end{center}
\vspace*{-11pt}
\caption{$T=0$, $k=1$. Wave height (left, see \eqref{def:H}) and momentum (right, see \eqref{def:P}) vs. wave speed. The red square corresponds to the limiting solution (see \eqref{def:stopping0}), for which $c\approx 0.76842127$. See Figure~\ref{fig2} for the profiles at the points labelled with (a) and (b) in the inset of the right panel.}
\label{fig1}
\end{figure}

We begin by taking $T=0$ and $k=1$. Figure~\ref{fig1} shows the wave height
\begin{equation}\label{def:H}
H=\max_{z\in[-\pi,\pi]}\phi(z)-\min_{z\in[-\pi,\pi]}\phi(z)
\end{equation}
and the momentum
\begin{equation}\label{def:P}
P=\frac12\int^\pi_{-\pi}\phi(z)^2~dz 
\end{equation}
from our numerical continuation of $2\pi$ periodic and even solutions of \eqref{eqn:phi} and \eqref{def:cT}. The result agrees, qualitatively, with \cite[Figure~4]{Kalisch;specVV} and others. 
We find that $H$ monotonically increases to $\approx 0.58156621$, highlighted with the red square, for which \eqref{def:stopping0} holds, whereas $P$ increases and then decreases, making one turning point. 
Also we find that the crest becomes sharper and the trough flatter as $H$ increases. The limiting solution must possess a cusp at the crest \cite{EW;peaking}. 

\begin{figure}[htp]
\begin{center}
\includegraphics[height=.21\textheight, angle =0]{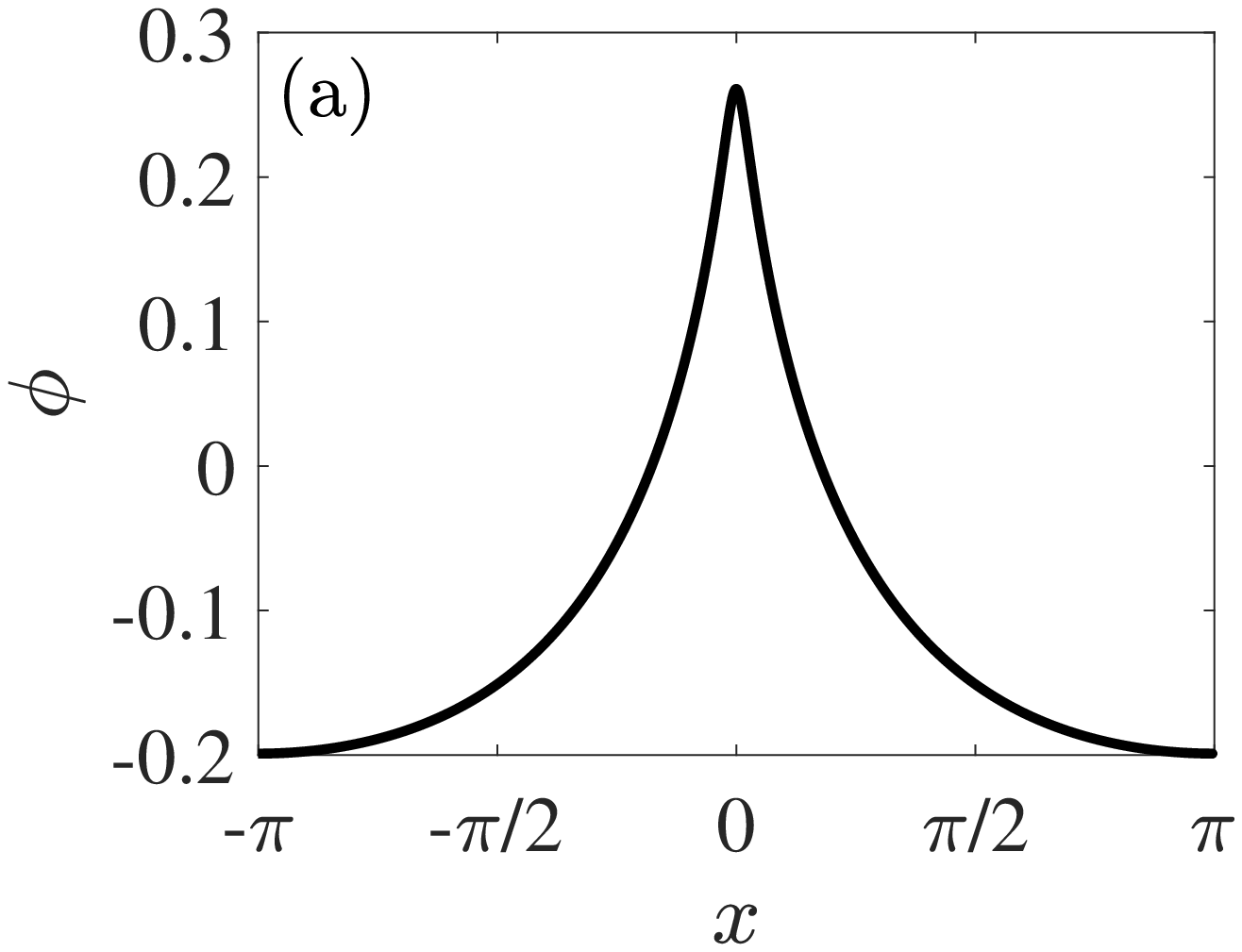}
\includegraphics[height=.21\textheight, angle =0]{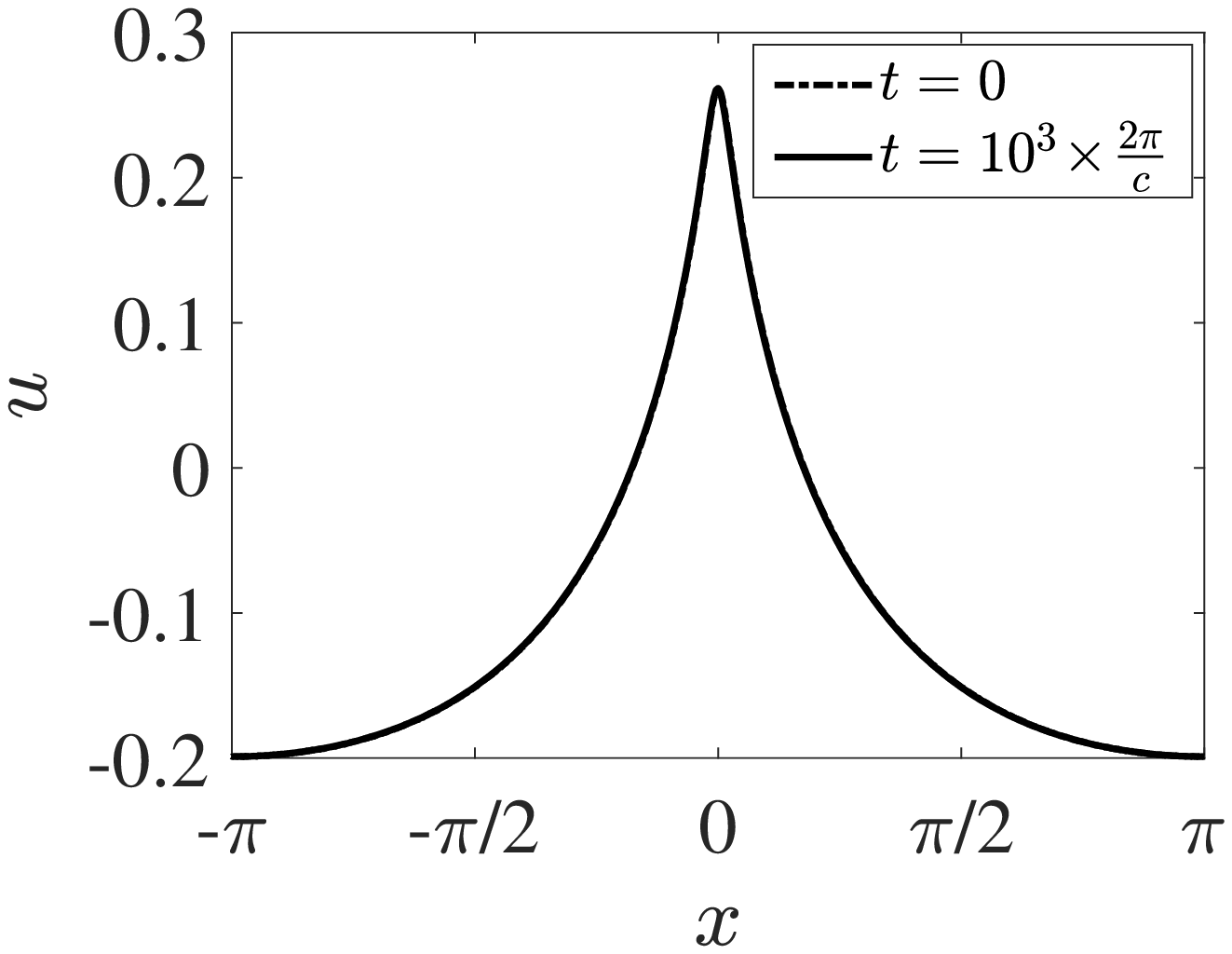}
\includegraphics[height=.21\textheight, angle =0]{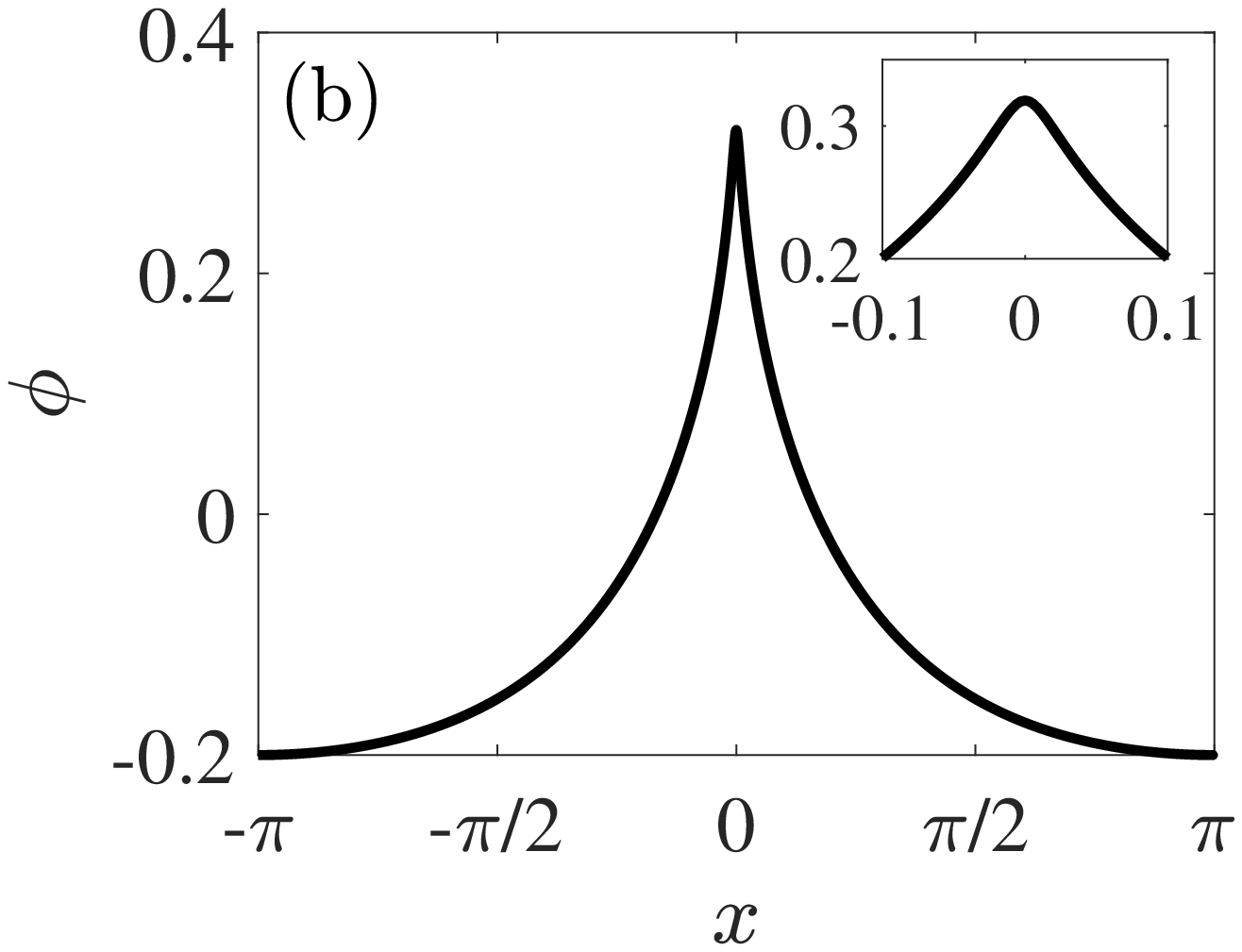}
\includegraphics[height=.21\textheight, angle =0]{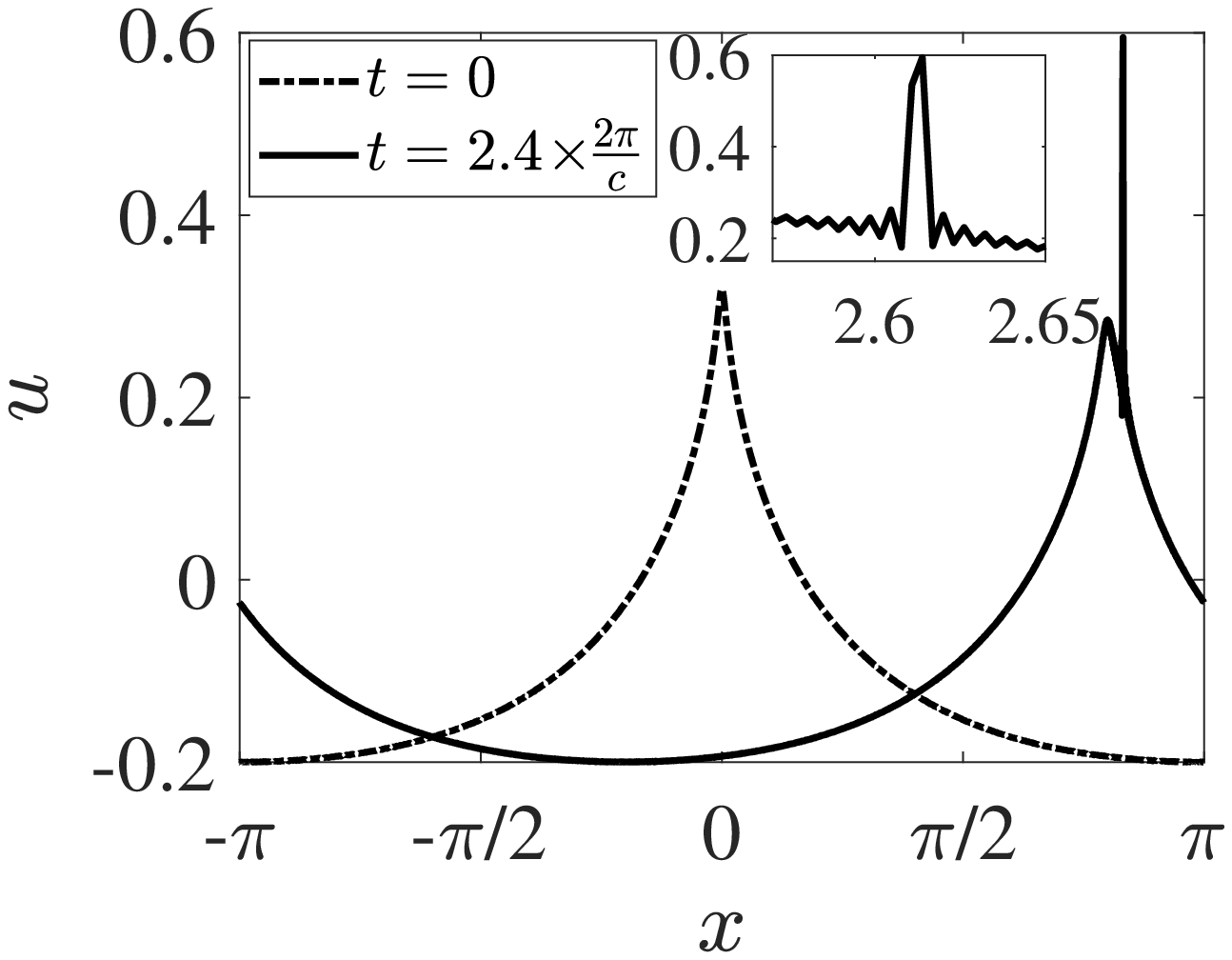}
\end{center}
\vspace*{-11pt}
\caption{
$T=0$, $k=1$. Left column: almost limiting waves at the points labelled with (a) and (b) in the inset of the right panel of Figure~\ref{fig1}, prior to (a) and past (b) the turning point of $P$, for which $c=0.767$. Right column: profiles perturbed by uniform random distributions of amplitudes $10^{-3}\times\|\phi\|_{L^\infty}$ at $t=0$ (dash-dotted), and of the solutions of \eqref{eqn:Whitham} at later instants of time (solid, see the legends), after translation of the $x$ axis (a). The insets in the bottom panels are close-ups near the crests.}
\label{fig2}
\end{figure}

The left column of Figure~\ref{fig2} shows almost limiting waves. The inset is a close-up near the crest, emphasizing smoothness. The right column shows the profiles of \eqref{eqn:IC}, namely periodic traveling waves of \eqref{eqn:Whitham} perturbed by uniformly distributed random noise of small amplitudes at $t=0$ (dash-dotted), and of the solutions of \eqref{eqn:Whitham} at later instants of time (solid). Wave (a), prior to the turning point of $P$, remains unchanged for $10^3$ time periods, after translation of the $x$ axis, implying orbital stability, whereas the inset reveals that wave (b), past the turning point, suffers from crest instability. Indeed, our numerical experiments point to transition from stability to instability at the turning point of $P$. But there is numerical evidence \cite{SKCK;MI} that waves (a) and (b) are (spectrally) modulationally unstable. 

\begin{figure}[htp]
\begin{center}
\includegraphics[height=.21\textheight, angle =0]{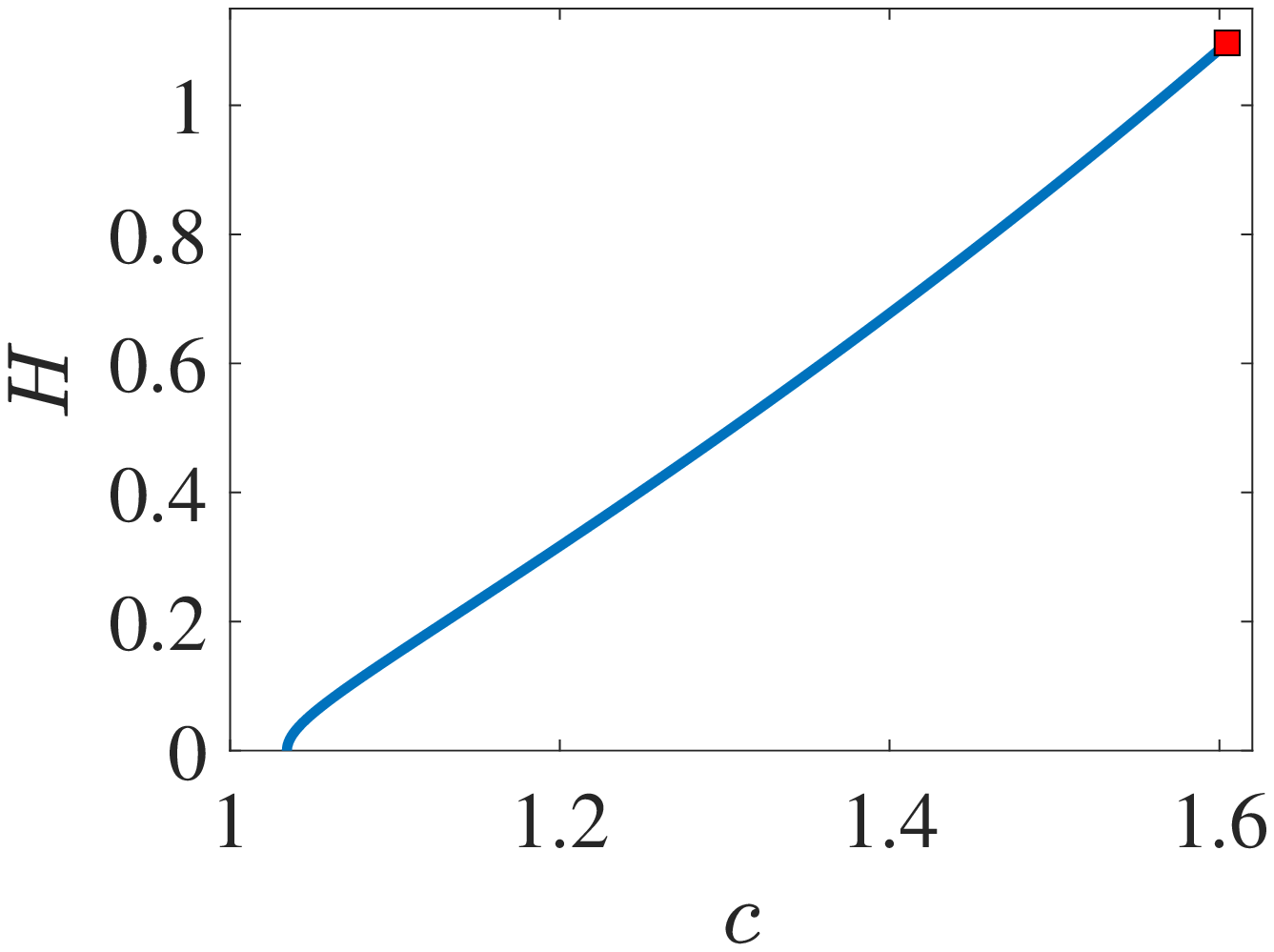}
\includegraphics[height=.21\textheight, angle =0]{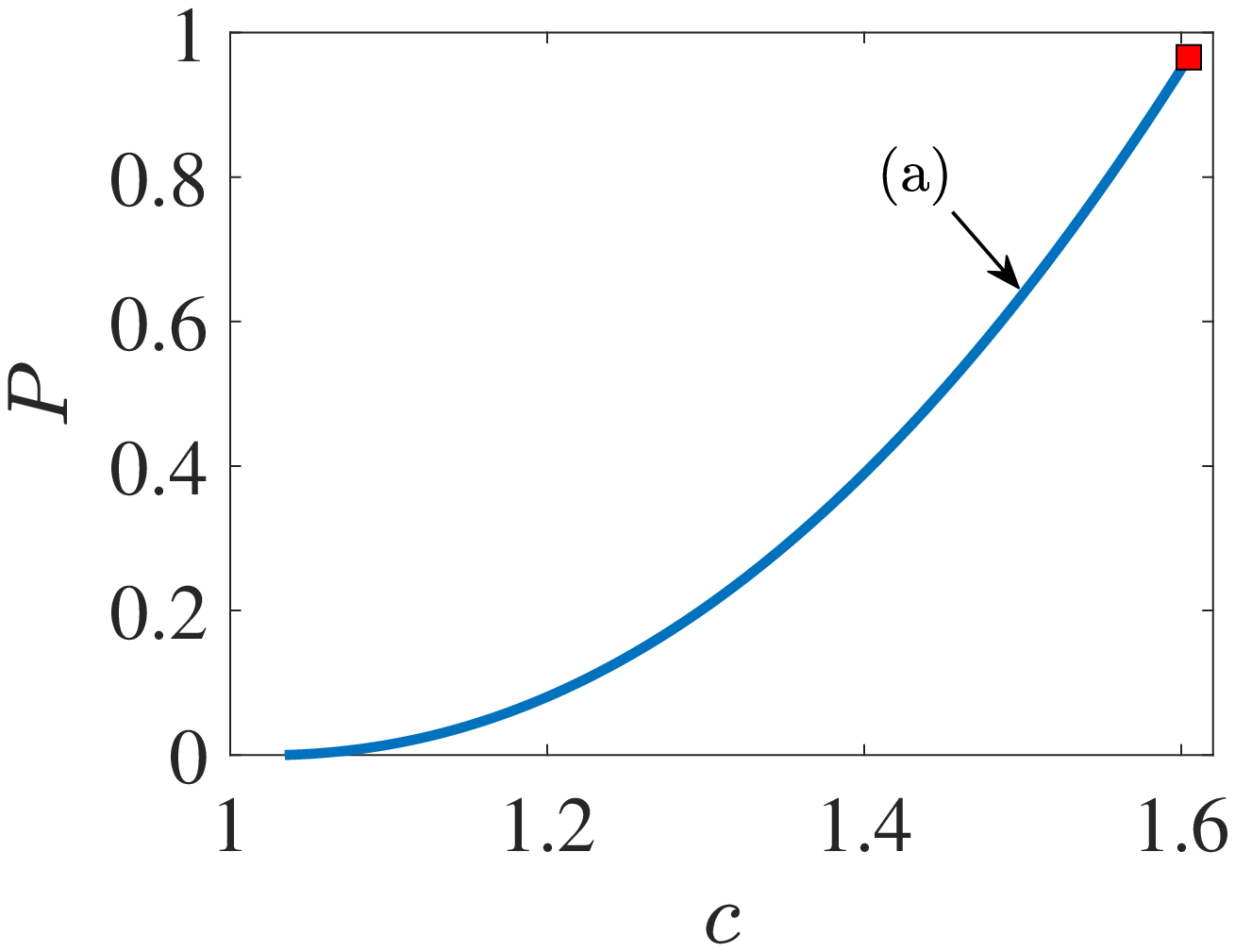}
\end{center}
\vspace*{-11pt}
\caption{
$T=4/\pi^2$, $k=1$. $H$ (left) and $P$ (right) vs. $c$. The red square corresponds to the limiting admissible solution (see \eqref{def:stopping}), for which $c\approx 1.6047287696$. See Figure~\ref{fig3a} for the profile at the point labelled with (a).}
\label{fig3}
\end{figure}

When $T=4/\pi^2$ and $k=1$, Figure~\ref{fig3} shows $H$ and $P$ versus $c$. Our numerical findings suggest that $H,P\to\infty$ monotonically. Also $\min_{z\in[-\pi,\pi]}\phi(z)\to -\infty$. The solution branch discontinues once \eqref{def:stopping} holds, highlighted with the red square, though, because the solution would be physically unrealistic, for the capillary-gravity Whitham equation models water waves in the finite depth. Our numerical continuation works well past the limiting admissible solution, nevertheless. We find that the crest becomes wider and flatter, the trough narrower and more peaked, as $H$ increases. But all solutions must be smooth \cite{EJ;WW}. 

\begin{figure}[htp]
\begin{center}
\includegraphics[height=.21\textheight, angle =0]{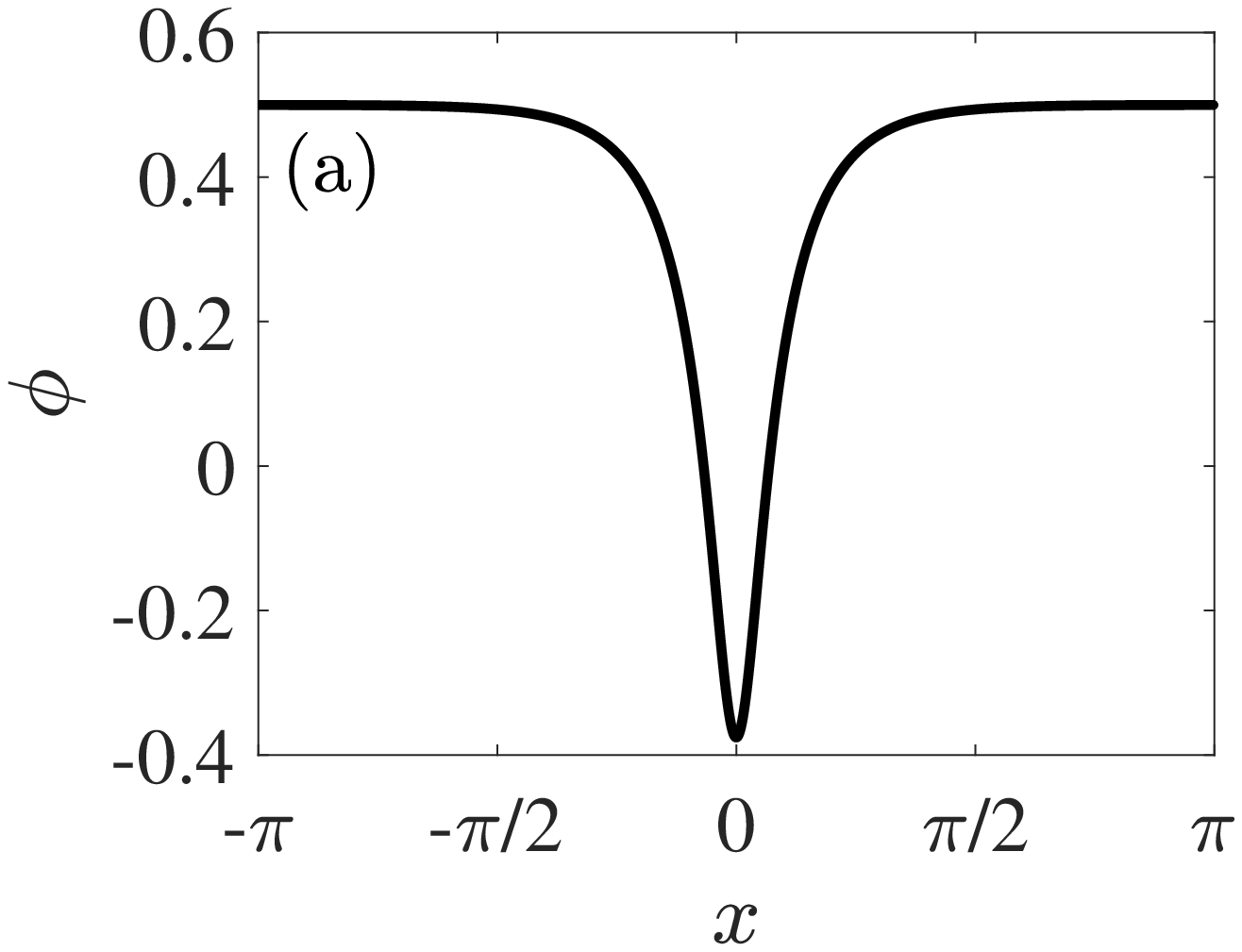}
\includegraphics[height=.21\textheight, angle =0]{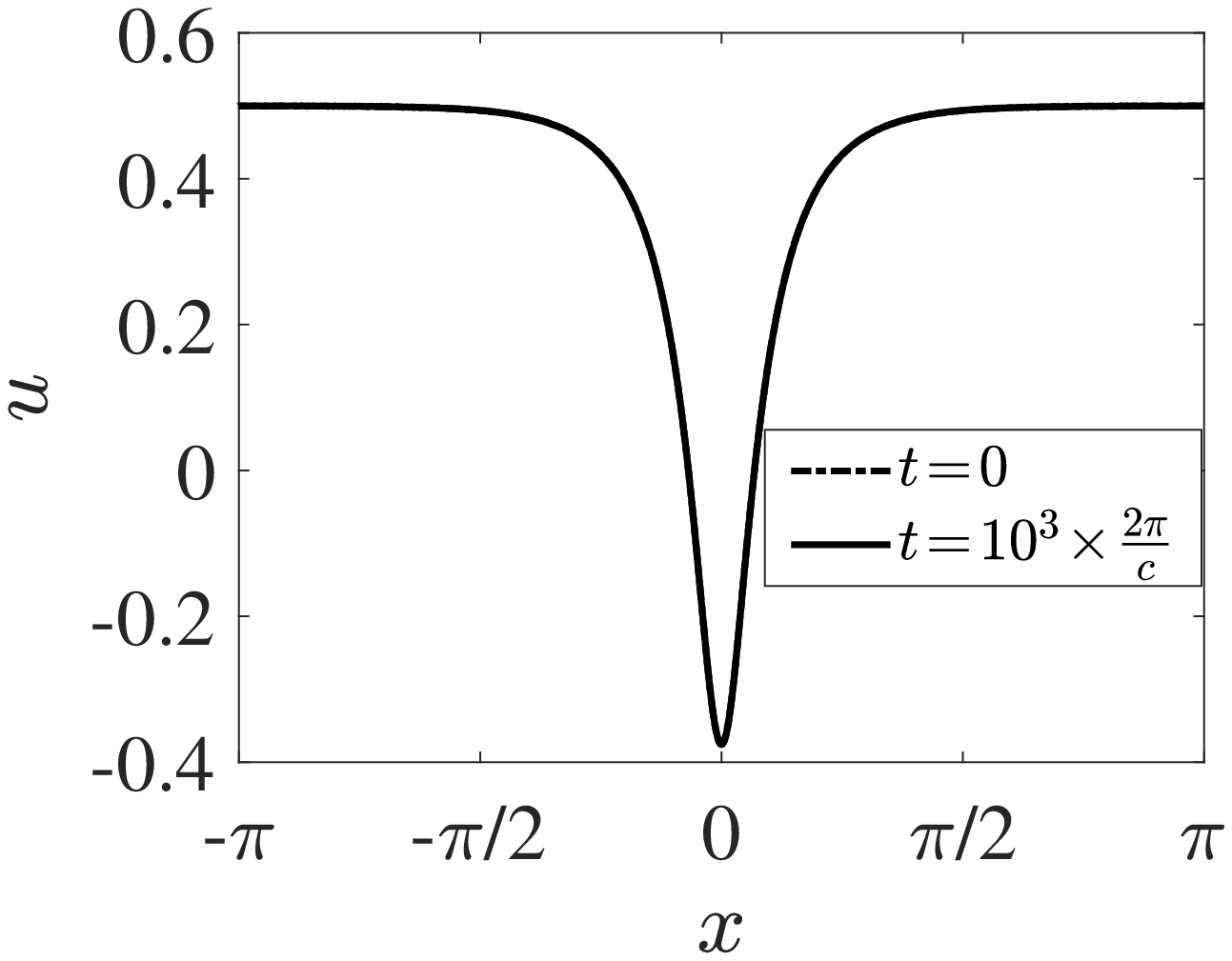}
\end{center}
\vspace*{-11pt}
\caption{
$T=4/\pi^2$, $k=1$. On the left, the profile at the point labelled with (a) in Figure~\ref{fig3}, for which $c=1.5$, and on the right, the profiles perturbed by uniformly distributed random noise of amplitude $10^{-3}\times\|\phi\|_{L^\infty}$ at $t=0$ (dash-dotted) and of the solution of \eqref{eqn:Whitham} at $t=10^3\times 2\pi/c$ (solid), after translation of the $x$ axis.}
\label{fig3a}
\end{figure}

The left panel of Figure~\ref{fig3a} shows an example wave. The right panel shows the profiles perturbed by small random noise at $t=0$ and of the solution of \eqref{eqn:Whitham} after $10^3$ time periods, after translation of the $x$ axis. Our numerical experiments suggest orbital stability for all wave height. But there is numerical evidence of modulational instability when $H$ is large \cite{carter;fluids}. 

Numerical evidence \cite{Kalisch;physD,carter;fluids} points to qualitatively the same results whenever $T\geq1/3$. 

\begin{figure}[htp]
\begin{center}
\includegraphics[height=.26\textheight, angle =0]{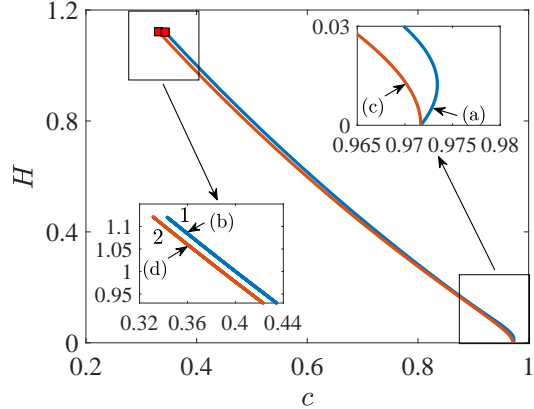}
\end{center}
\vspace*{-11pt}
\caption{
$T=T(1.2)$ (see \eqref{def:T(k1,k2)}). $H$ vs. $c$ for $k=1$ (blue) and $k=2$ (orange). The red squares correspond to the limiting admissible solutions for the $k=1$ and $k=2$ branches, for which $c\approx 0.3429722287$ and $c\approx 0.33120532402$, respectively. Insets are closeups near the beginning and end of numerical continuations. See Figure~\ref{fig5} for the profiles at the points labelled with (a) through (d) in the insets.}
\label{fig4}
\end{figure}

\begin{figure}[htp]
\begin{center}
\includegraphics[height=.21\textheight, angle =0]{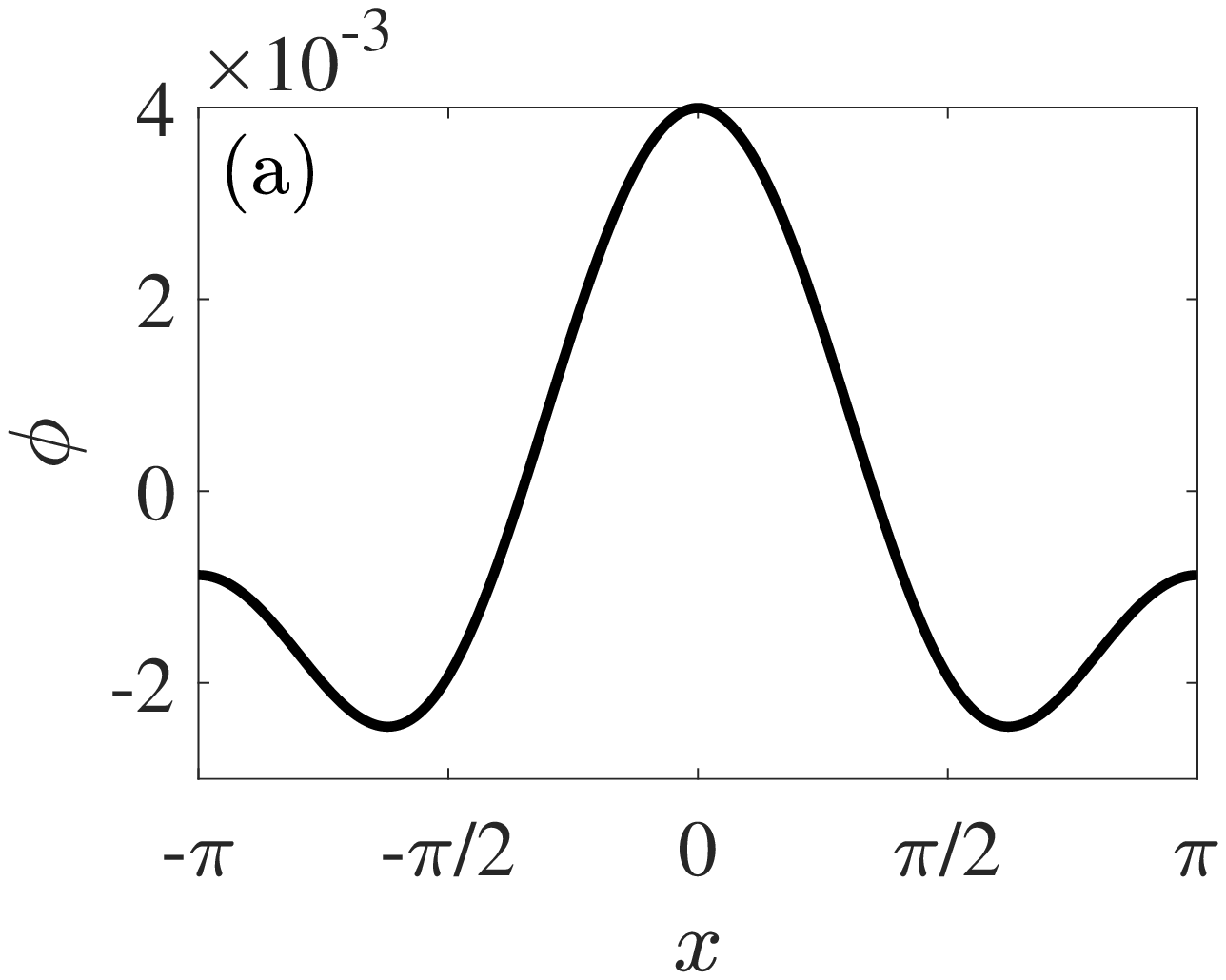}
\includegraphics[height=.21\textheight, angle =0]{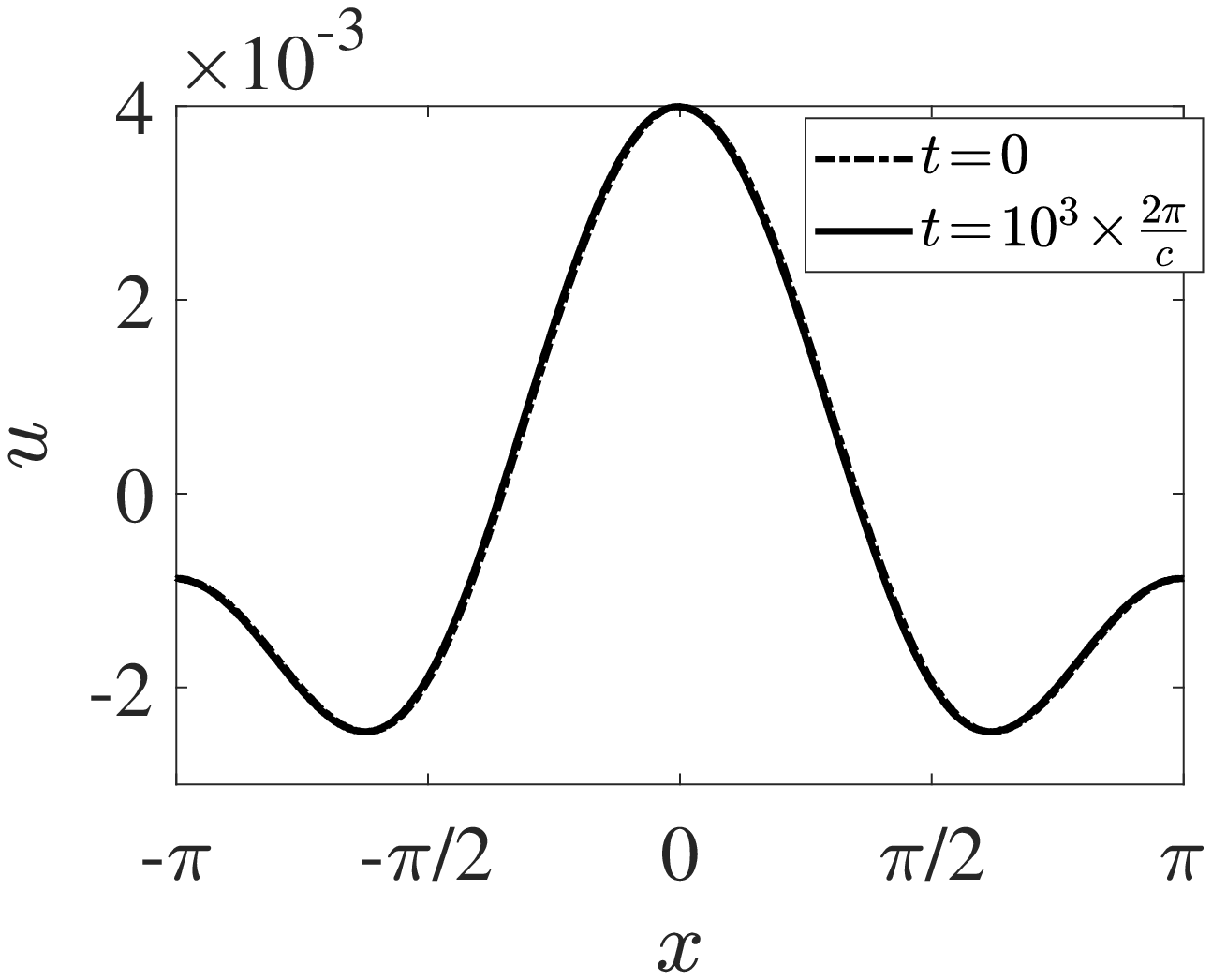}\\
\includegraphics[height=.21\textheight, angle =0]{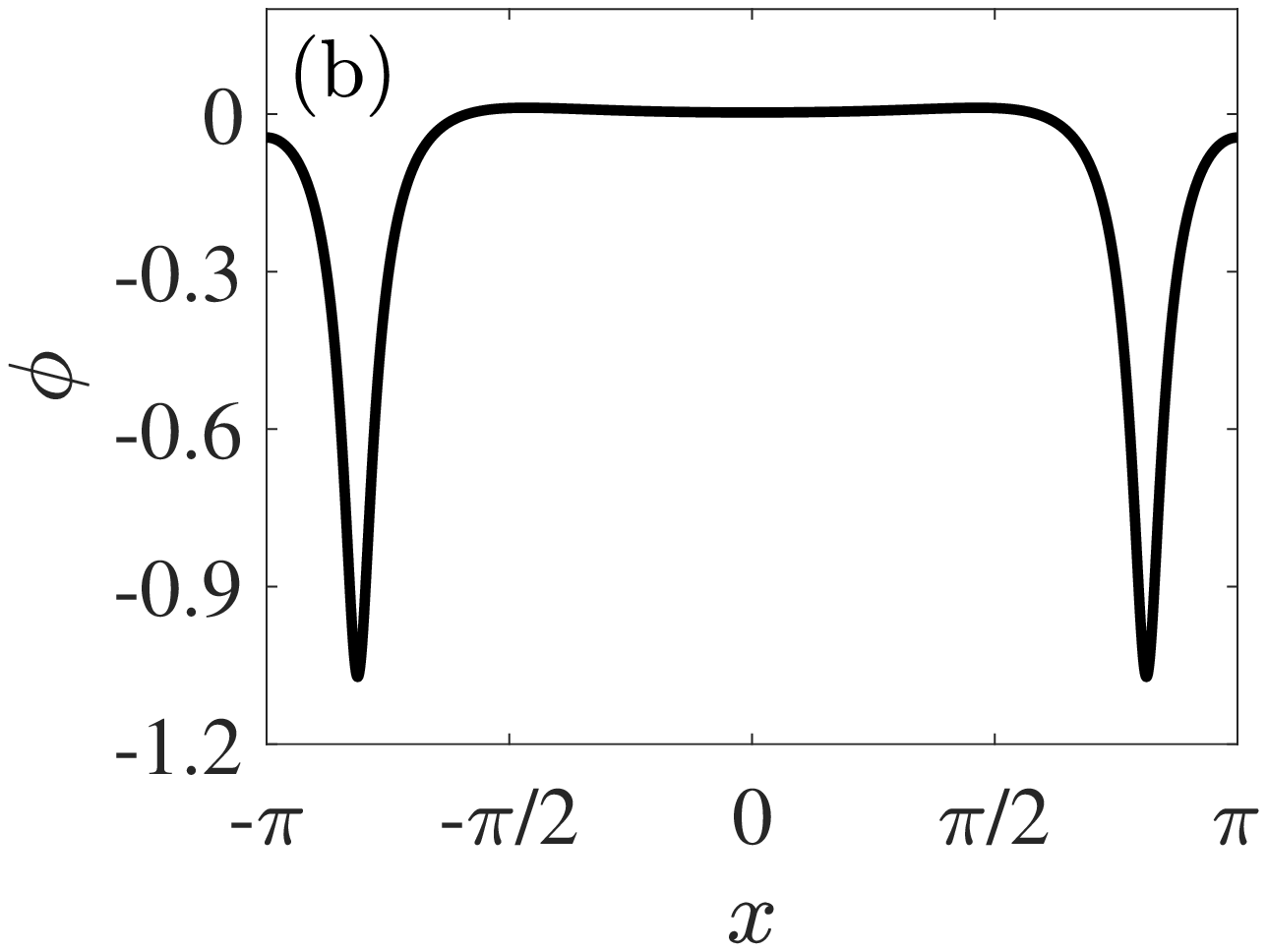}
\includegraphics[height=.21\textheight, angle =0]{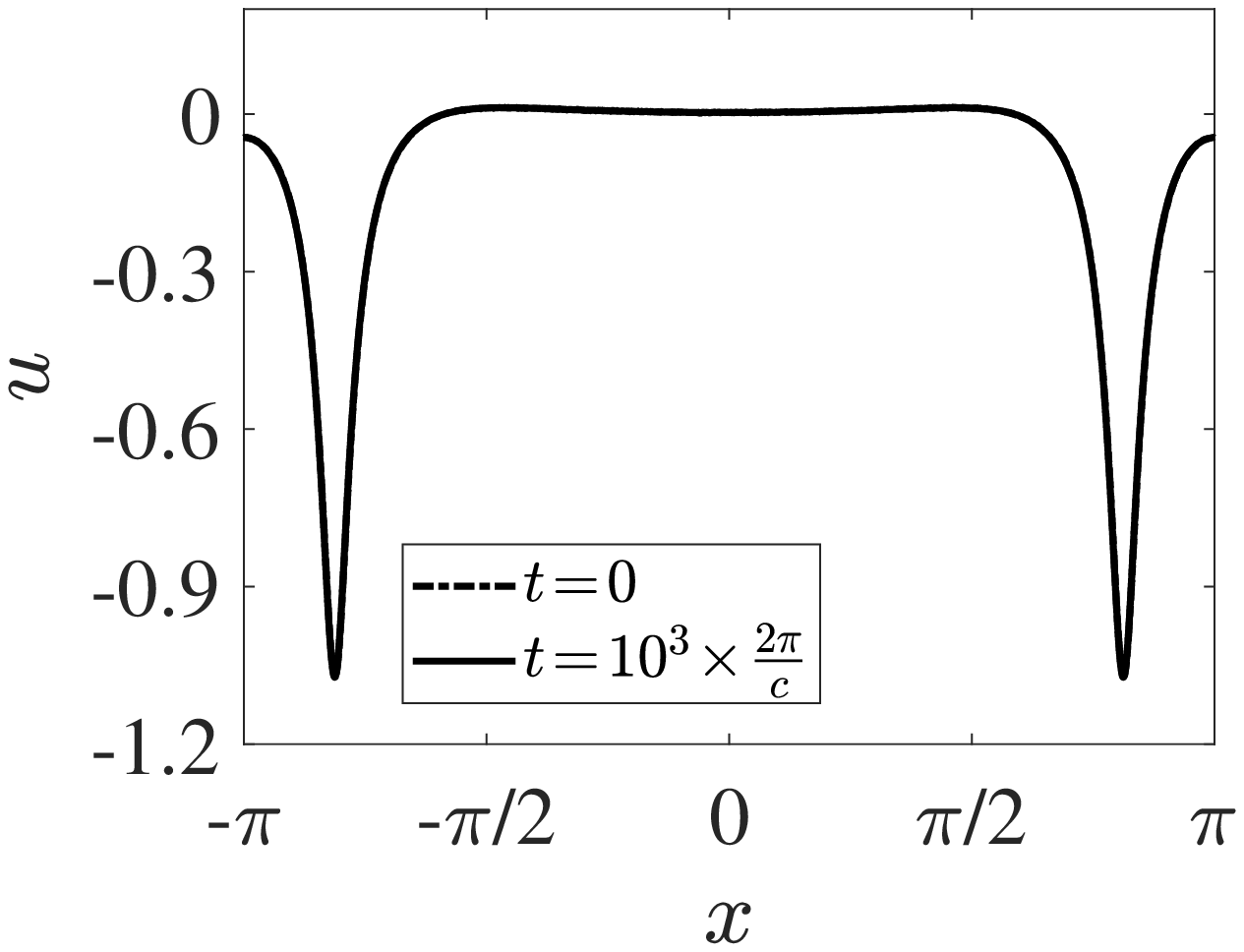}\\
\includegraphics[height=.21\textheight, angle =0]{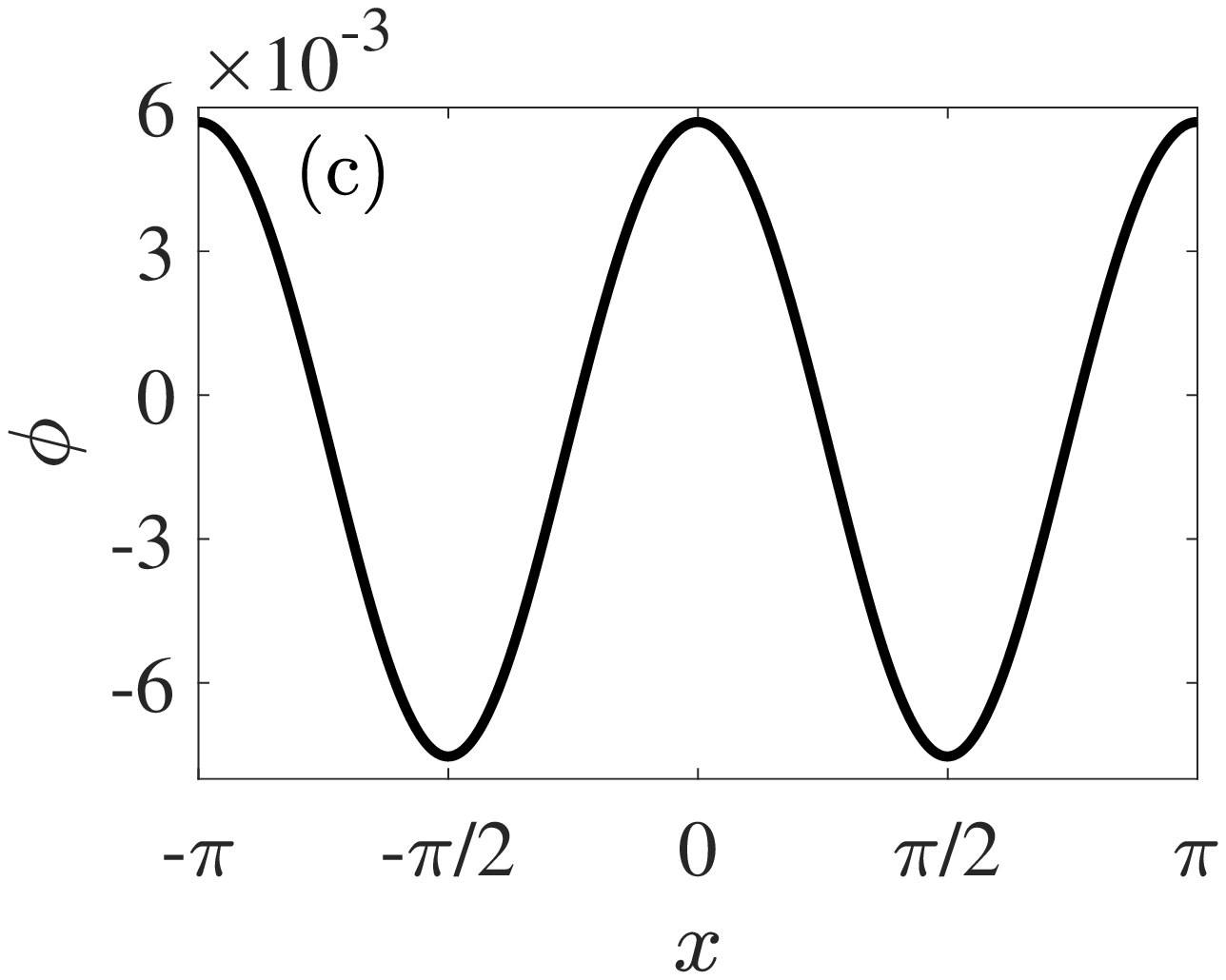}
\includegraphics[height=.21\textheight, angle =0]{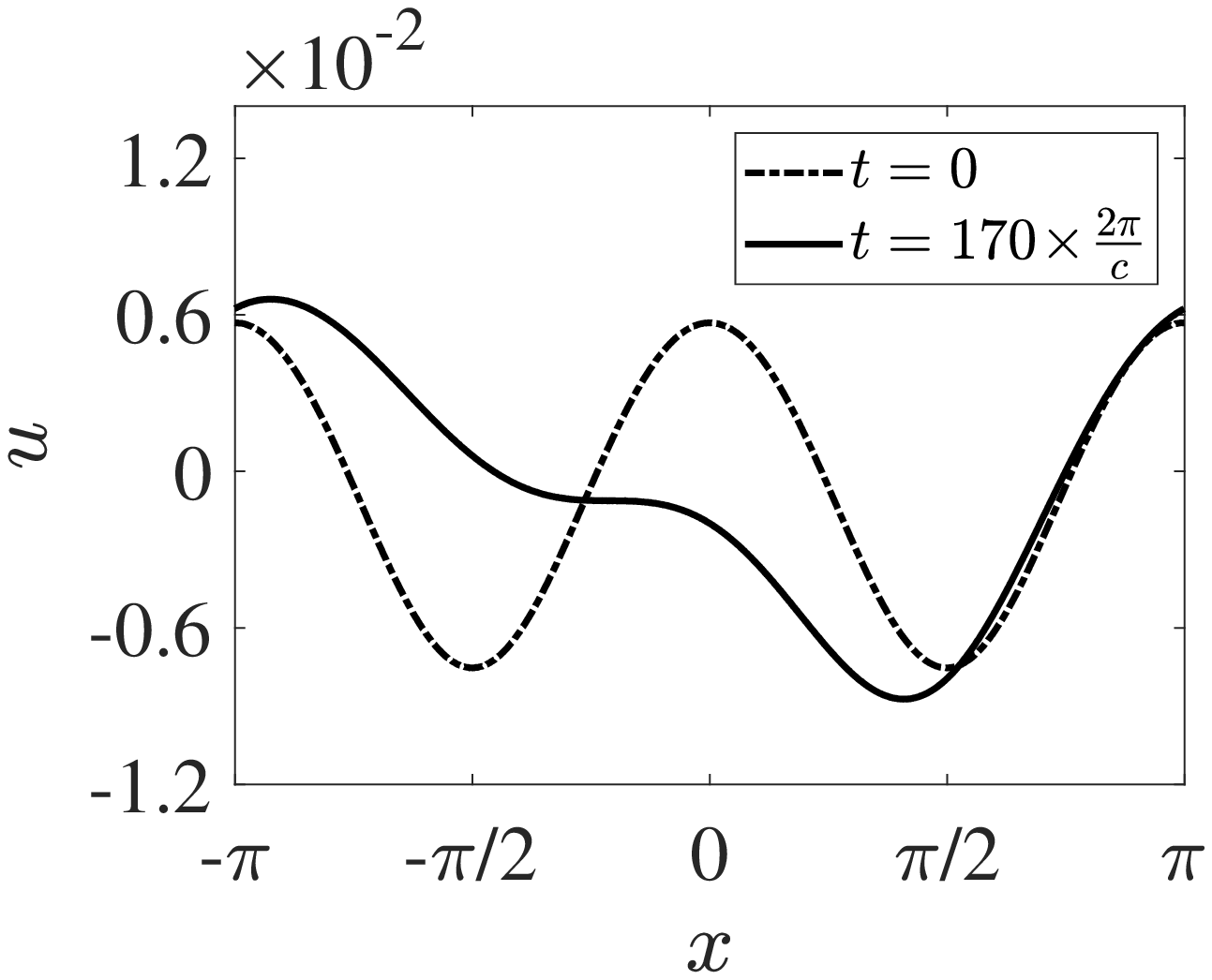}\\
\includegraphics[height=.21\textheight, angle =0]{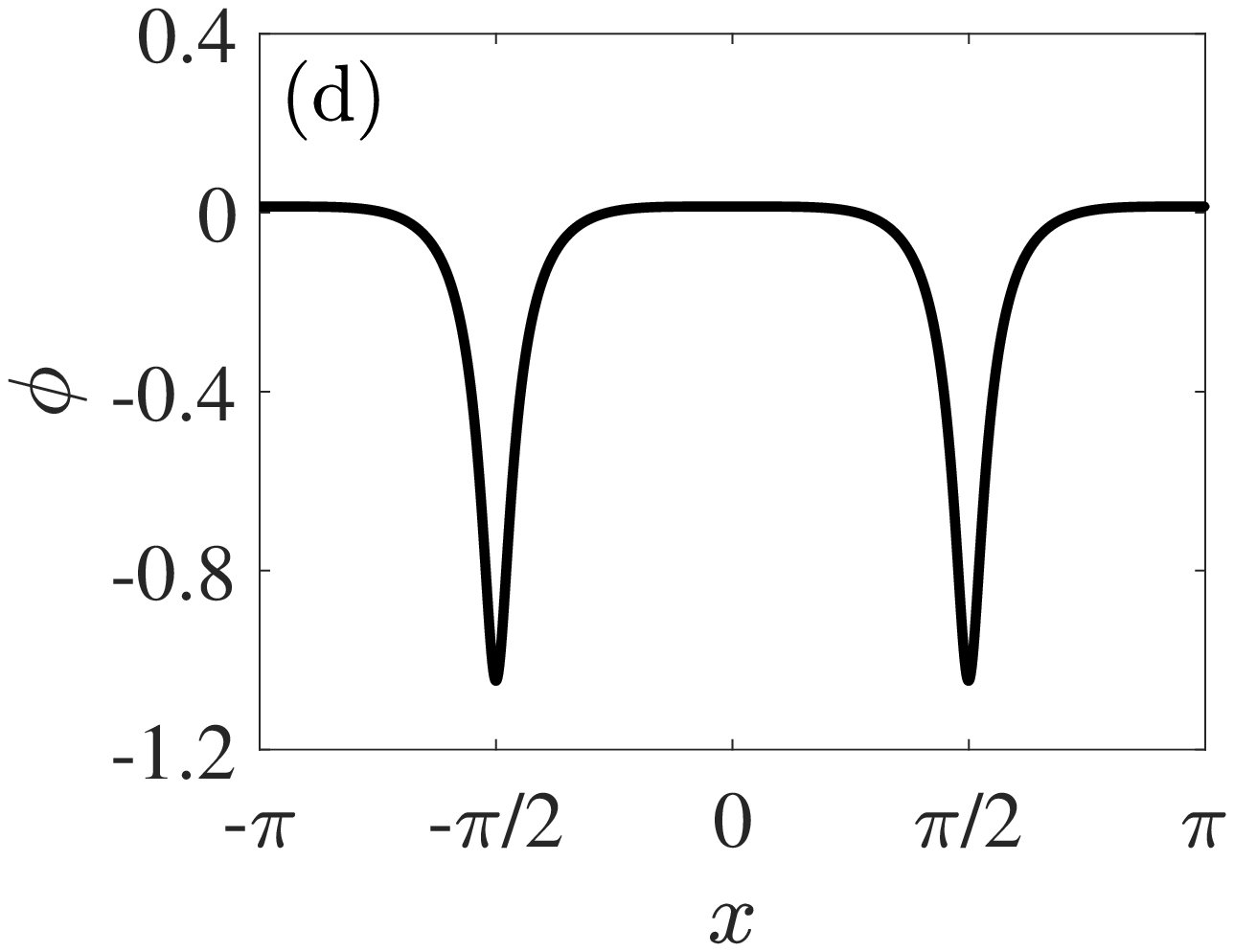}
\includegraphics[height=.21\textheight, angle =0]{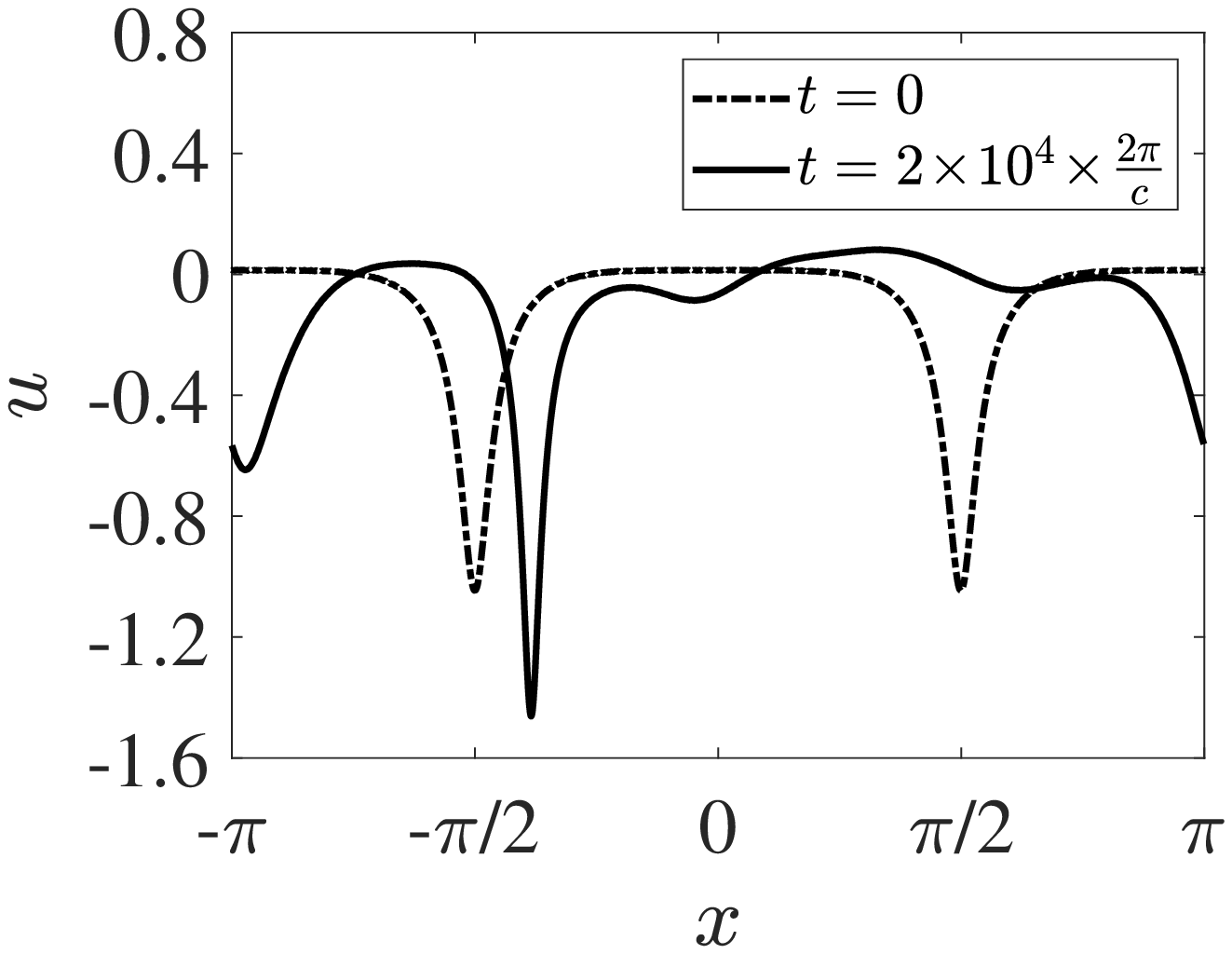}
\end{center}
\vspace*{-11pt}
\caption{
$T=T(1,2)$, $k=1,2$. Left column: profiles at the points labelled with (a),(b) ($k=1$) and (c),(d) ($k=2$) in the insets of Figure~\ref{fig4}, for which $c=0.973$~(a), $0.97$~(c), and $0.36$ (b,d). Right column: profiles perturbed by small random noise at $t=0$ (dash-dotted) and of the solutions at later instants of time (solid, see the legends), after translation of the $x$ axis (a,b). Notice different vertical scales.}
\label{fig5}
\end{figure}

\begin{figure}[htp]
\begin{center}
\includegraphics[height=.26\textheight, angle =0]{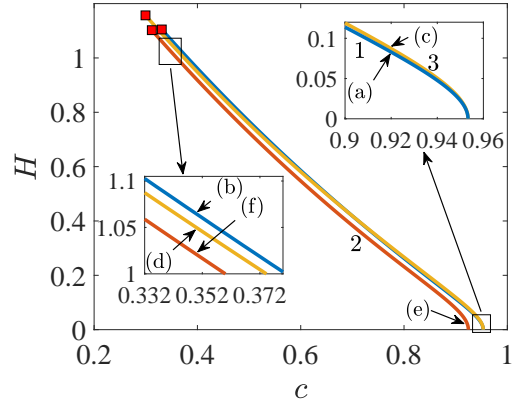}
\end{center}
\vspace*{-11pt}
\caption{
$T=T(1,3)$. $H$ vs. $c$ for $k=1$ (blue), $k=3$ (yellow), and $k=2$ (orange). The red squares correspond to the limiting admissible solutions, for which $c\approx0.3310243764$ ($k=1$), $c\approx0.2994866669$ ($k=3$), and $c\approx0.3117165220$ ($k=2$). See Figures~\ref{fig7} and \ref{fig8} for the profiles at the points labelled with (a)-(f) in the insets.}
\label{fig6}
\end{figure}

We turn the attention to $T=T(1,2)\approx 0.2396825$ (see \eqref{def:T(k1,k2)}). There exists a two-parameter sheet of nontrivial, periodic and even solutions of \eqref{eqn:phi} and \eqref{def:cT} in the vicinity of $\phi=0$ and $c=c_\text{ww}(1,T)=c_\text{ww}(2,T)\approx 0.97166609$ \cite{EJ;WW}. See also Section~\ref{sec:prelim}. Figure~\ref{fig4} shows $H$ versus $c$ for the $k=1$ and $k=2$ branches, all the way up to the limiting admissible solutions. There are no\footnote{We observe a turning point of $P$ for greater wave height, but the solution is inadmissible.} turning points of $P$. The left column of Figure~\ref{fig5} shows waves in the $k=1$ and $k=2$ branches for small and large $H$. The small height result agrees with \cite[Figure~6]{Kalisch;physD}. 

Observe `bimodal' waves in the $k=1$ branch. Indeed, there cannot exist $2\pi$ periodic and unimodal waves, whose profile monotonically decreases from a single crest to the trough over the period \cite{EJ;WW}. See also Section~\ref{sec:prelim}. For small wave height, the fundamental mode seems dominant, so that there is one crest over the period~$=2\pi$, but the fundamental and second modes are resonant, whereby a much smaller wave breaks up the trough into two. See the left panel of Figure~\ref{fig5}(a). As $H$ increases, the effects of the second mode seem more pronounced, so that the wave separating the troughs becomes higher. See the left of Figure~\ref{fig5}(b). Observe, to the contrary, $\pi$ periodic and unimodal waves in the $k=2$ branch. See the left of Figure~\ref{fig5}(c,d). We find that the crests become wider and flatter, the troughs narrower and more peaked, as $H$ increases in the $k=1$ and $k=2$ branches. See the left of Figure~\ref{fig5}(b,d). 

Our numerical experiments suggest orbital stability for the $k=1$ branch  (see the right panels of Figure~\ref{fig5}(a,b)) versus instability for $k=2$ (see the right of Figure~\ref{fig5}(c,d)).

\begin{figure}[htp]
\begin{center}
\includegraphics[height=.21\textheight, angle =0]{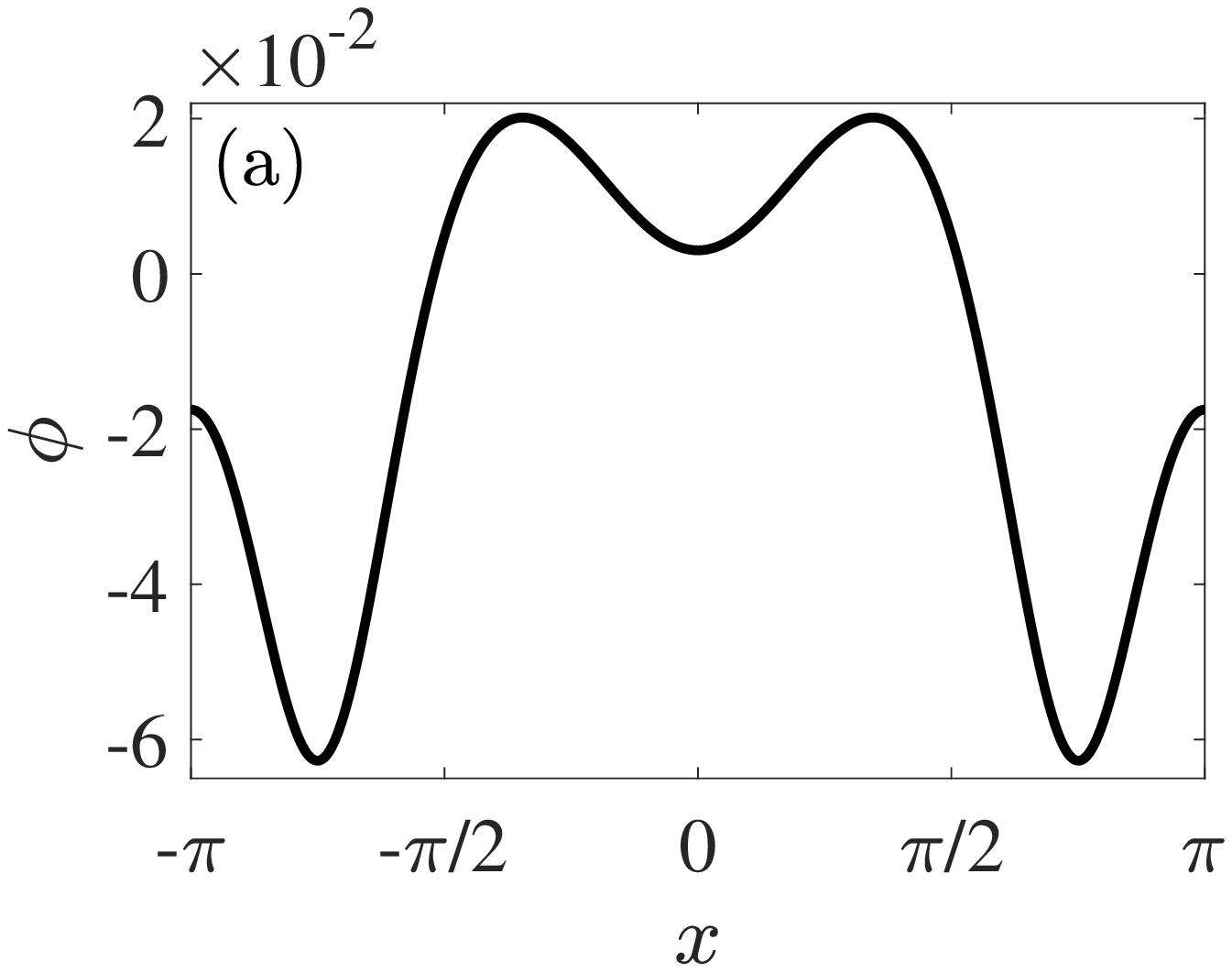}
\includegraphics[height=.21\textheight, angle =0]{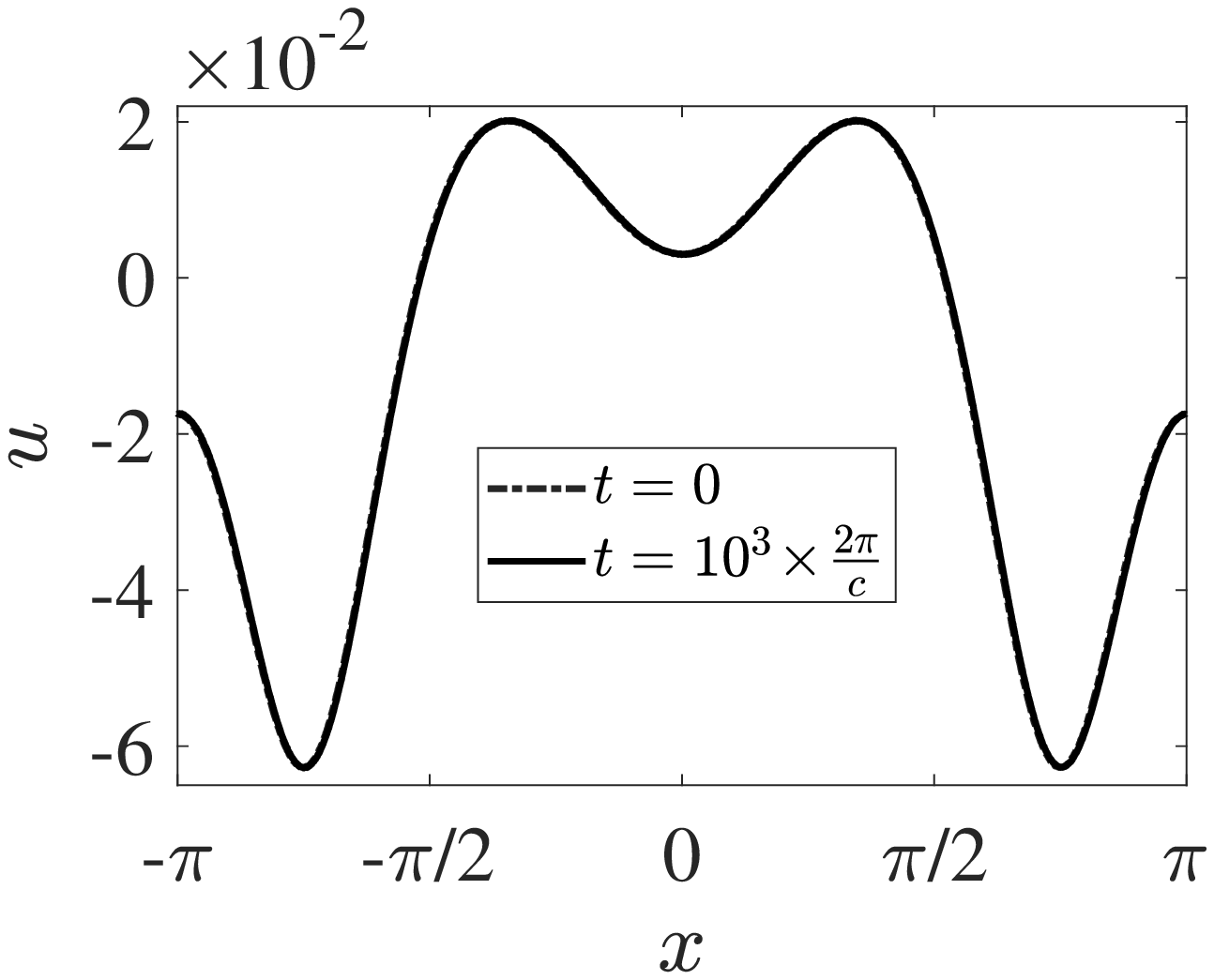}\\
\includegraphics[height=.21\textheight, angle =0]{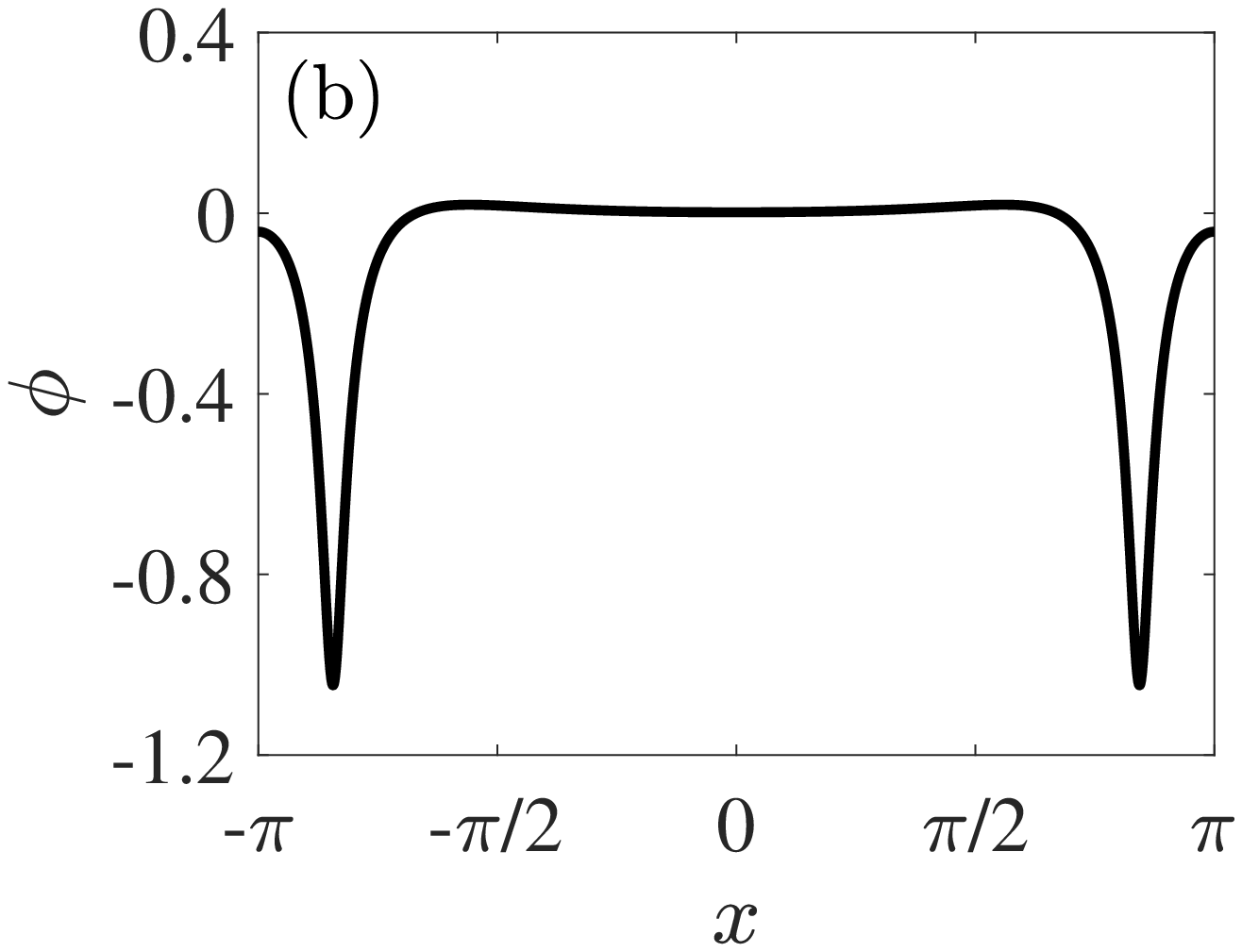}
\includegraphics[height=.21\textheight, angle =0]{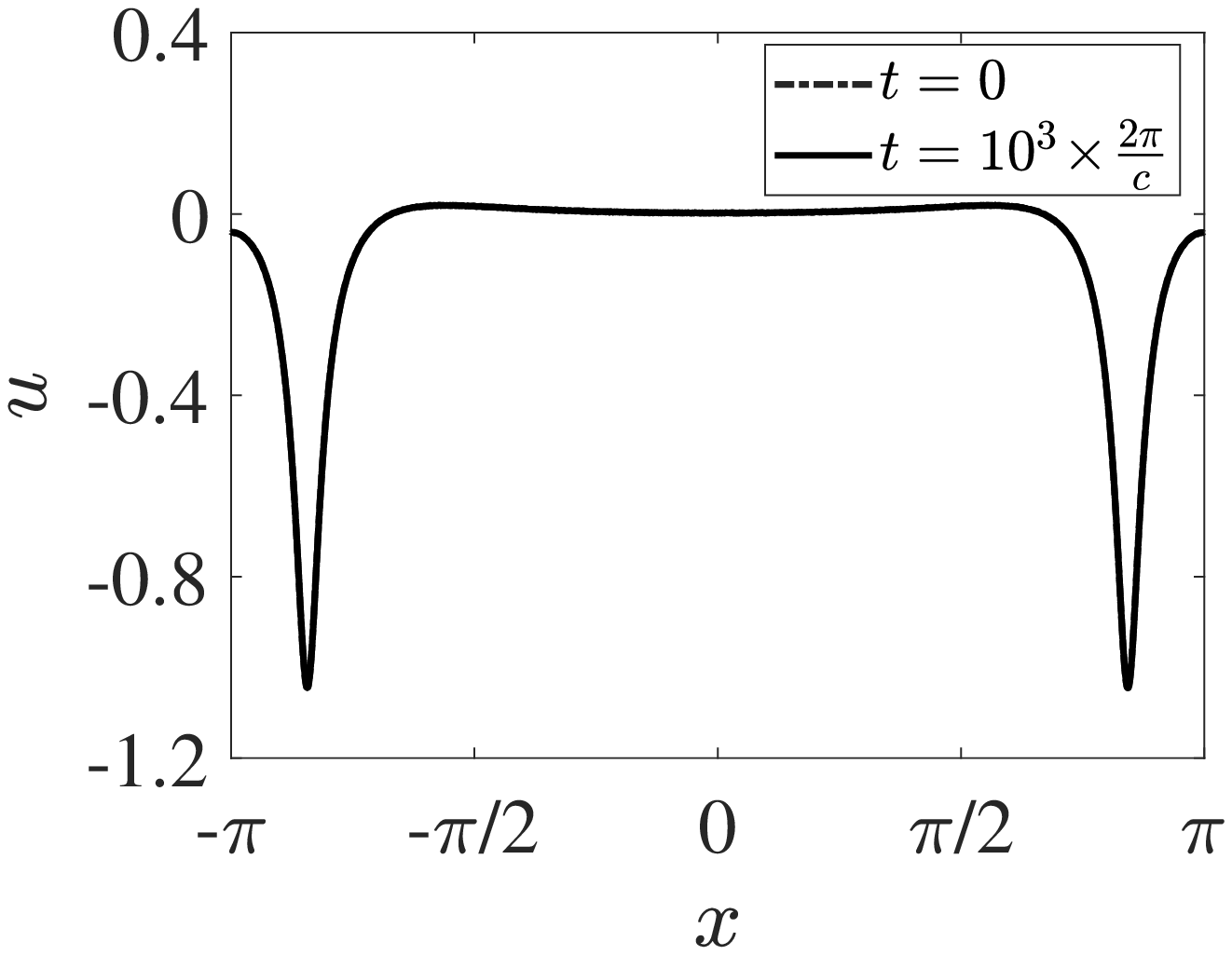}
\end{center}
\vspace*{-11pt}
\caption{
T=T(1,3), $k=1$. Left column: profiles at the points labelled with (a) and (b) in the insets of Figure~\ref{fig6}, for which $c=0.92$ and $0.35$. Right column: profiles perturbed by small random noise at $t=0$ (dash-dotted) and of the solutions after $10^3$ periods (solid), after translation of the $x$ axis (a).}
\label{fig7}
\end{figure}

\begin{figure}[htp]
\begin{center}
\includegraphics[height=.21\textheight, angle =0]{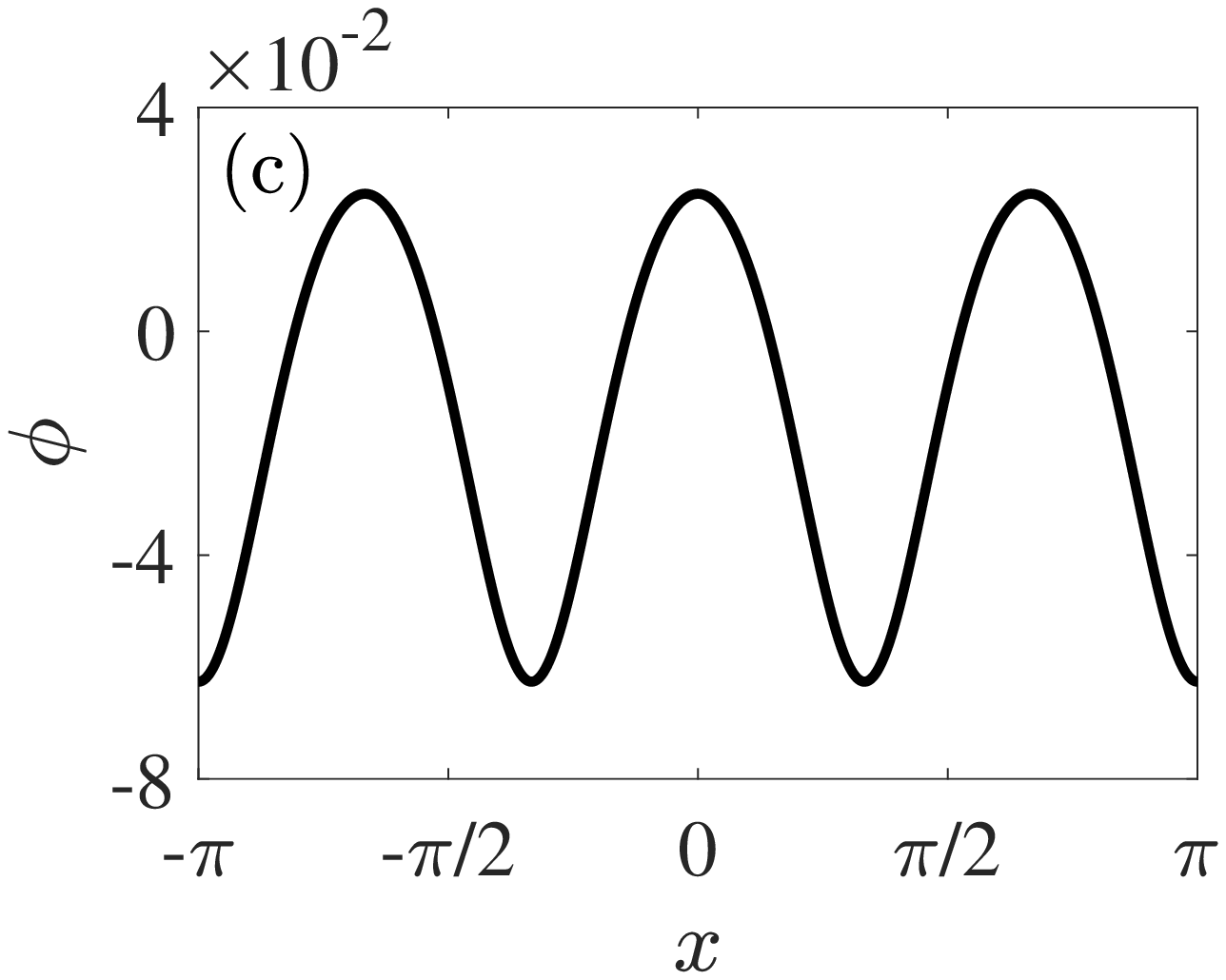}
\includegraphics[height=.21\textheight, angle =0]{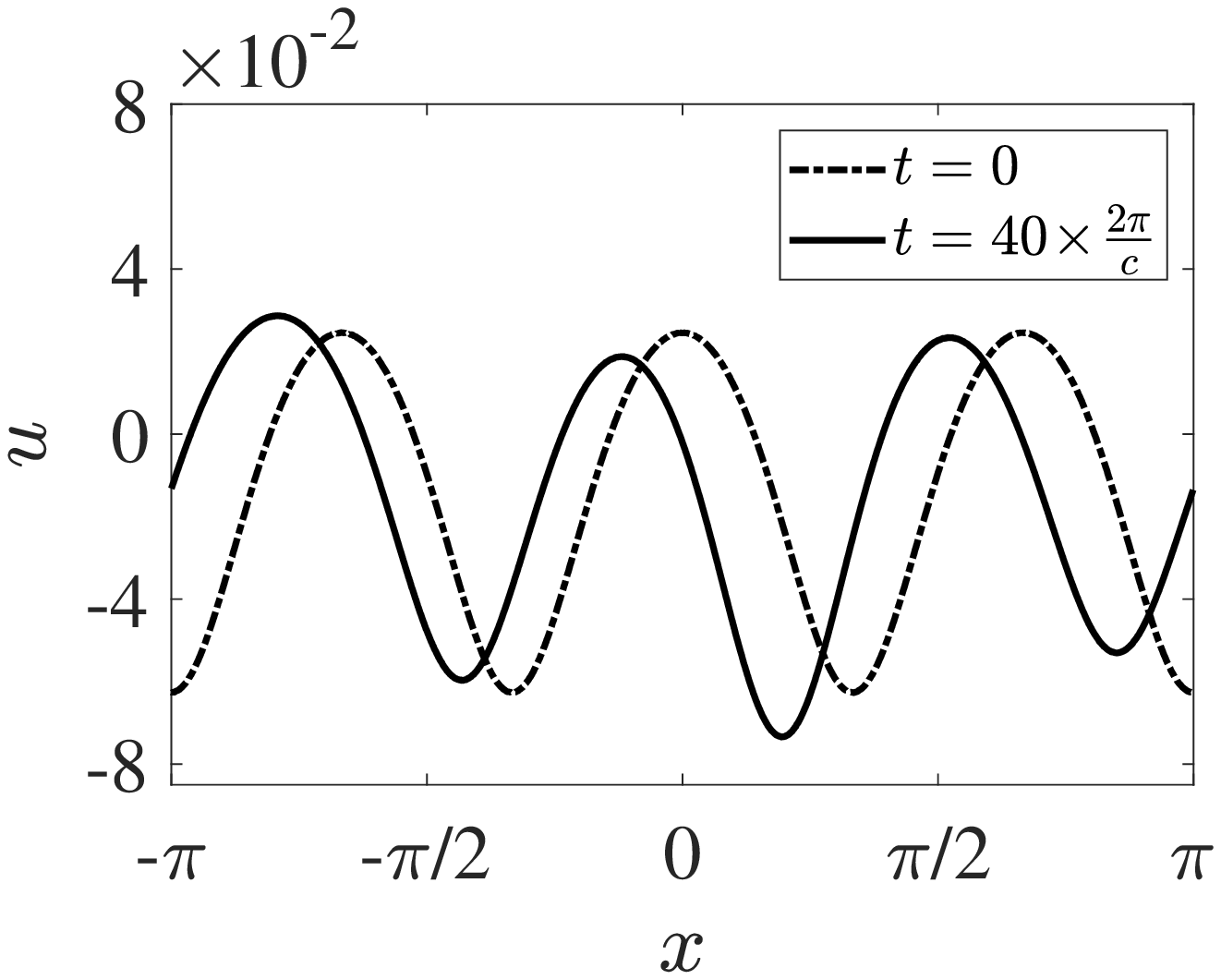}\\
\includegraphics[height=.21\textheight, angle =0]{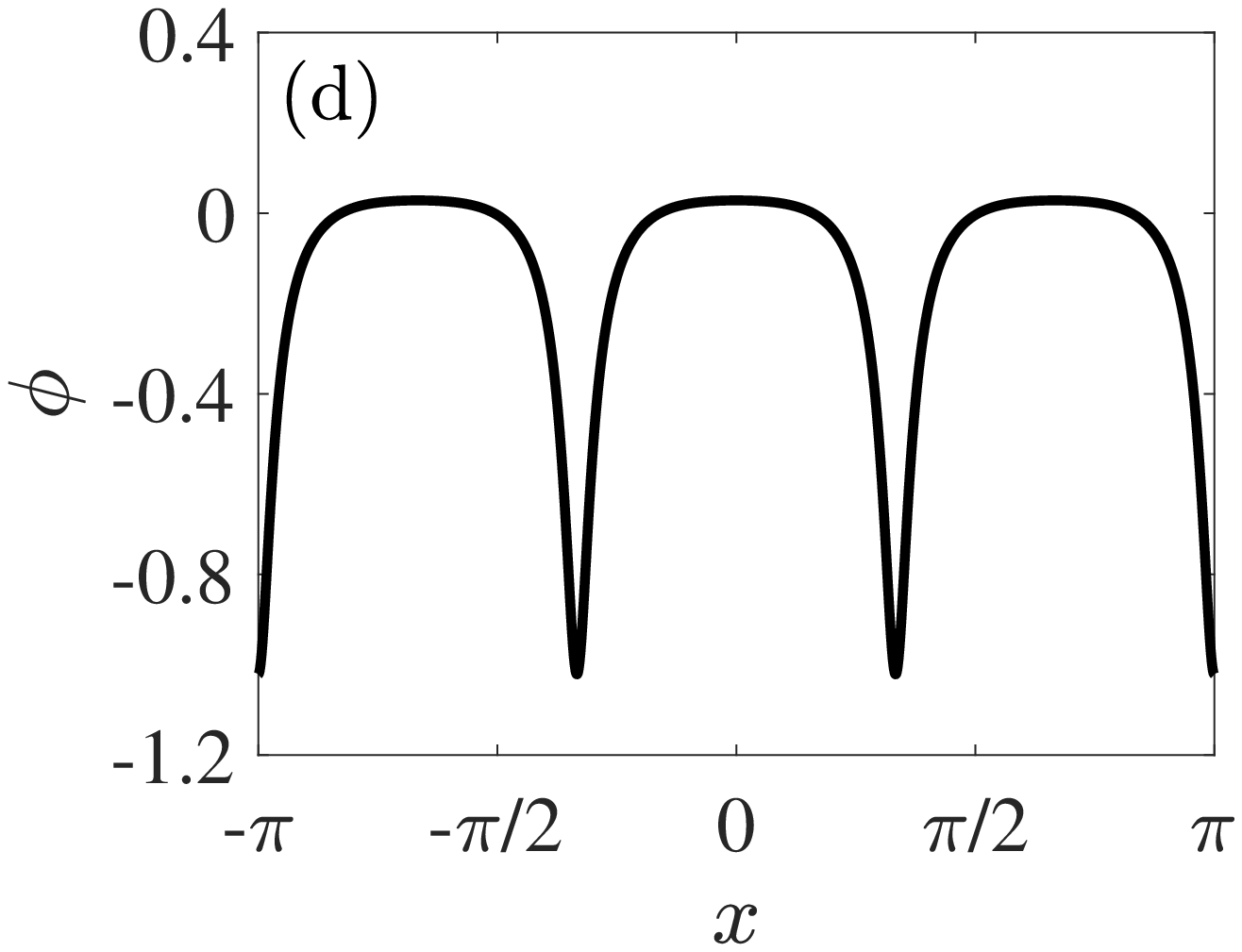}
\includegraphics[height=.21\textheight, angle =0]{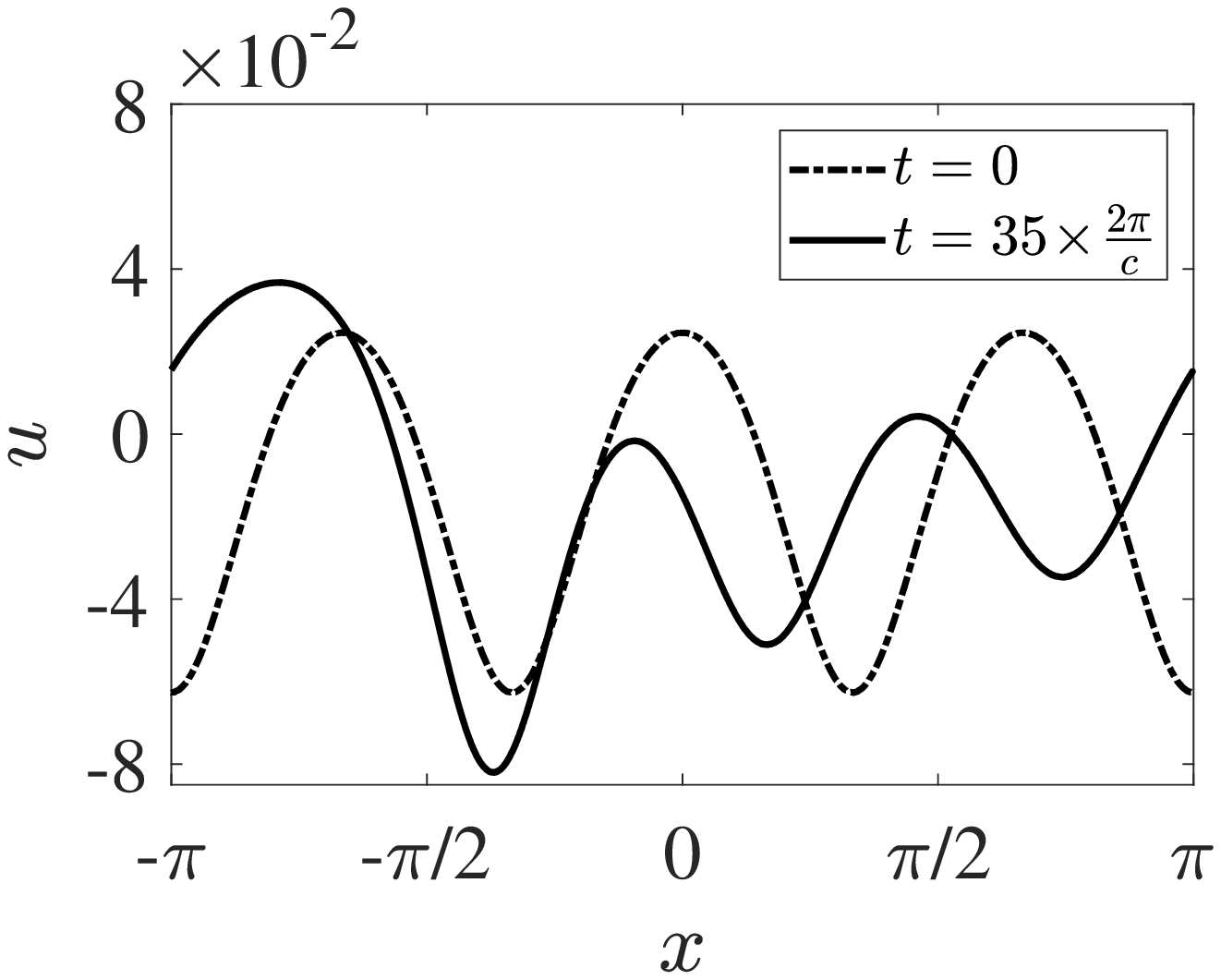}\\
\includegraphics[height=.21\textheight, angle =0]{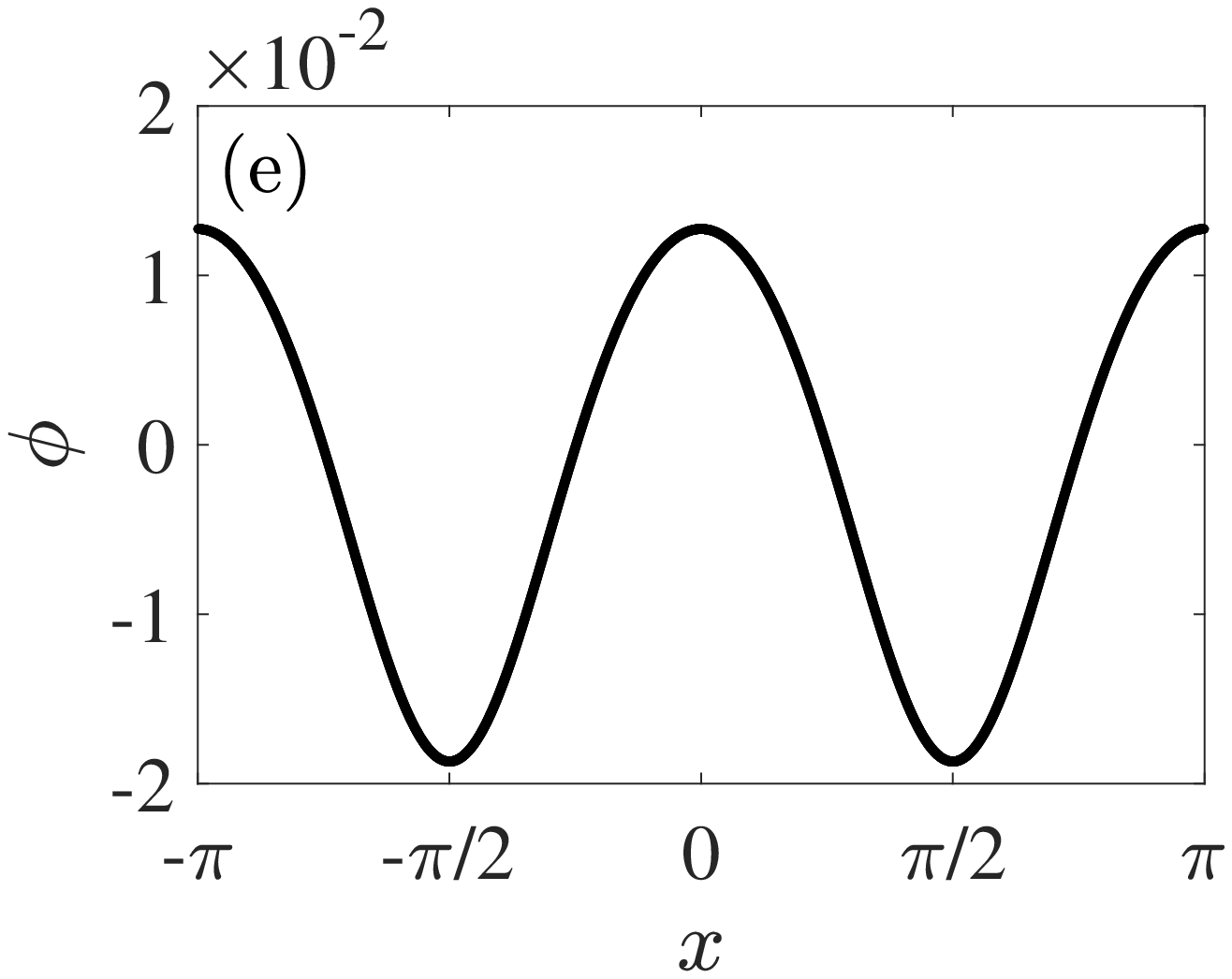}
\includegraphics[height=.21\textheight, angle =0]{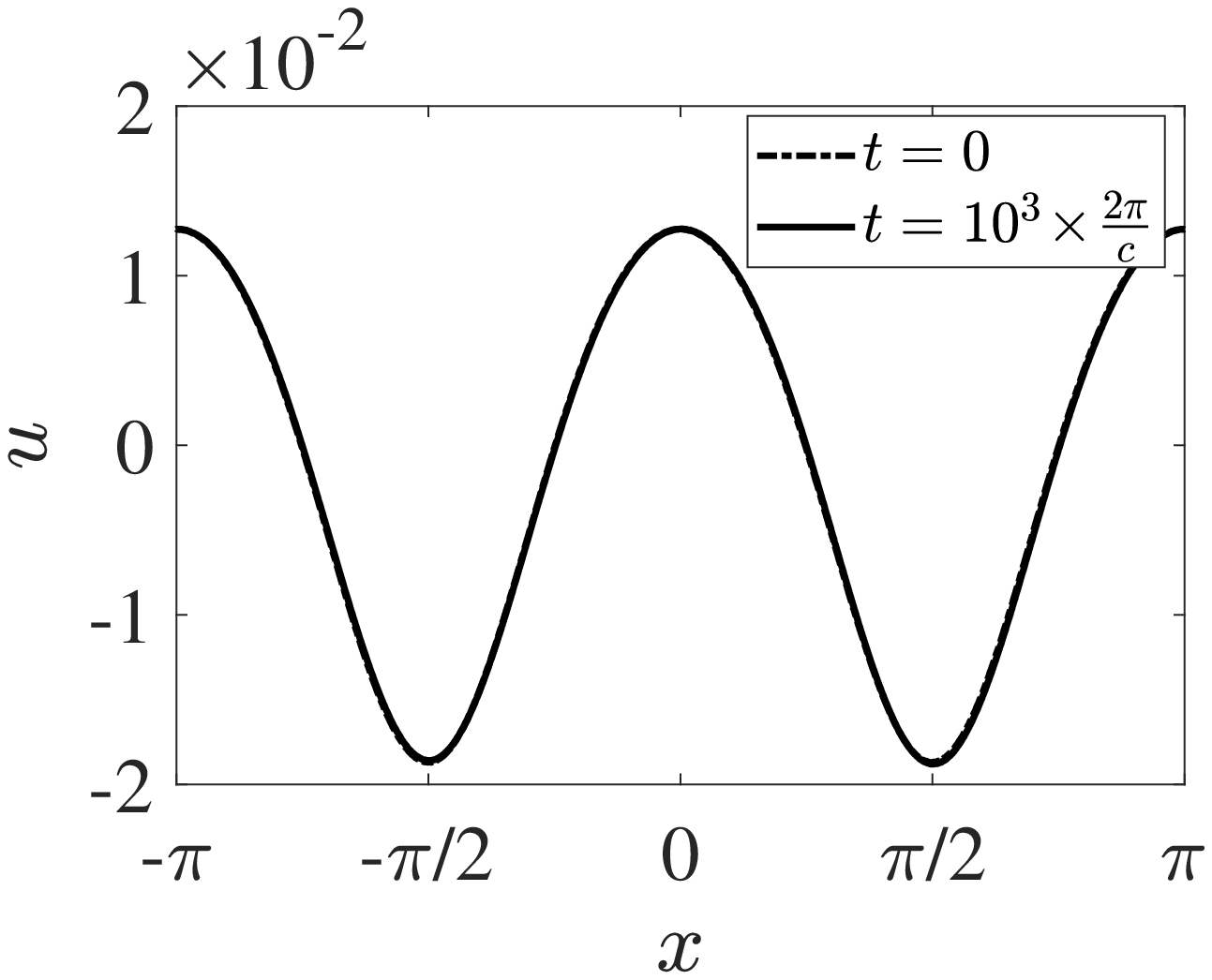}\\
\includegraphics[height=.21\textheight, angle =0]{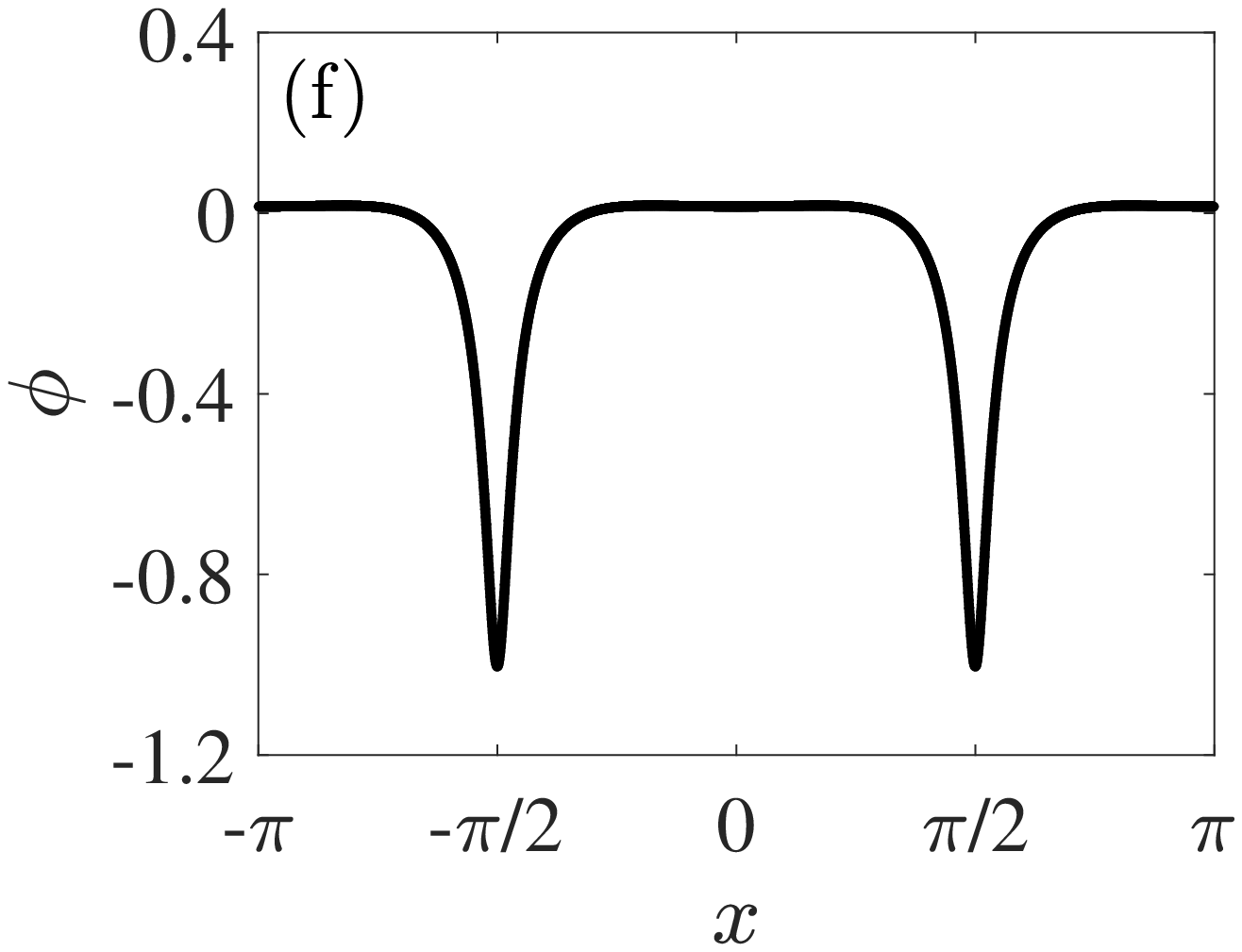}
\includegraphics[height=.21\textheight, angle =0]{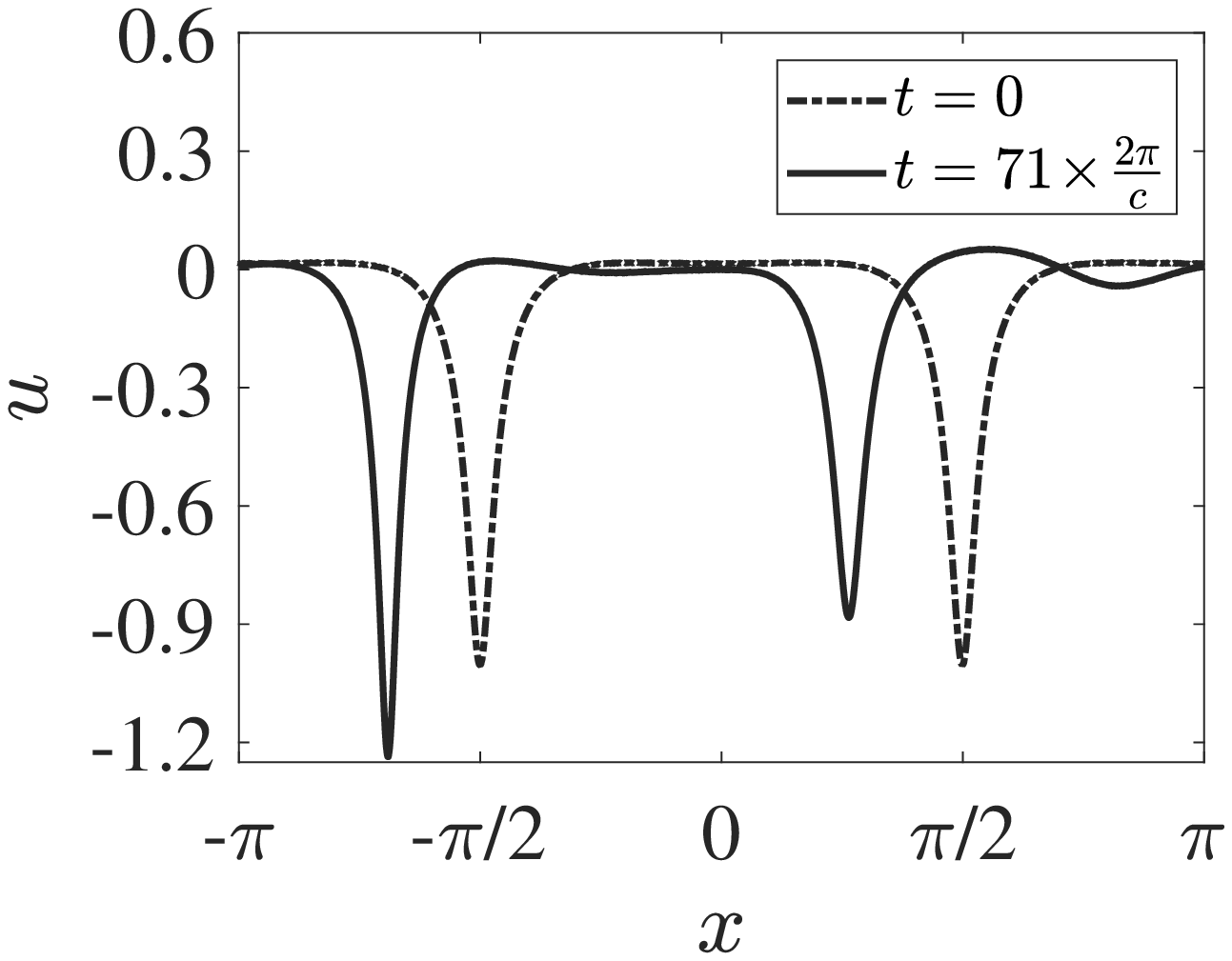}
\end{center}
\vspace*{-11pt}
\caption{
$T=T(1,3)$. Similar to Figure~\ref{fig7} but $k=3$ (c,d) and $k=2$ (e,f), $c=0.92$~(c,e) and $0.35$~(d,f). On the right, solid curves are the solution profiles at later times (see the legends), after translation of the $x$ axis (e).}
\label{fig8}
\end{figure}

We take matters further to $T=T(1,3)$. See Figures~\ref{fig6}, \ref{fig7} and \ref{fig8}. The results for the $k=1$ and $k=3$ branches are similar to those when $T=T(1,2)$ and $k=1,2$. We pause to remark that in the $k=1$ branch, for small $H$, a smaller wave breaks up the trough into two, and a much smaller wave breaks up the crest into two. See the left panel of Figure~\ref{fig7}(a). As $H$ increases, the wave separating the crests becomes lower, transforming into one wide and flat crest, whereas the wave separating the troughs become higher (see the left of Figure~\ref{fig7}(b)), whereby resembling those when $T=T(1,2)$ and $k=1$. Observe $\pi$ periodic and unimodal waves in the $k=2$ branch, orbitally stable for small $H$ versus unstable for large $H$. See Figure~\ref{fig8}(e,f). 

When $T=T(2,3)$, to the contrary, we find $\pi$ and $2\pi/3$ periodic unimodal waves in the $k=2$ and $k=3$ branches, respectively, corroborating a local bifurcation theorem (see \cite{EJ;WW}, for instance). 
See also Section~\ref{sec:prelim}. We find that the crests become wider and flatter, the troughs narrower and more peaked, as $H$ increases. See the left column of Figure~\ref{fig10}. Our numerical experiments (see the right of Figure~\ref{fig10}) suggest 
orbital instability for the $k=2$ branch for large $H$ and for $k=3$ for all $H$. 

When $T=T(2,5)$, on the other hand, the left column of Figure~\ref{fig12} shows bimodal waves in the $k=2$ branch. The local bifurcation theorem \cite{EJ;WW} dictates $\pi$ periodic and unimodal waves, but we numerically find that they are not in the $k=2$ branch. For small $H$, for wave (a), for instance, a smaller wave breaks up the trough into two over the half period. As $H$ increases, for wave (b), for instance, the troughs become narrower and more peaked. Our numerical experiments (see the right of Figures~\ref{fig12} and \ref{fig13}) suggest orbital stability for the $k=2$ branch  for small $H$ versus instability for $k=2$ for large $H$ and for $k=5$ for all $H$. 

To proceed, when $T=T(1,4)$, the $k=1$ branch lies above and to the right of the $k=4$ branch at least for small $H$ (not shown), but as $T$ increases, $c_\text{ww}(4;T)$ increases more rapidly than $c_\text{ww}(1;T)$, so that when $T=T(1,4)+0.001$, for instance, the $k=1$ branch crosses the $k=4$ branch. Figure~\ref{fig14} shows $H$ and $P$ versus $c$ in the $k=1$ and $k=4$ branches for all admissible solutions. The small height result agrees with \cite[Figure~10(a)]{Kalisch;physD}. We find that in the $k=1$ branch, $H$ and $P$ turn to connect the $k=4$ branch, whereas in the $k=4$ branch, $H$ and $P$ monotonically increase to the limiting admissible solution, highlighted by the red square in the insets. 

The left panels of Figures~\ref{fig15}, \ref{fig16} and \ref{fig17} show several profiles along the $k=1$ and $k=4$ branches. In the $k=1$ branch, for small $H$, wave (a), for instance, is $2\pi$ periodic and unimodal. After the $k=1$ branch crosses the $k=4$ branch, on the other hand, wave (b), for instance, becomes bimodal, resembling those when $T=T(1,4)$ and $k=1$. Continuing along the $k=1$ branch, for waves (c) and (d), for instance, high frequency ripples of $k=4$ ride a carrier wave of $k=1$. 
When the $k=1$ and $k=4$ branches almost connect, wave (e) in the $k=1$ branch and wave (f) for $k=4$ are almost the same. In the $k=4$ branch, wave (g), for instance, is $\pi/2$ periodic and unimodal. 

Our numerical experiments (see the right panels of Figures~\ref{fig15}, \ref{fig16} and \ref{fig17}) suggest orbital stability for the $k=1$ branch for all $H$ and for $k=4$ for small $H$, versus instability for $k=4$ for large $H$. Particularly, stability and instability do not change at the turning point of $P$.  

Last but not least, when $T=T(1,5)+0.0001$, Figure~\ref{fig18} shows $H$ and $P$ versus $c$ for the $k=1$ and $k=5$ branches. The small height result agrees with \cite[Figure~10(b)]{Kalisch;physD}. We find that the $k=1$ branch crosses and connects the $k=5$ branch, like when $T=T(1,4)+0.0001$ and $k=1,4$, but it continues after connecting all the wave up to the limiting admissible solution. See the insets. The left panels of Figures~\ref{fig19}, \ref{fig20} and \ref{fig21} show several profiles along the $k=1$ and $k=5$ branches. The results for the $k=1$ branch up to connecting and for $k=5$ are similar to those when $T=T(1,4)+0.0001$ and $k=1,4$. In the $k=1$ branch, after it connects the $k=4$ branch at the point (c), the results are similar to those when $T=T(1,5)$ and $k=1$. See waves (d), (e) and (f). 
Our numerical experiments (see the right panels of Figures~\ref{fig19}, \ref{fig20} and \ref{fig21}) suggest orbital stability for the $k=1$ branch for all $H$ and for $k=5$ for small $H$, versus instability for $k=5$ for large $H$. 

We emphasize orbital stability for the $k=1$ branch for all $T>0$ for all wave height throughout our numerical experiments.

\section{Discussion}\label{sec:discussion}

Here we employ efficient and highly accurate numerical methods for computing periodic traveling waves of \eqref{eqn:Whitham} and experimenting with their (nonlinear) orbital stability and instability. Our findings suggest, among many others, stability whenever $T>0$ for the $k=1$ branch versus instability when $0<T<1/3$ for $k\geq2,\in\mathbb{N}$ branches at least for large wave height. Currently under investigation is to take matters further to all $T\geq0$ to all $k$, to classify the orbital stability and instability of all periodic traveling waves. It will be interesting to devise other numerical continuation methods, for instance, deflated continuation techniques \cite{PhysRevA.102.053307,CHARALAMPIDIS2020105255}, for detecting {\em disconnected} solution branches and others. Also interesting will be to numerically investigate (spectral) modulational stability and instability of orbitally stable waves. Our methodology will be useful for exploring the nonlinear dynamics of modulationally unstable waves. Also it can help tackle the capillary-gravity wave problem and other nonlinear dispersive equations. Of course, it is of great importance to rigorously prove the numerical results. 

\newpage

\subsection*{Acknowledgement}

The authors are grateful to Henrik Kalisch for helpful discussions. EGC acknowledges the hospitality of the Department of Mathematics at the University of Illinois at Urbana-Champaign where early stages of this work took place.

\bibliographystyle{amsplain}
\bibliography{Whitham}

\newpage
\clearpage


\begin{figure}[htp]
\begin{center}
\includegraphics[height=.26\textheight, angle =0]{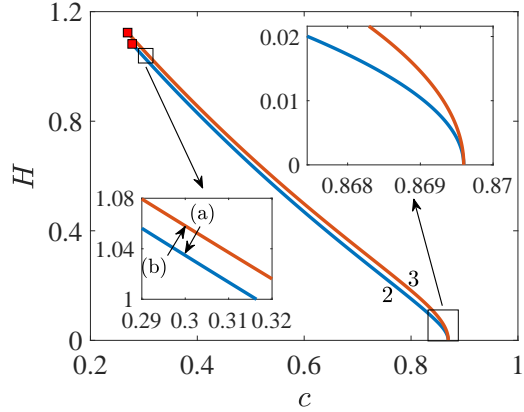}
\end{center}
\vspace*{-11pt}
\caption{
%
%
$T=T(2,3)$. $H$ vs. $c$ for $k=2$ (blue) and $k=3$ (orange). The red squares correspond to the limiting admissible solutions, for which $c \approx 0.27804634505$ ($k=2$) and $c \approx 0.26958276662$ ($k=3$). See Figure~\ref{fig10} for the profiles at the points labelled with (a) and (b) in the inset.}
\label{fig9}
\end{figure}

\begin{figure}[htp]
\begin{center}
\includegraphics[height=.21\textheight, angle =0]{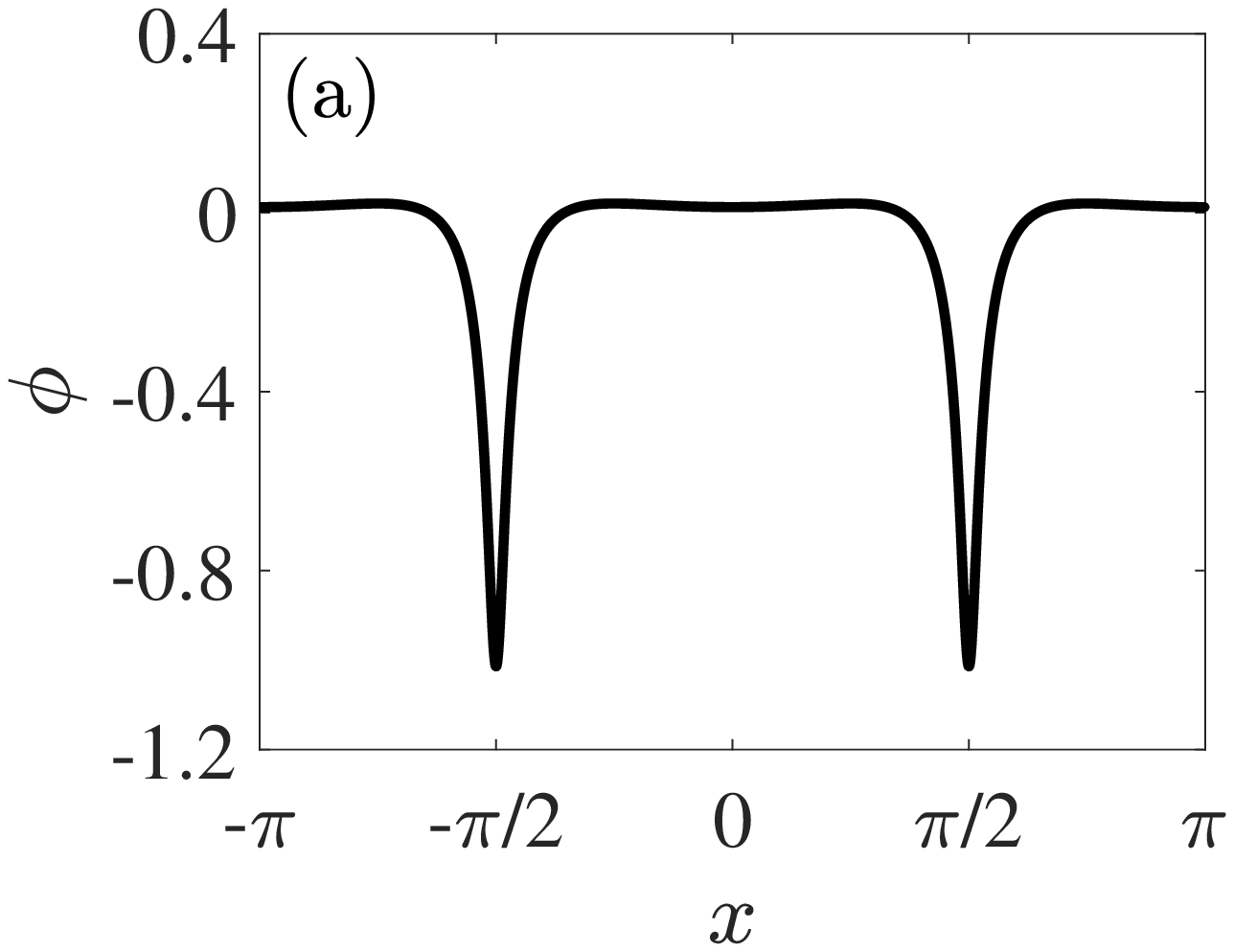}
\includegraphics[height=.21\textheight, angle =0]{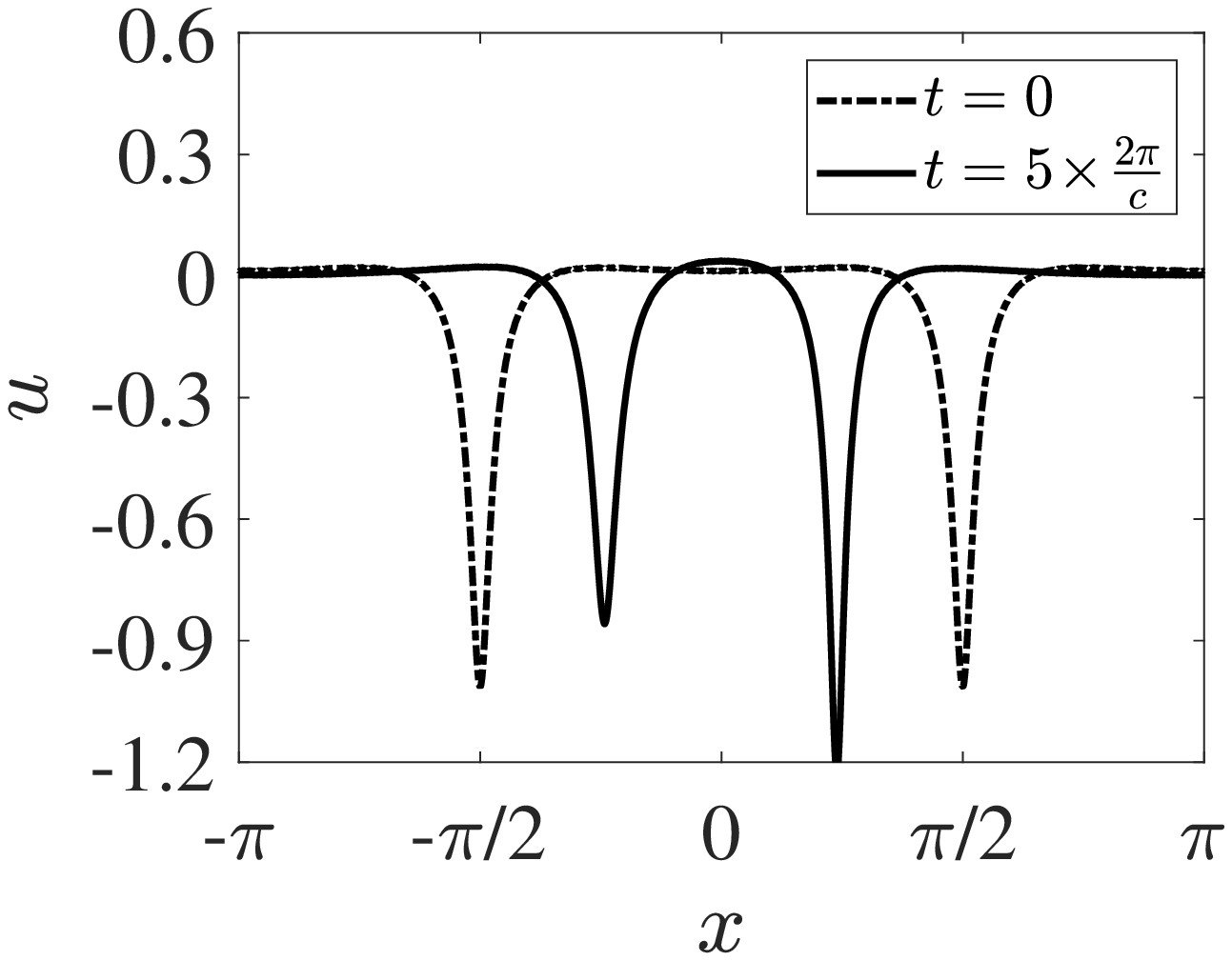}\\
\includegraphics[height=.21\textheight, angle =0]{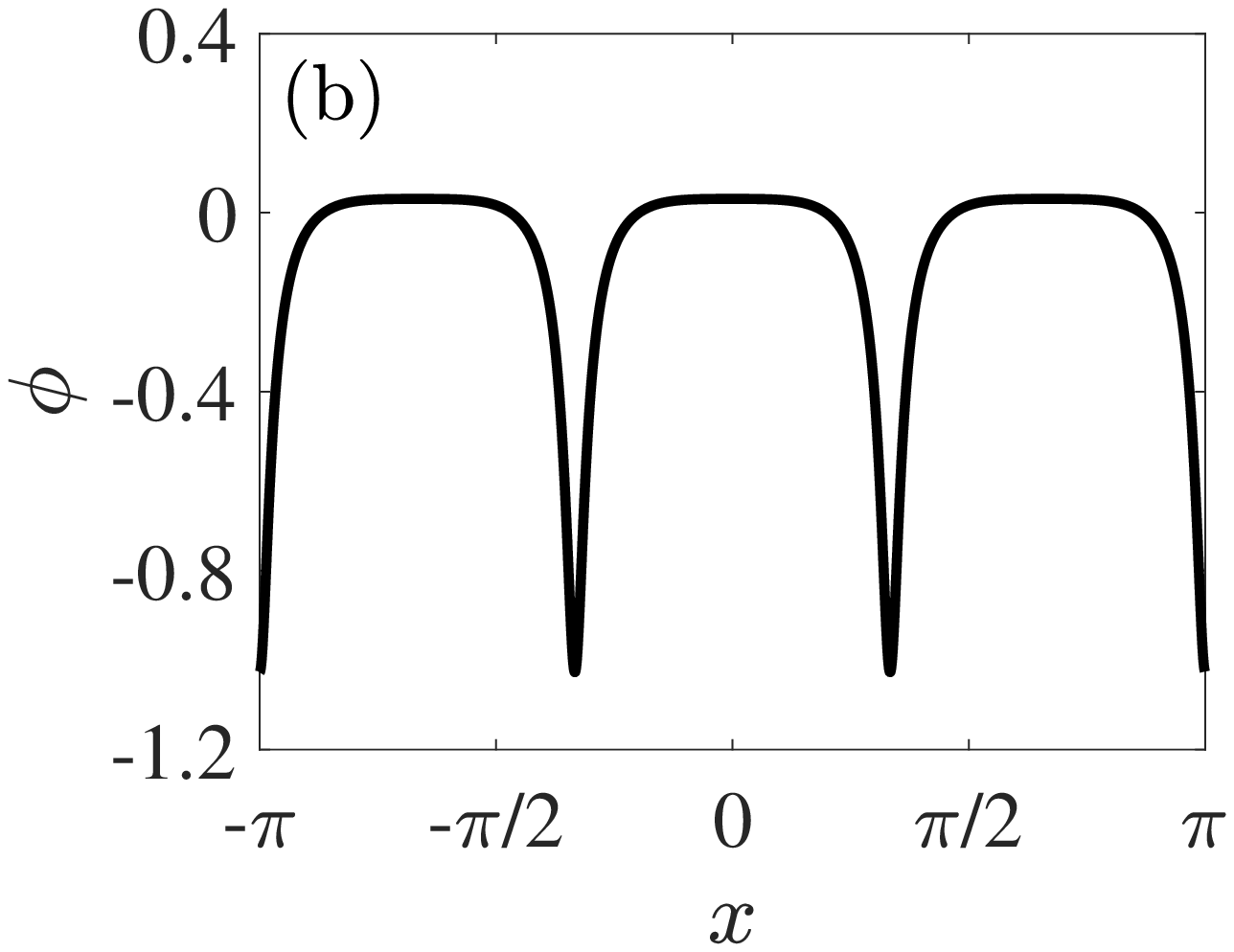}
\includegraphics[height=.21\textheight, angle =0]{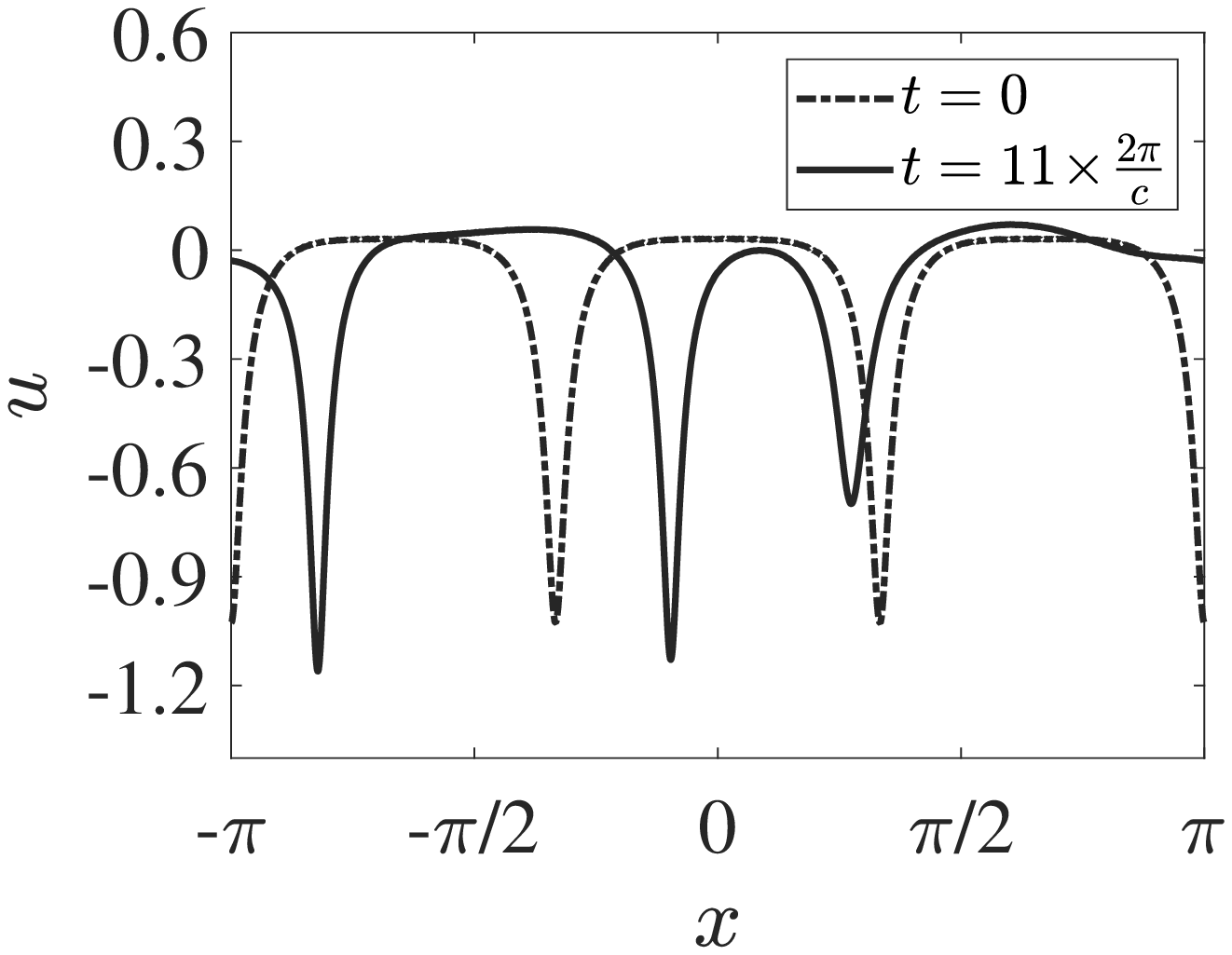}
\end{center}
\vspace*{-11pt}
\caption{
$T=T(2,3)$. Left column: profiles at the points labelled with (a) ($k=2$) and (b) ($k=3$) in Figure~\ref{fig9}, for which $c=0.3$. Right column: profiles perturbed by small random noise at $t=0$ (dash-dotted) and of the solutions at later times (solid, see the legends).}
\label{fig10}
\end{figure}

\begin{figure}[htp]
\begin{center}
\includegraphics[height=.26\textheight, angle =0]{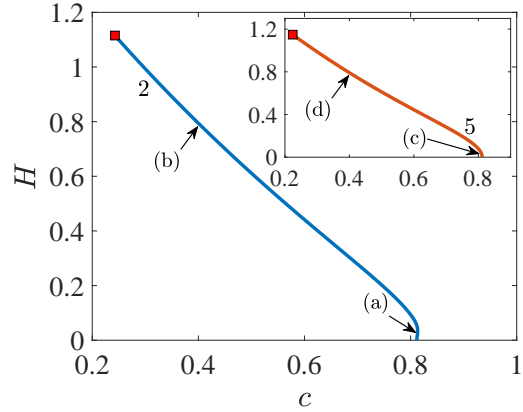}
\end{center}
\vspace*{-11pt}
\caption{
%
%
$T=T(2,5)$. $H$ vs. $c$ for $k=2$ (blue) and $k=5$ (inset, orange). The red squares correspond to the limiting admissible solutions, for which $c\approx 0.24281537129$ ($k=2$) and $c\approx 0.22417409072$ ($k=5$). See Figures~\ref{fig12} and \ref{fig13} for the profiles at the points labelled with (a)-(d).}
\label{fig11}
\end{figure}

\begin{figure}[htp]
\begin{center}
\includegraphics[height=.21\textheight, angle =0]{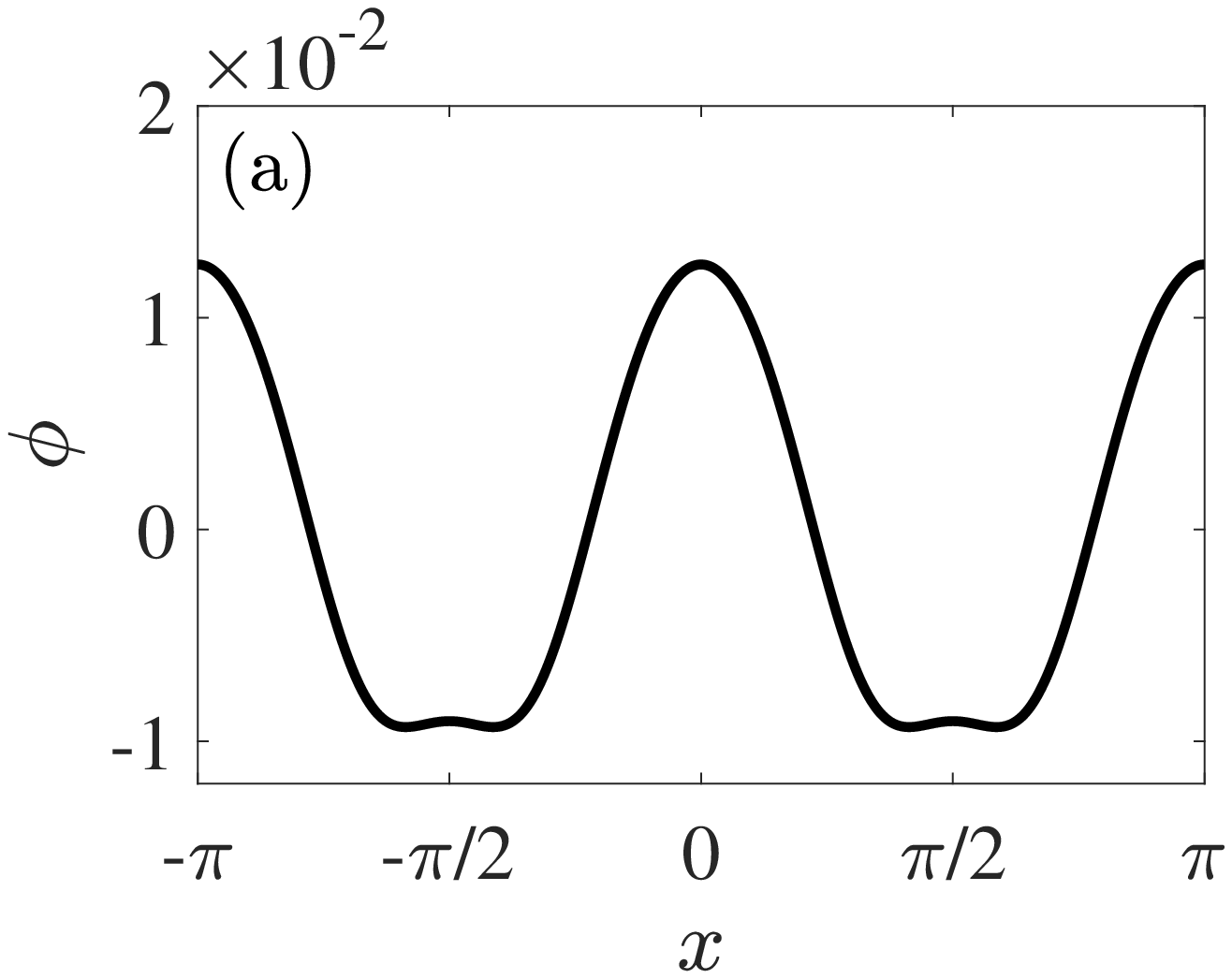}
\includegraphics[height=.21\textheight, angle =0]{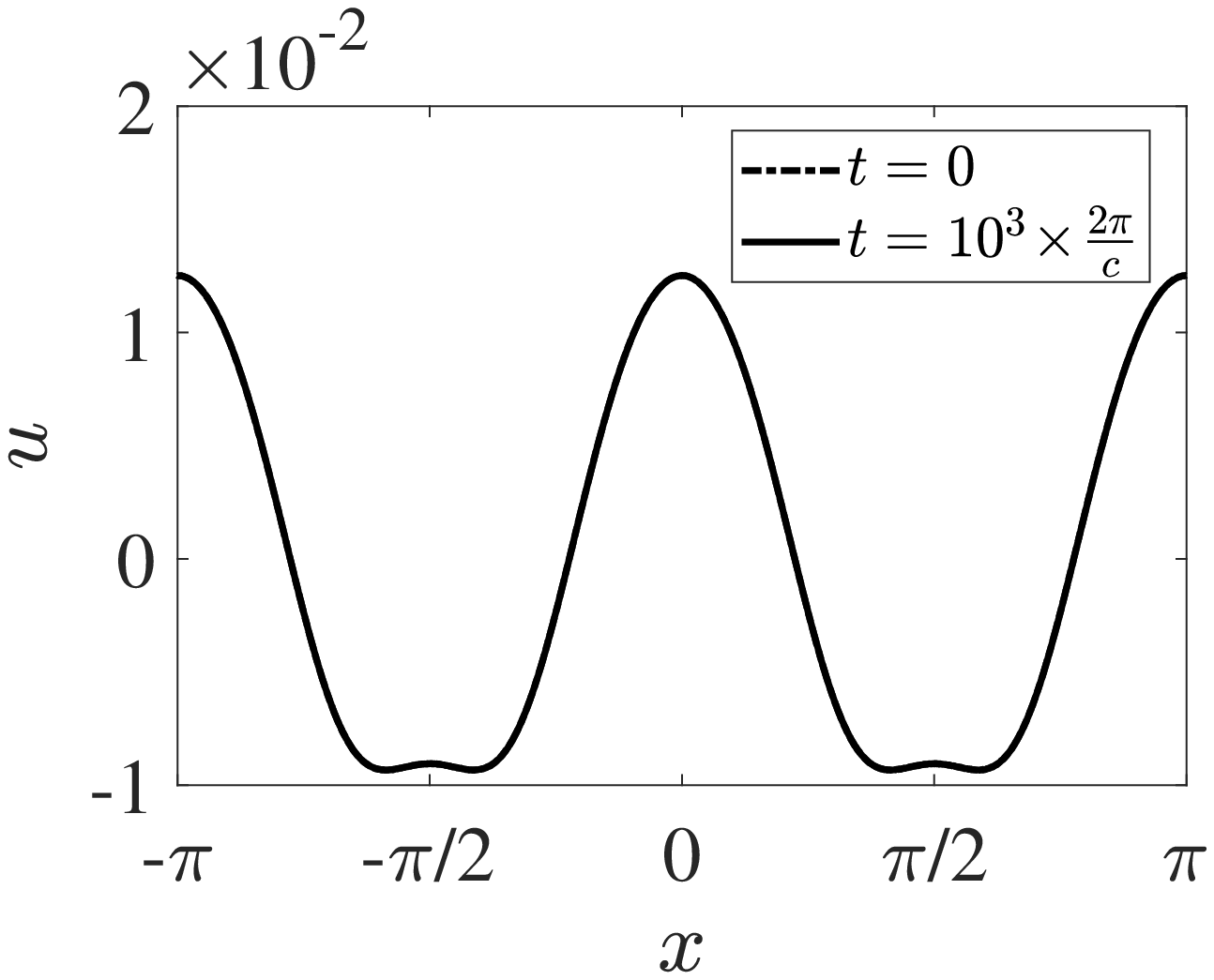}\\
\includegraphics[height=.21\textheight, angle =0]{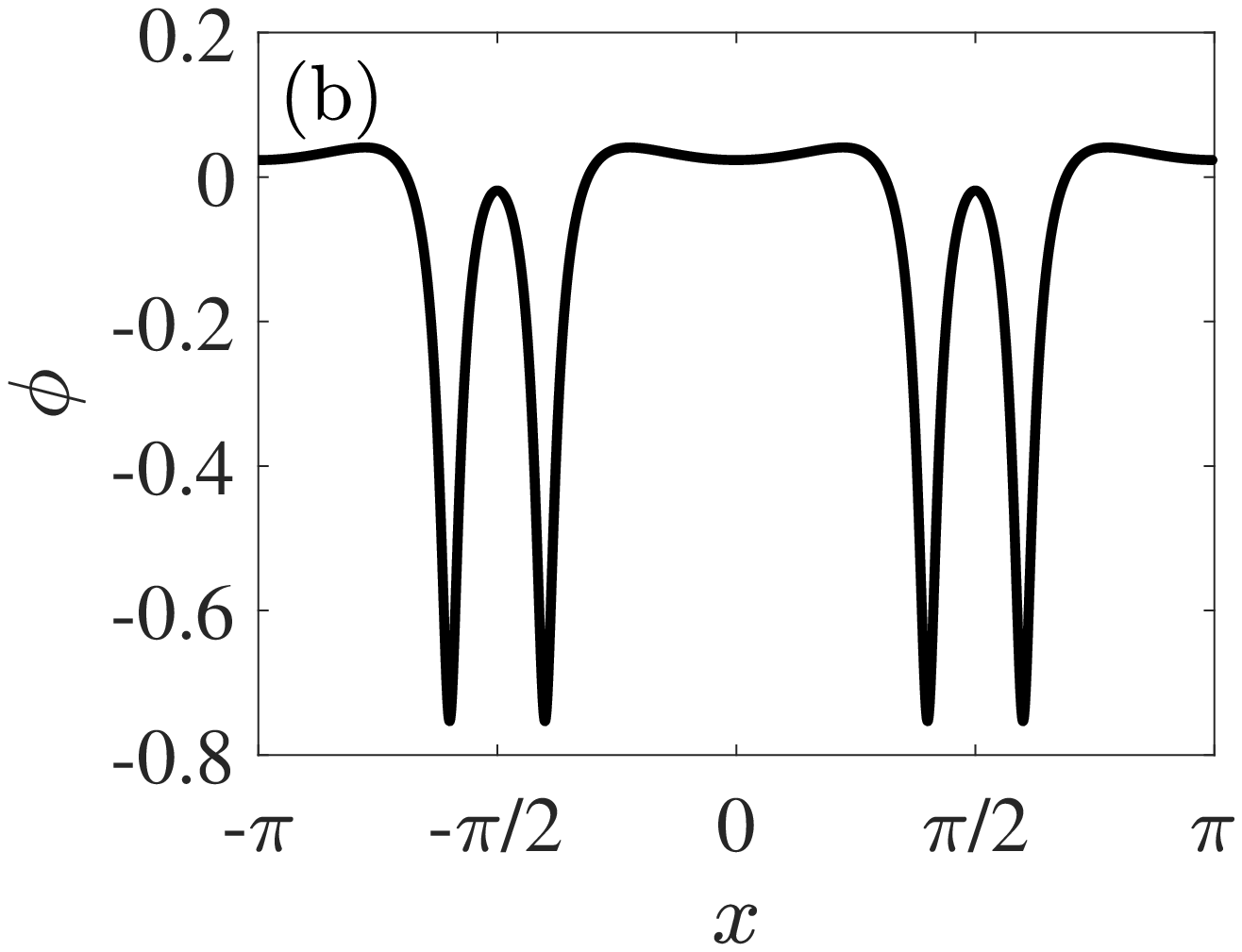}
\includegraphics[height=.21\textheight, angle =0]{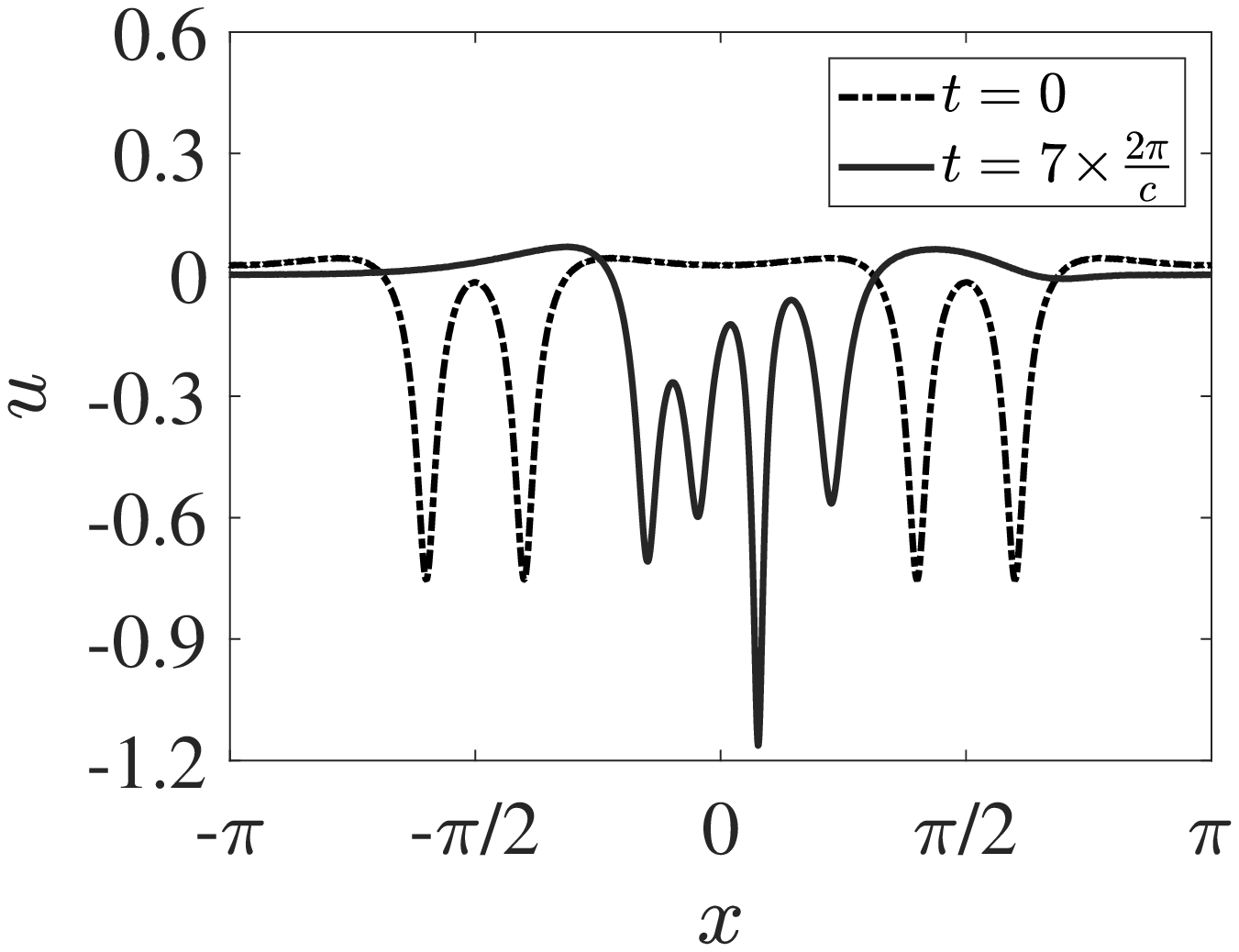}
\end{center}
\vspace*{-11pt}
\caption{
%
$T=T(2,5)$, $k=2$. Left column: profiles at the points labelled with (a) and (b) in Figure~\ref{fig11}, for which $c=0.813$ (a) and $0.4$~(b). Right column: profiles perturbed by small random noise at $t=0$ (dash-dotted) and of the solutions at later times (solid), after translation of the $x$ axis (a).}
\label{fig12}
\end{figure}

\begin{figure}[htp]
\begin{center}
\includegraphics[height=.21\textheight, angle =0]{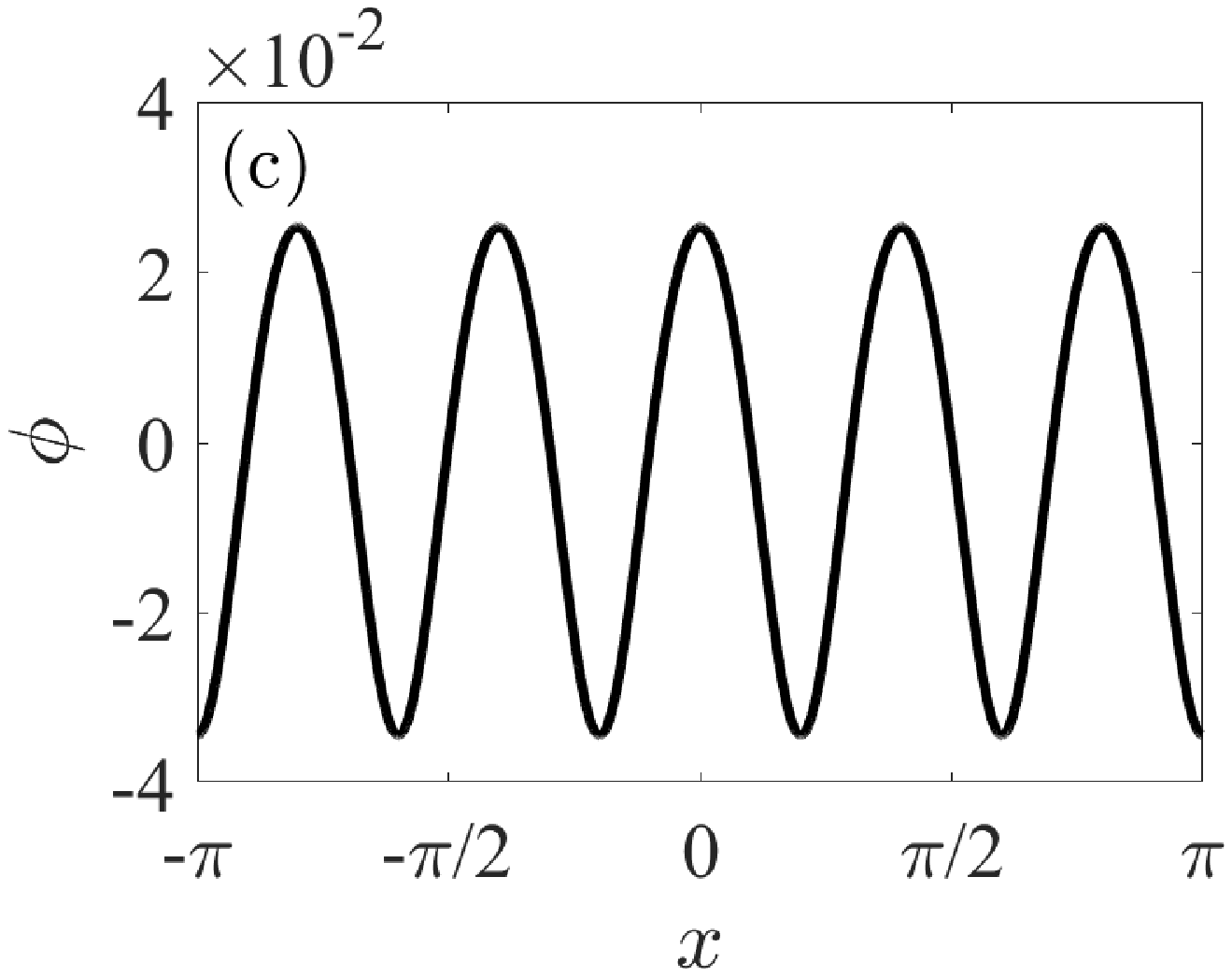}
\includegraphics[height=.21\textheight, angle =0]{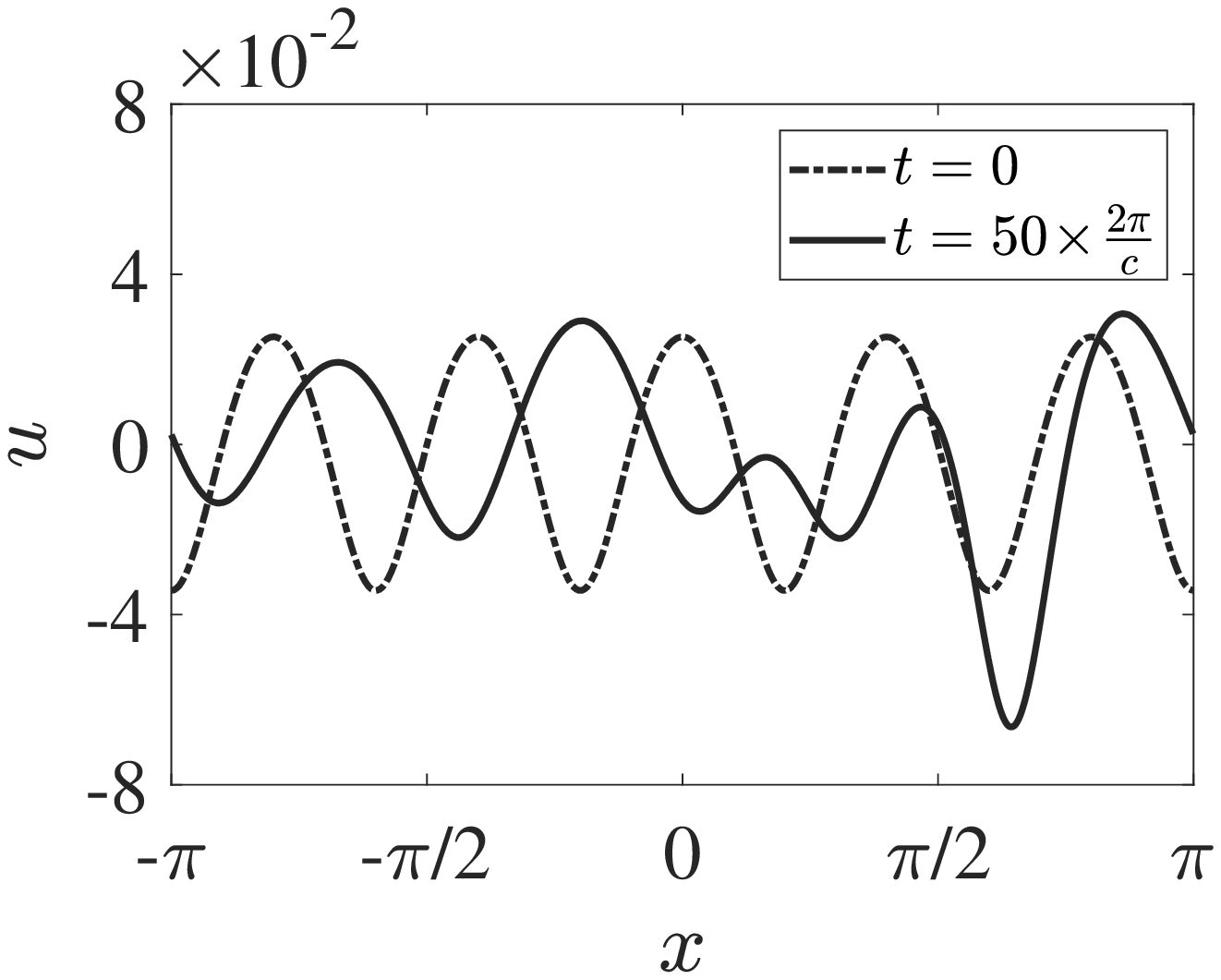}\\
\includegraphics[height=.21\textheight, angle =0]{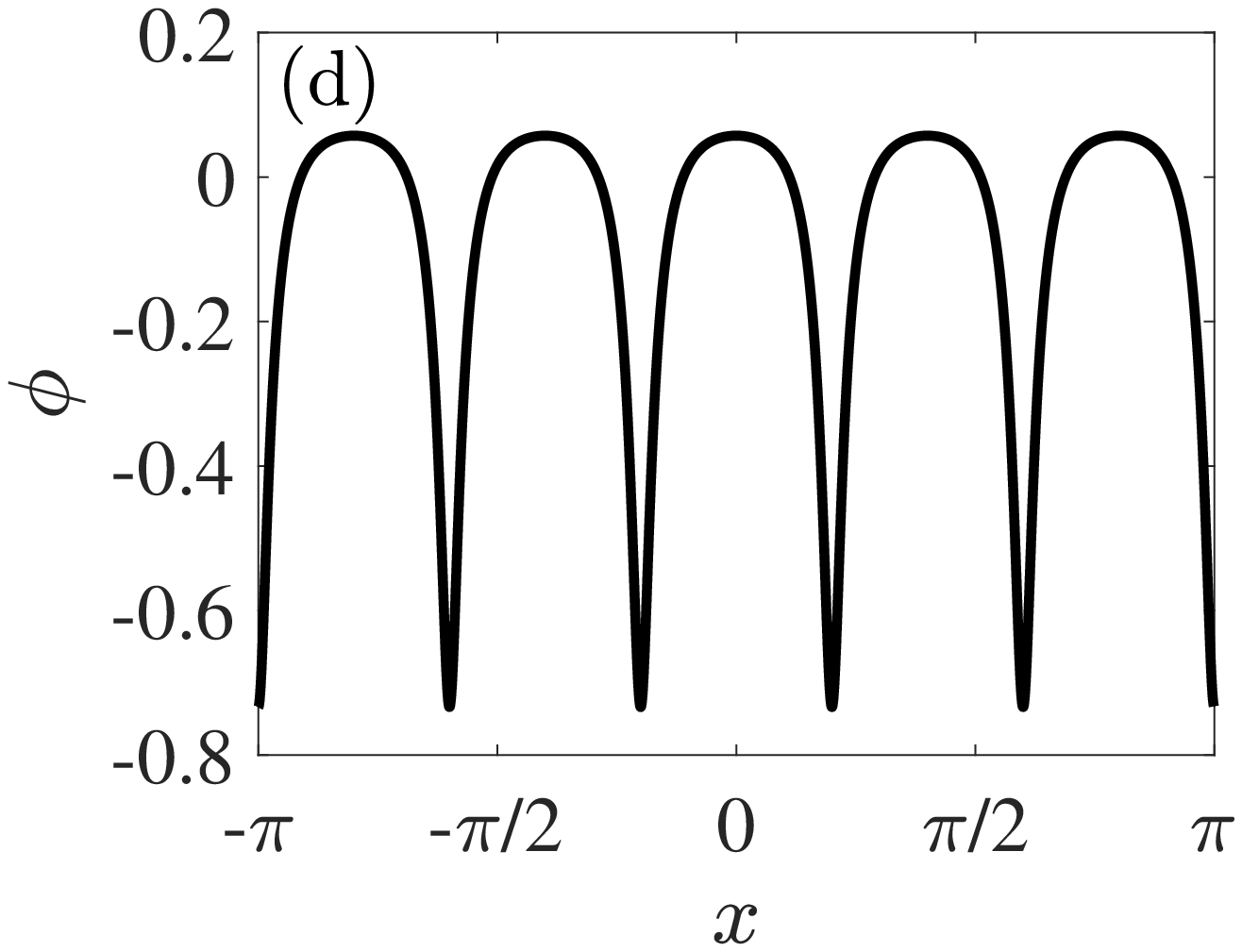}
\includegraphics[height=.21\textheight, angle =0]{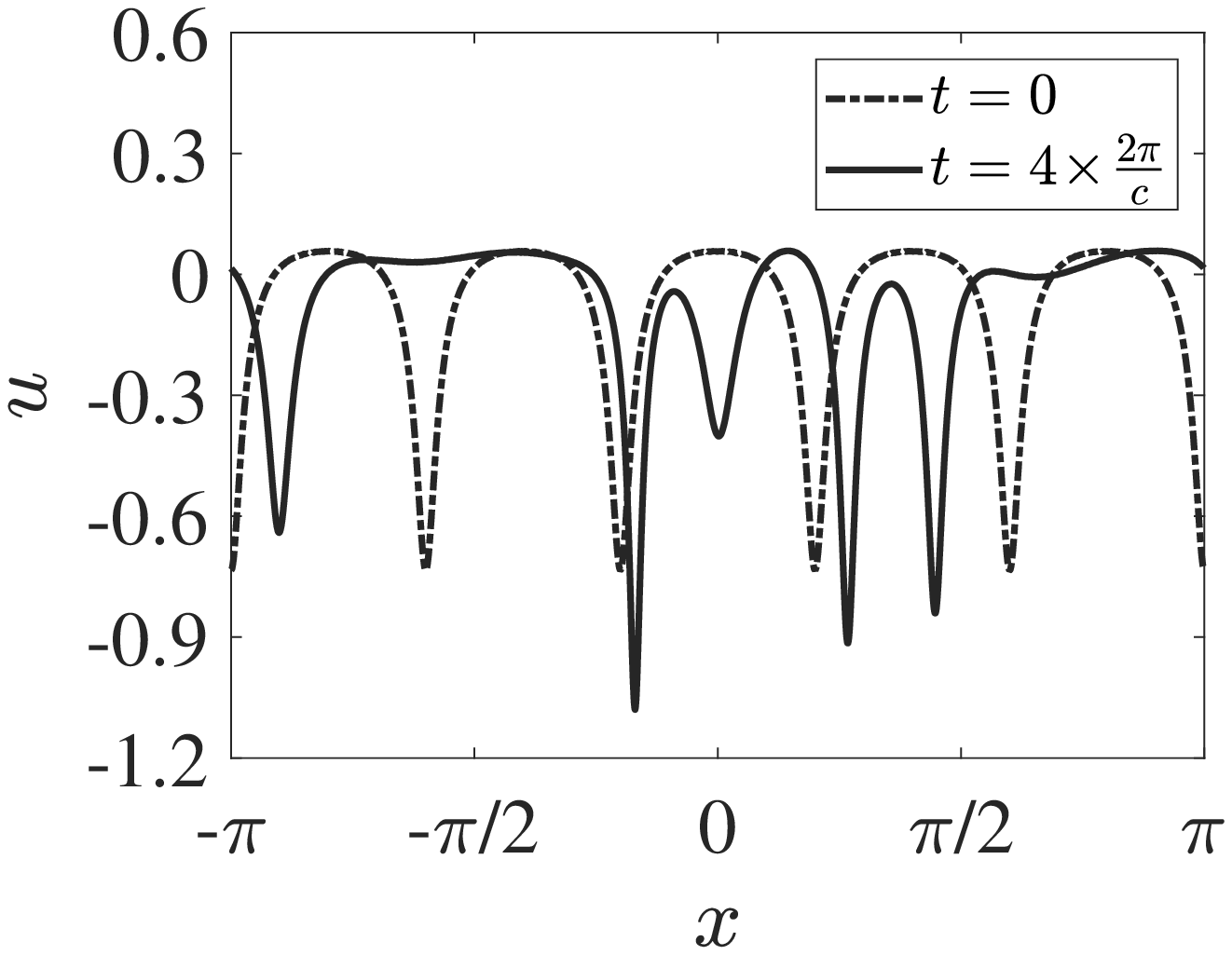}
\end{center}
\vspace*{-11pt}
\caption{
$T=T(2,5)$. Similar to Figure~\ref{fig12} but $k=5$, $c=0.805$ (c) and $0.4$ (d).}
\label{fig13}
\end{figure}

\begin{figure}[htp]
\begin{center}
\includegraphics[height=.26\textheight, angle =0]{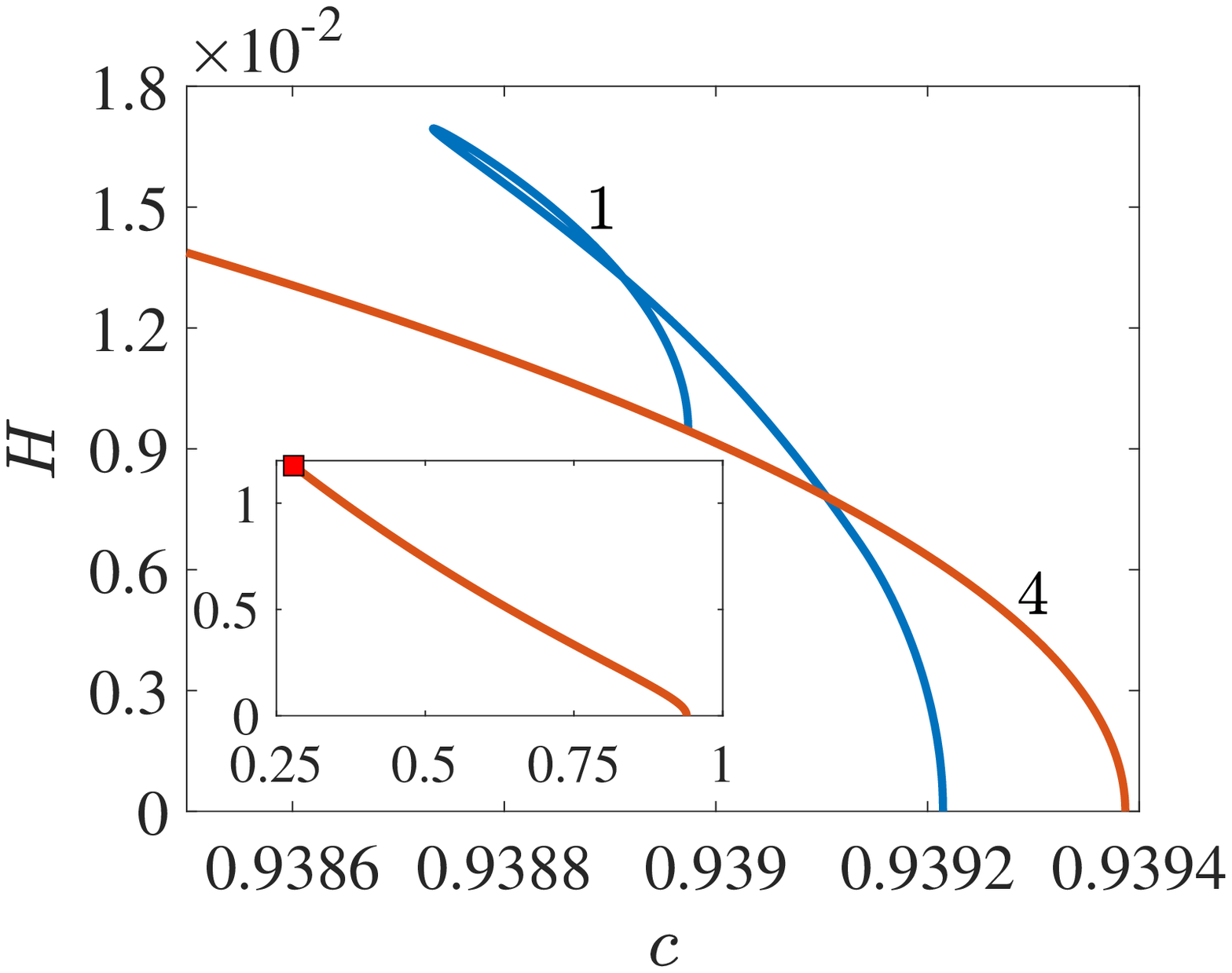}
\includegraphics[height=.26\textheight, angle =0]{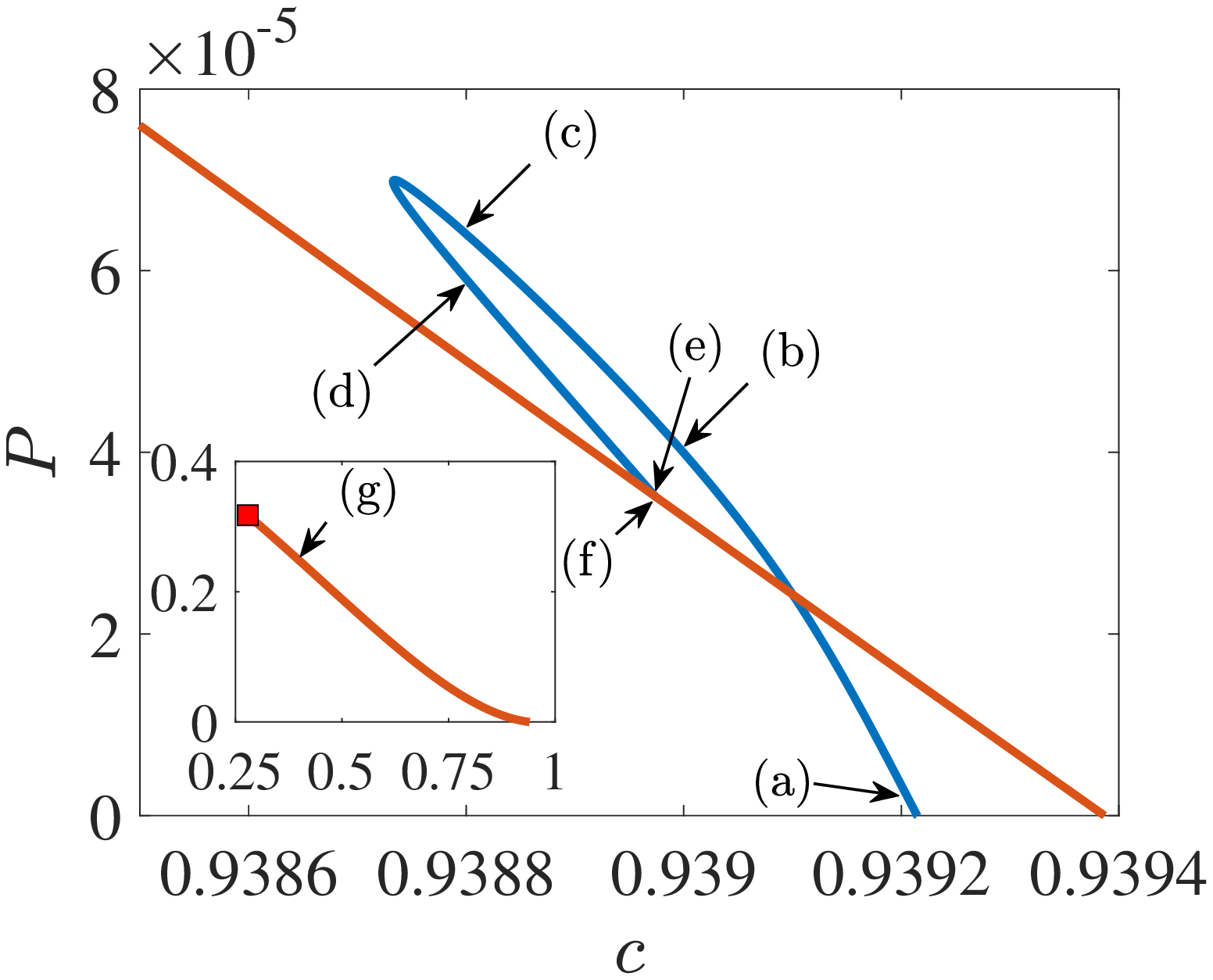}
\end{center}
\vspace*{-11pt}
\caption{
$T=T(1,4)+0.0001$. $H$ (left) and $P$ (right) vs. $c$ for $k=1$ (blue) and $k=4$ (orange). The red square in the insets corresponds to the limiting admissible solution for $k=4$, for which $c \approx 0.2788888416$. See Figures~\ref{fig15}, \ref{fig16}, \ref{fig17} for the profiles at the points labelled with (a)-(g) in the right panel.}
\label{fig14}
\end{figure}

\begin{figure}[htp]
\begin{center}
\includegraphics[height=.21\textheight, angle =0]{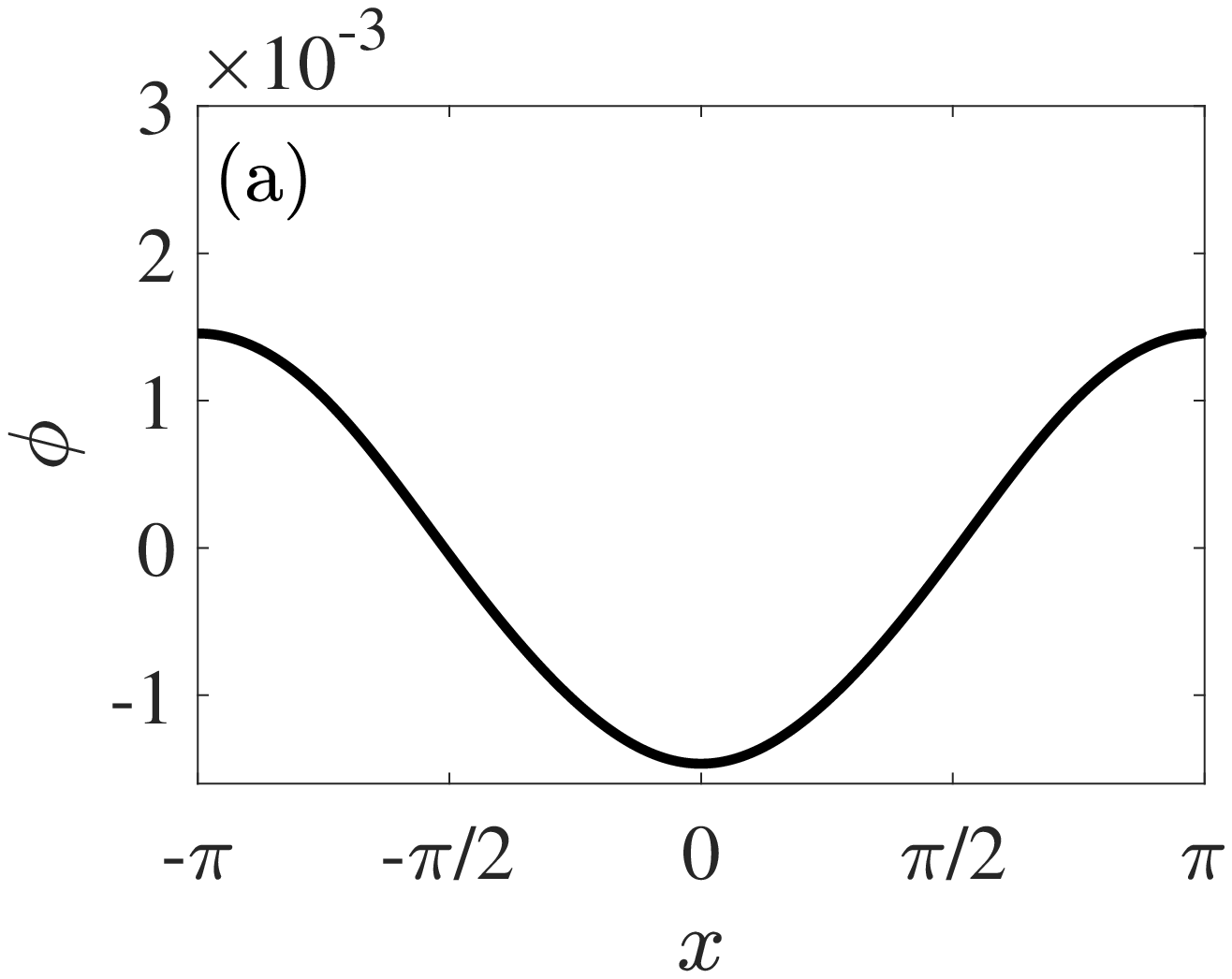}
\includegraphics[height=.21\textheight, angle =0]{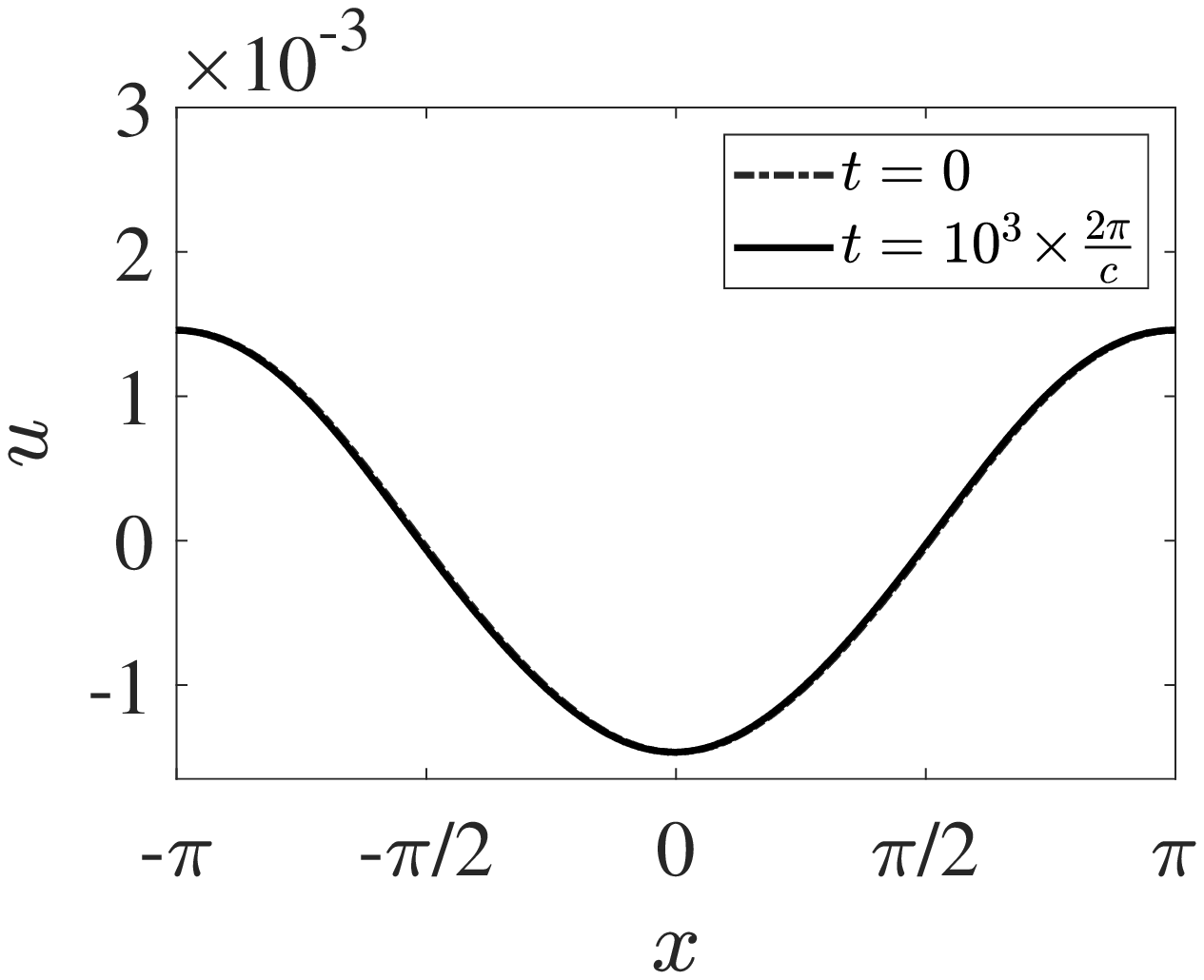}\\
\includegraphics[height=.21\textheight, angle =0]{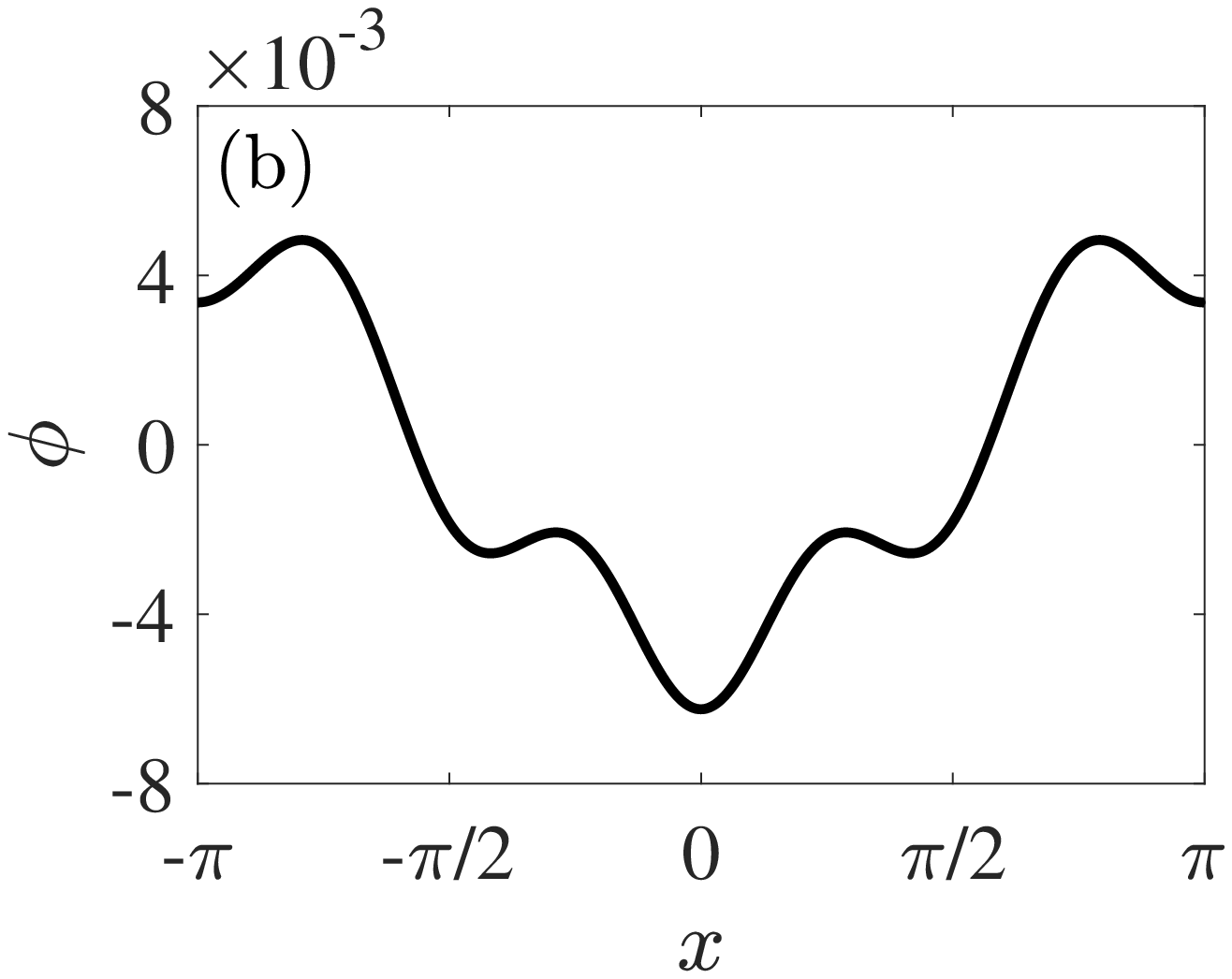}
\includegraphics[height=.21\textheight, angle =0]{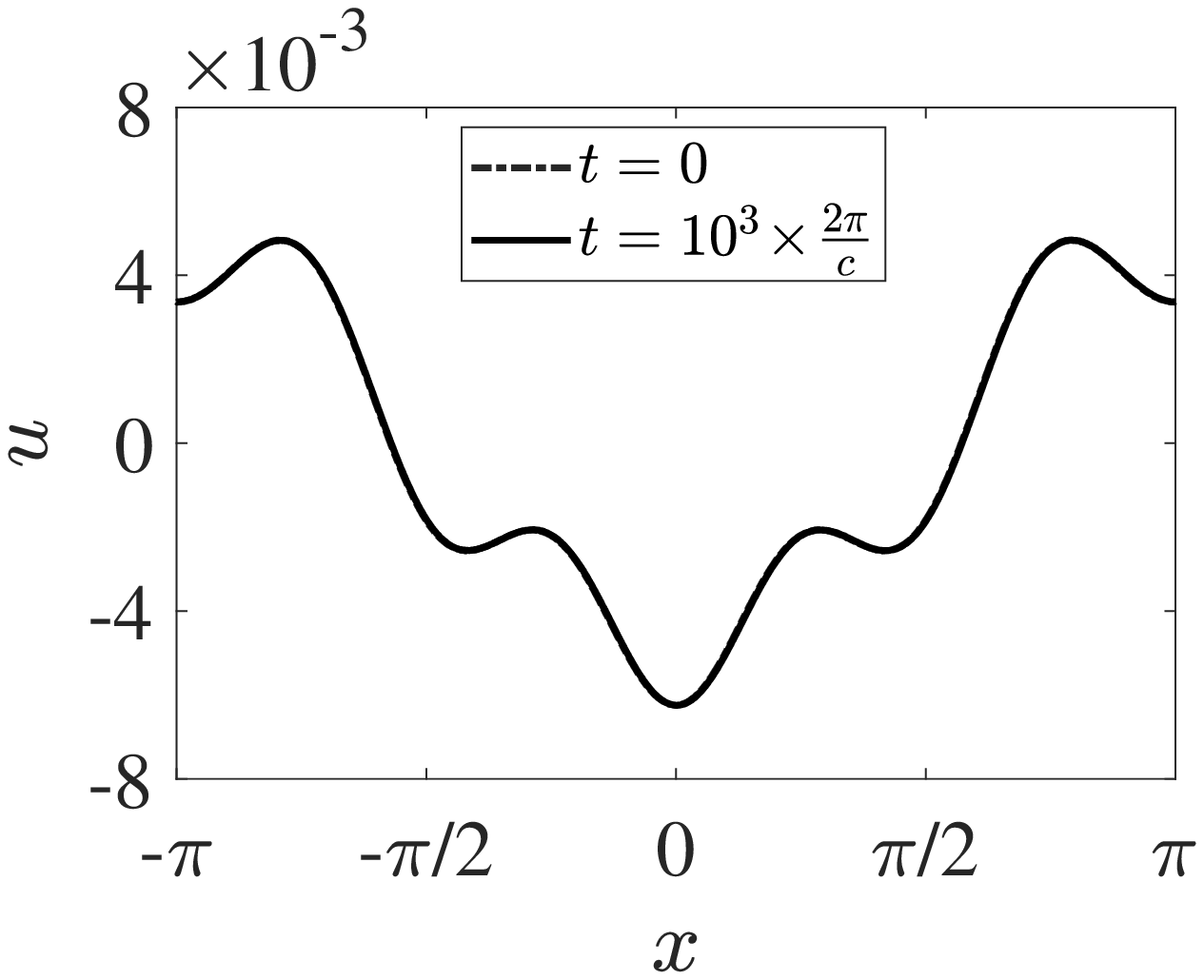}\\
\includegraphics[height=.21\textheight, angle =0]{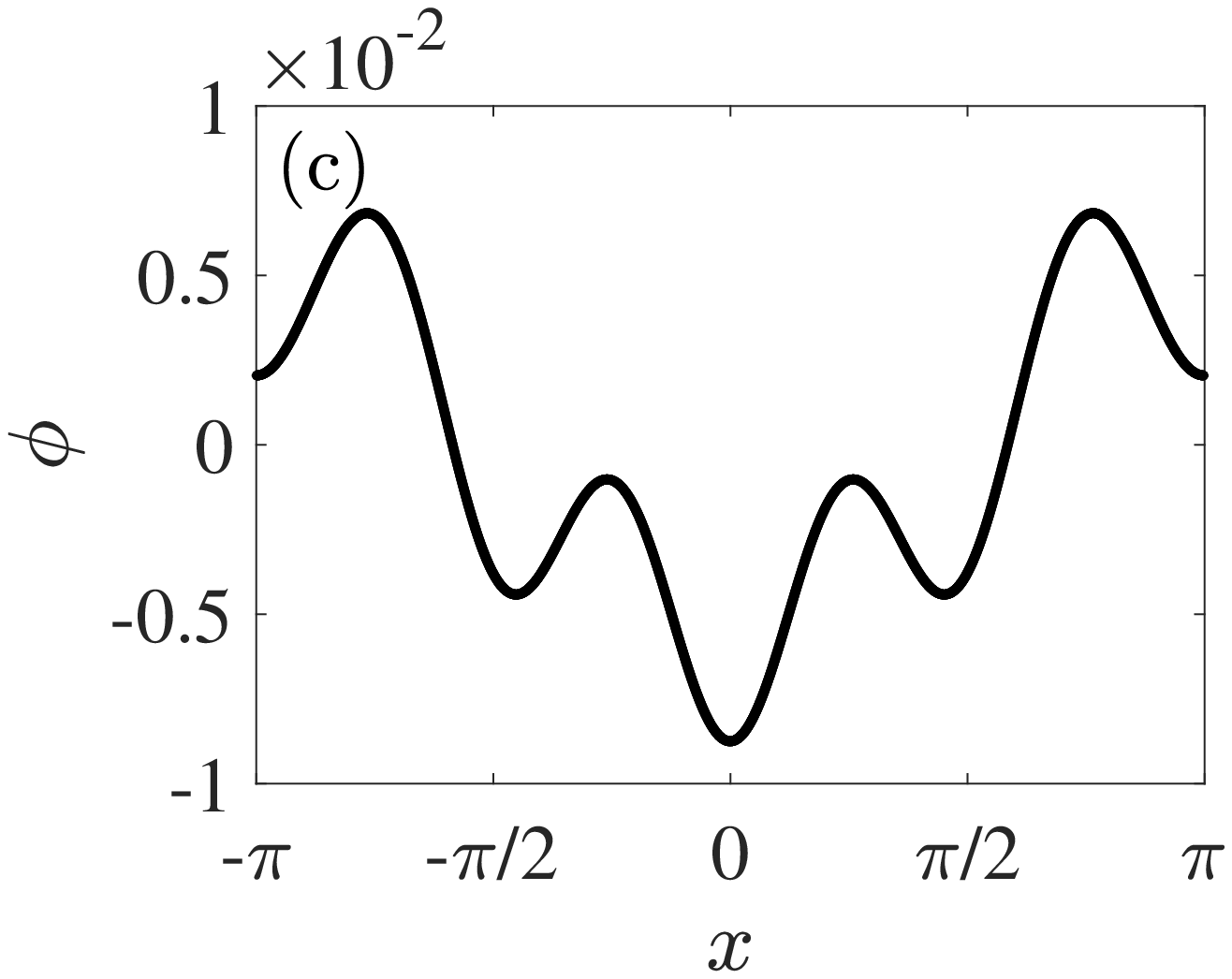}
\includegraphics[height=.21\textheight, angle =0]{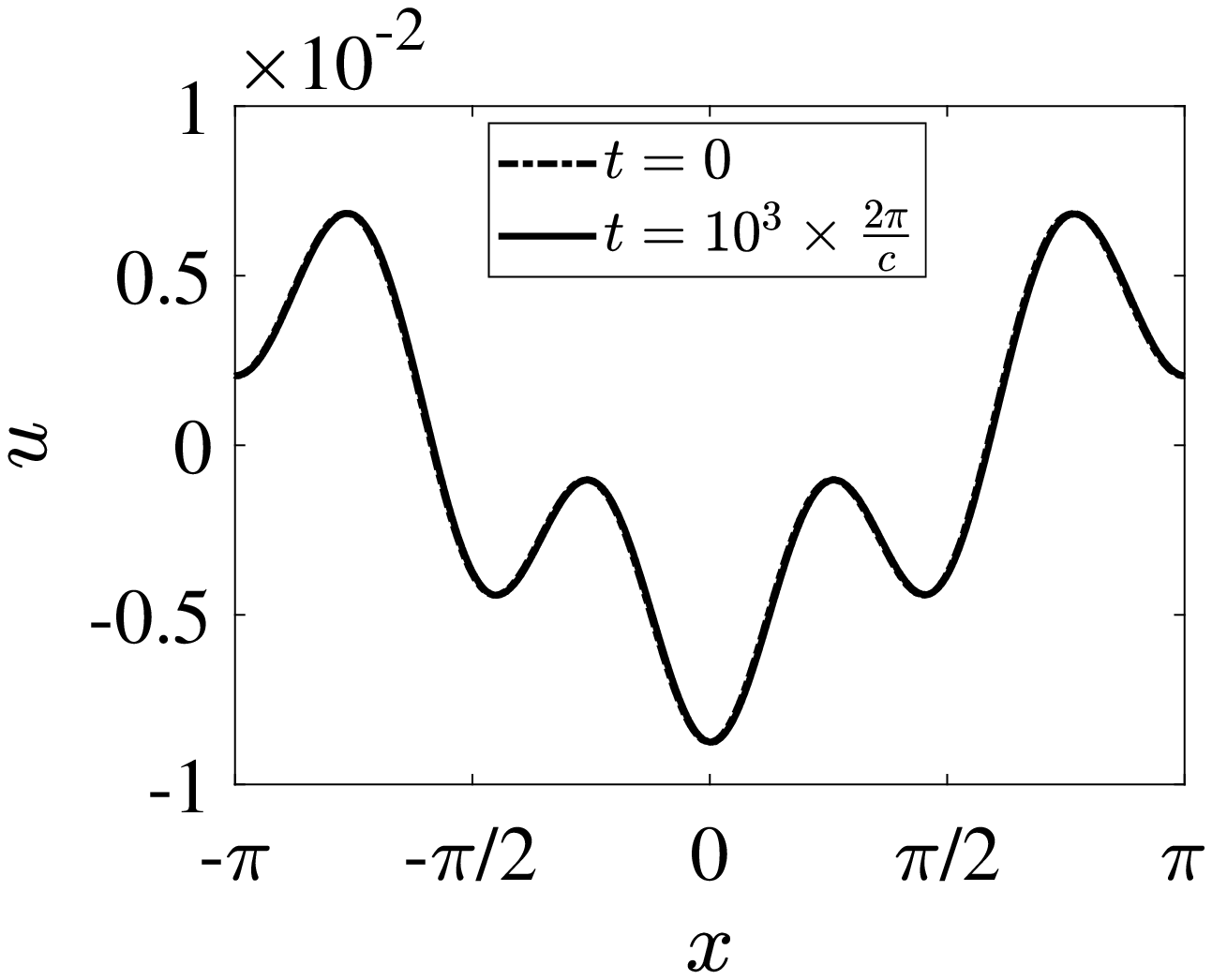}\\
\includegraphics[height=.21\textheight, angle =0]{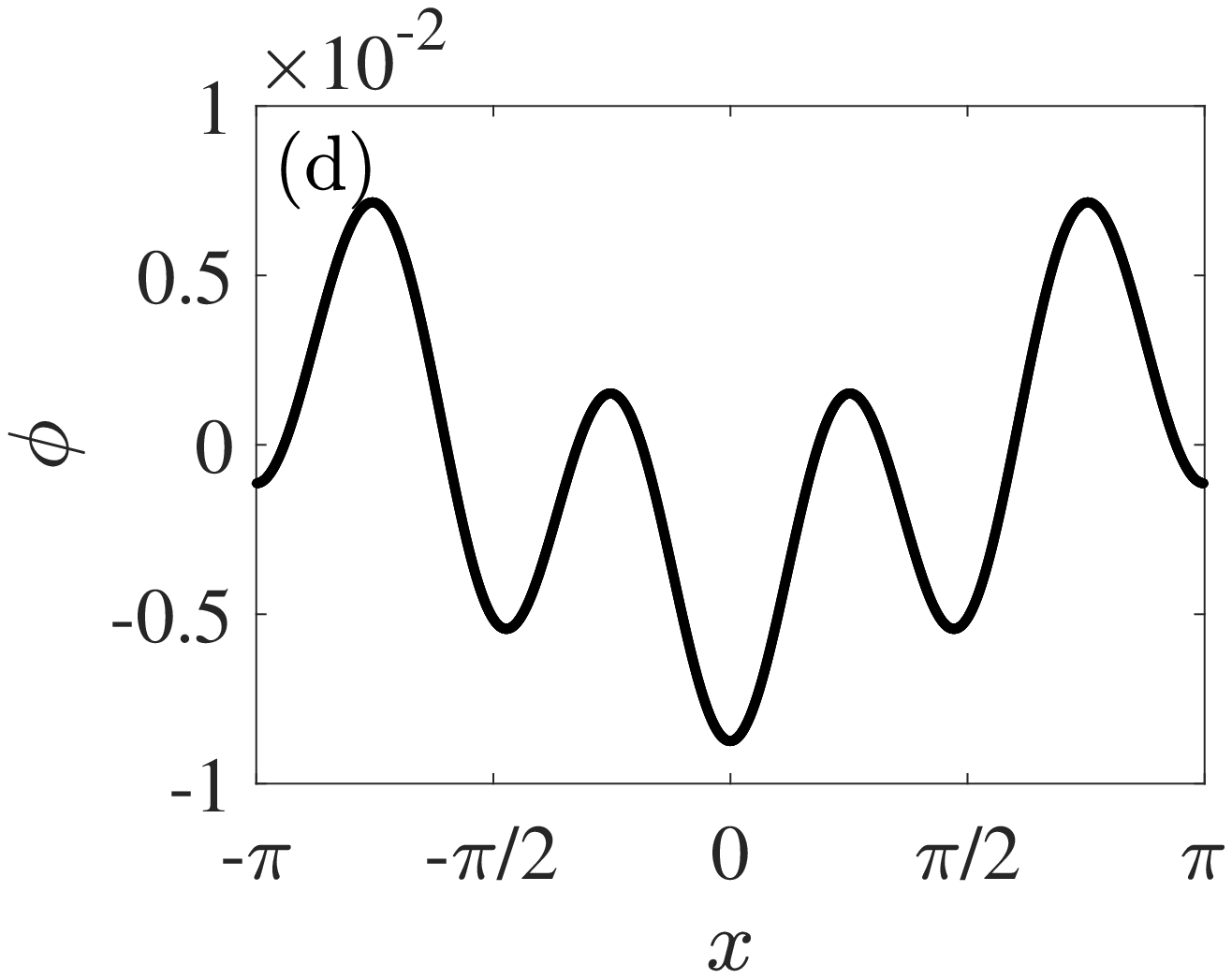}
\includegraphics[height=.21\textheight, angle =0]{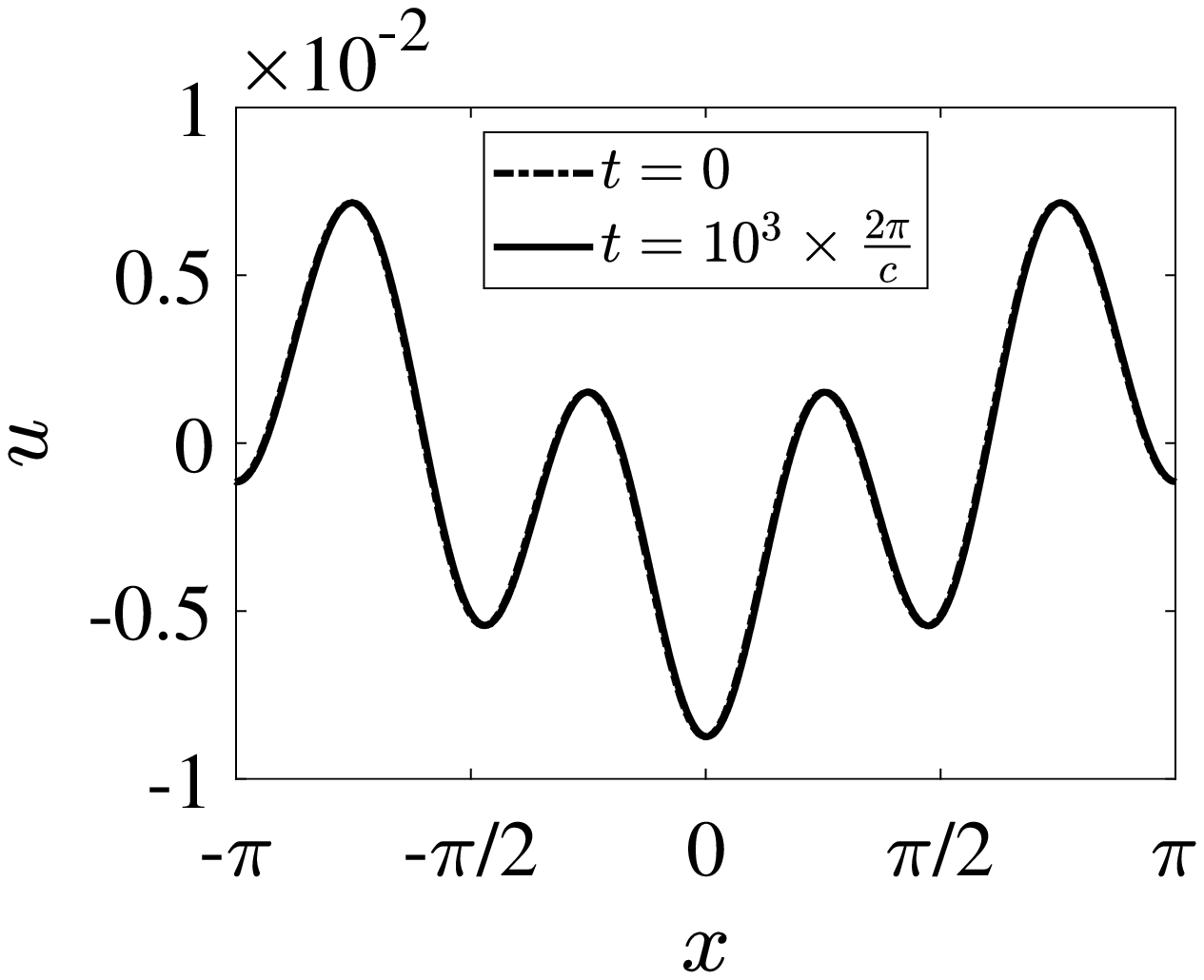}
\end{center}
\vspace*{-11pt}
\caption{
$T=T(1,4)+0.0001$, $k=1$. Left column: profiles at the points labelled with (a)-(d) in the right panel of Figure~\ref{fig14}, prior to (a) and past (b) crossing the $k=4$ branch, and prior to (c) and past (d) the turning point of $P$, for which $c=0.9392$ (a), $0.939$ (b), and $0.9388$ (c,d). Right column: profiles perturbed by small random noise at $t=0$ (dash-dotted) and of the solutions after $10^3$ periods (solid), after translation of the $x$ axis.}
\label{fig15}
\end{figure}

\begin{figure}[htp]
\begin{center}
\includegraphics[height=.21\textheight, angle =0]{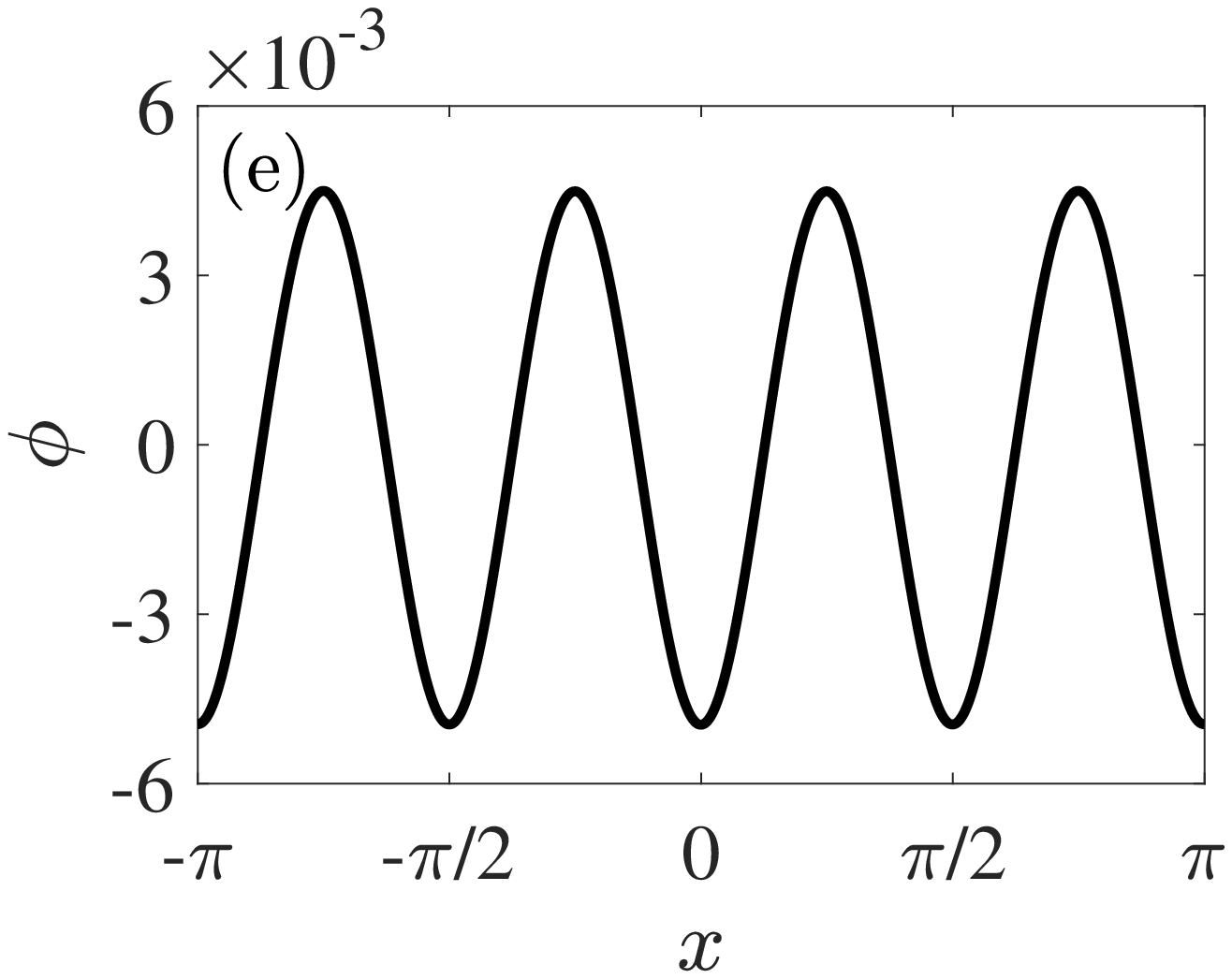}
\includegraphics[height=.21\textheight, angle =0]{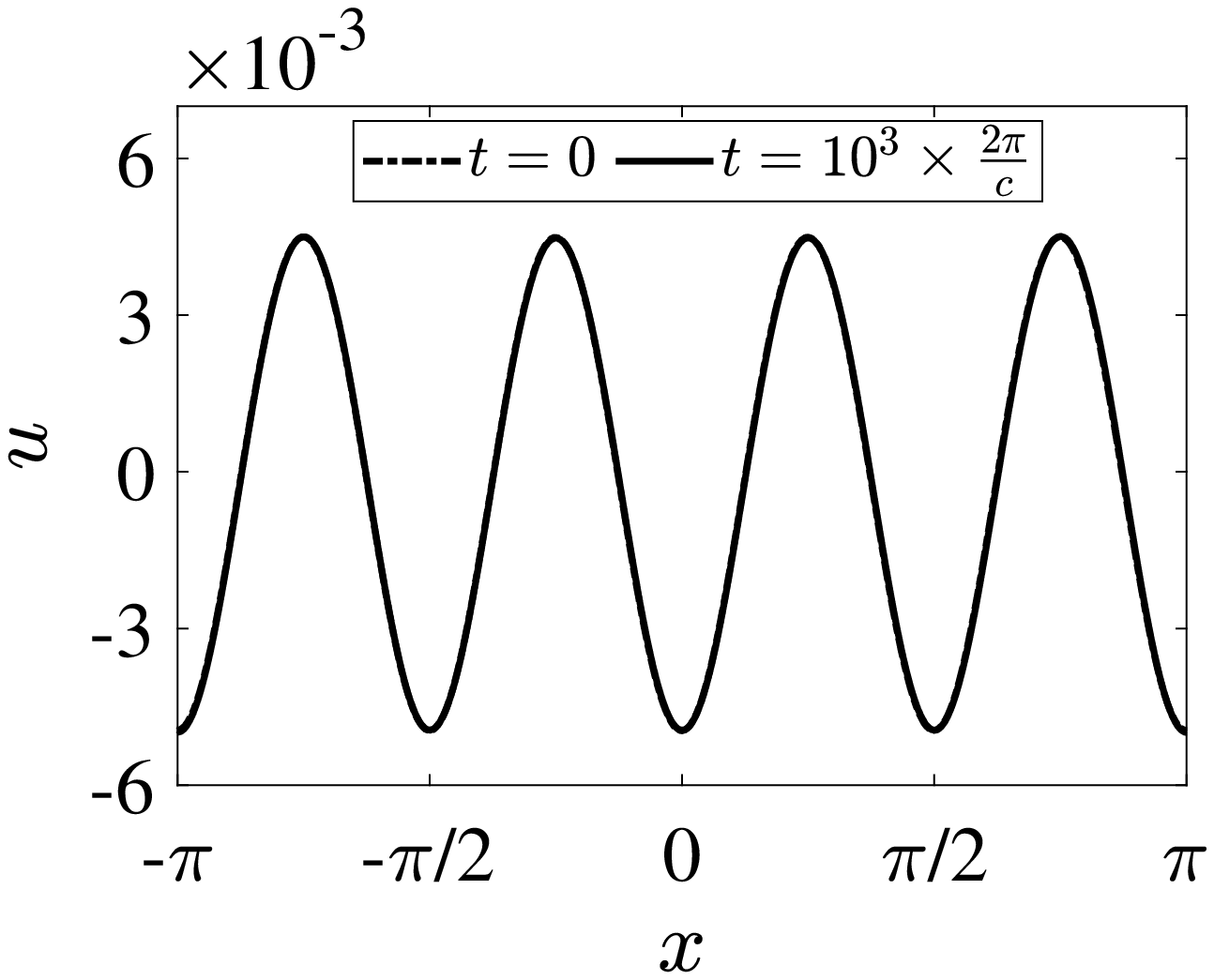}\\
\end{center}
\vspace*{-11pt}
\caption{
$T=T(1,4)+0.0001$, $k=1$. Similar to Figure~\ref{fig15} but almost connecting the $k=4$ branch, for which $c\approx0.93897389482$.}
\label{fig16}
\end{figure}

\begin{figure}[htp]
\begin{center}
\includegraphics[height=.21\textheight, angle =0]{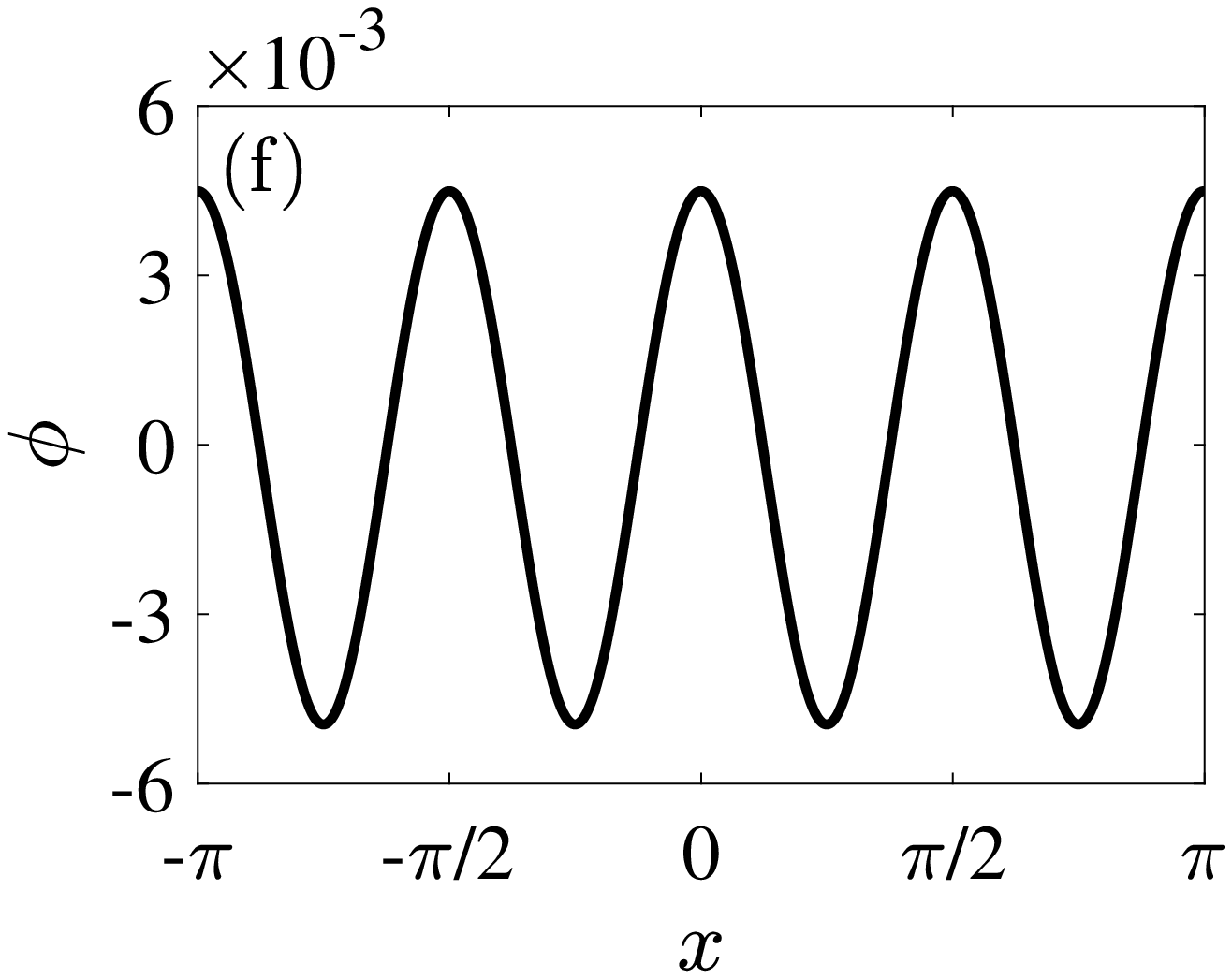}
\includegraphics[height=.21\textheight, angle =0]{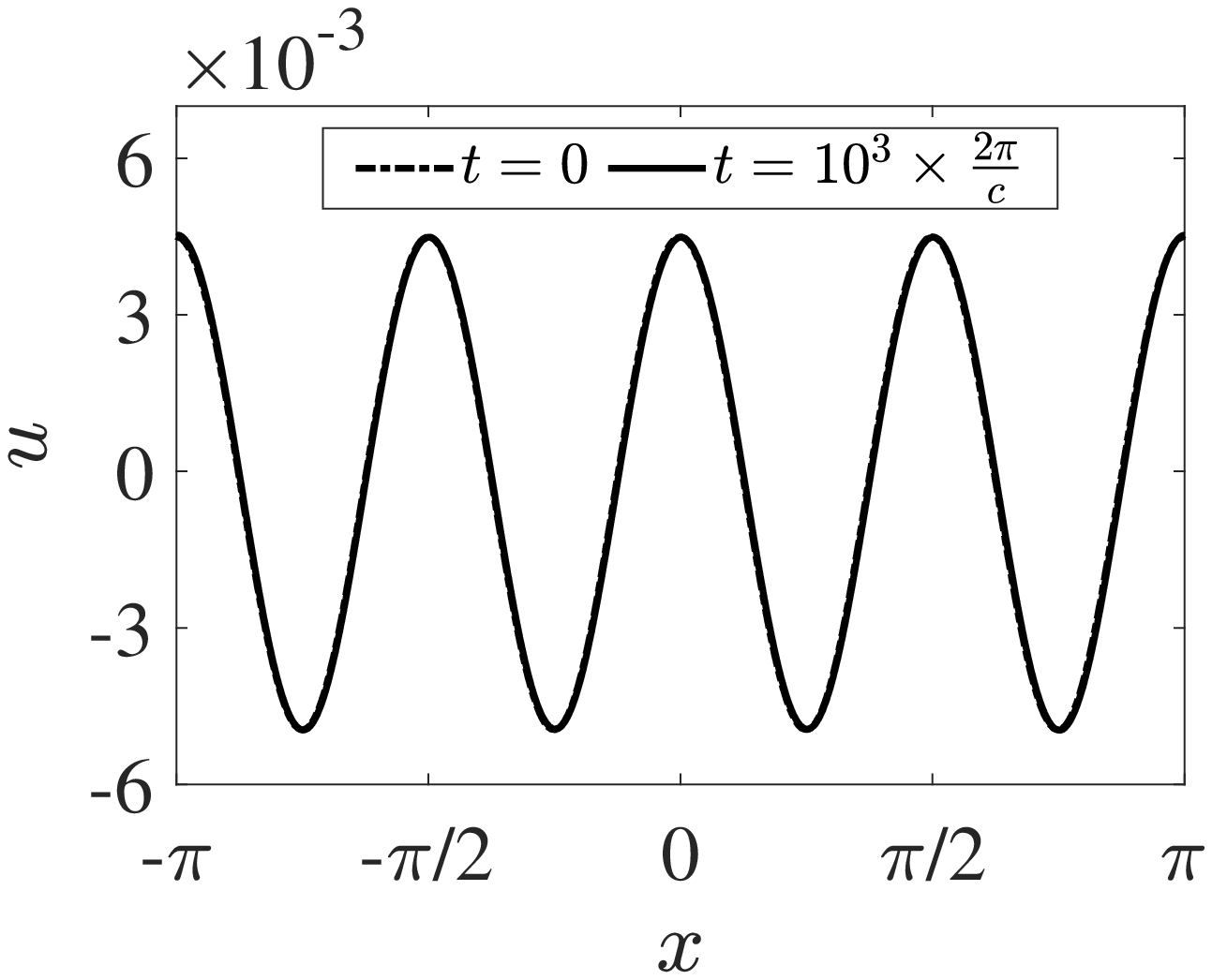}\\
\includegraphics[height=.21\textheight, angle =0]{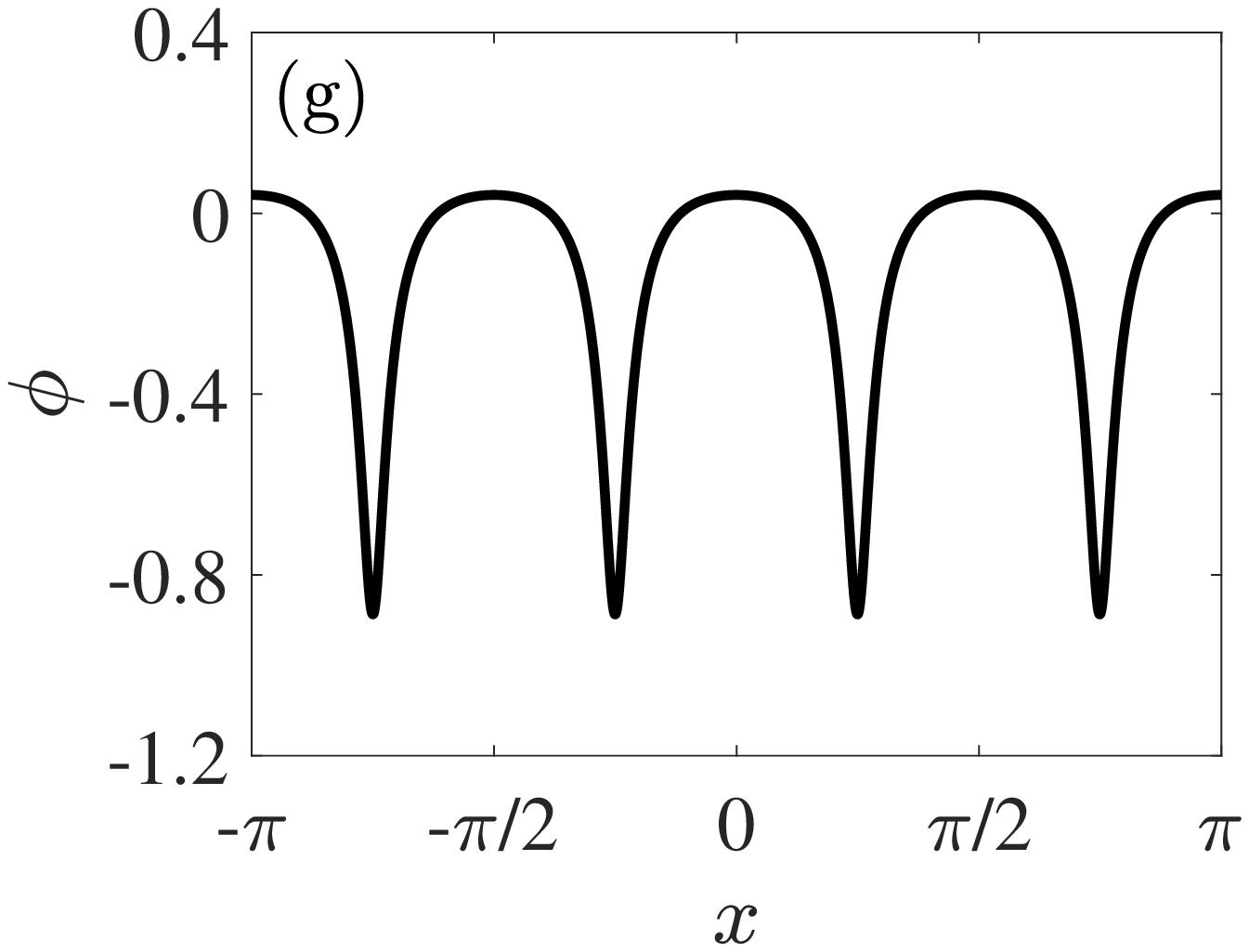}
\includegraphics[height=.21\textheight, angle =0]{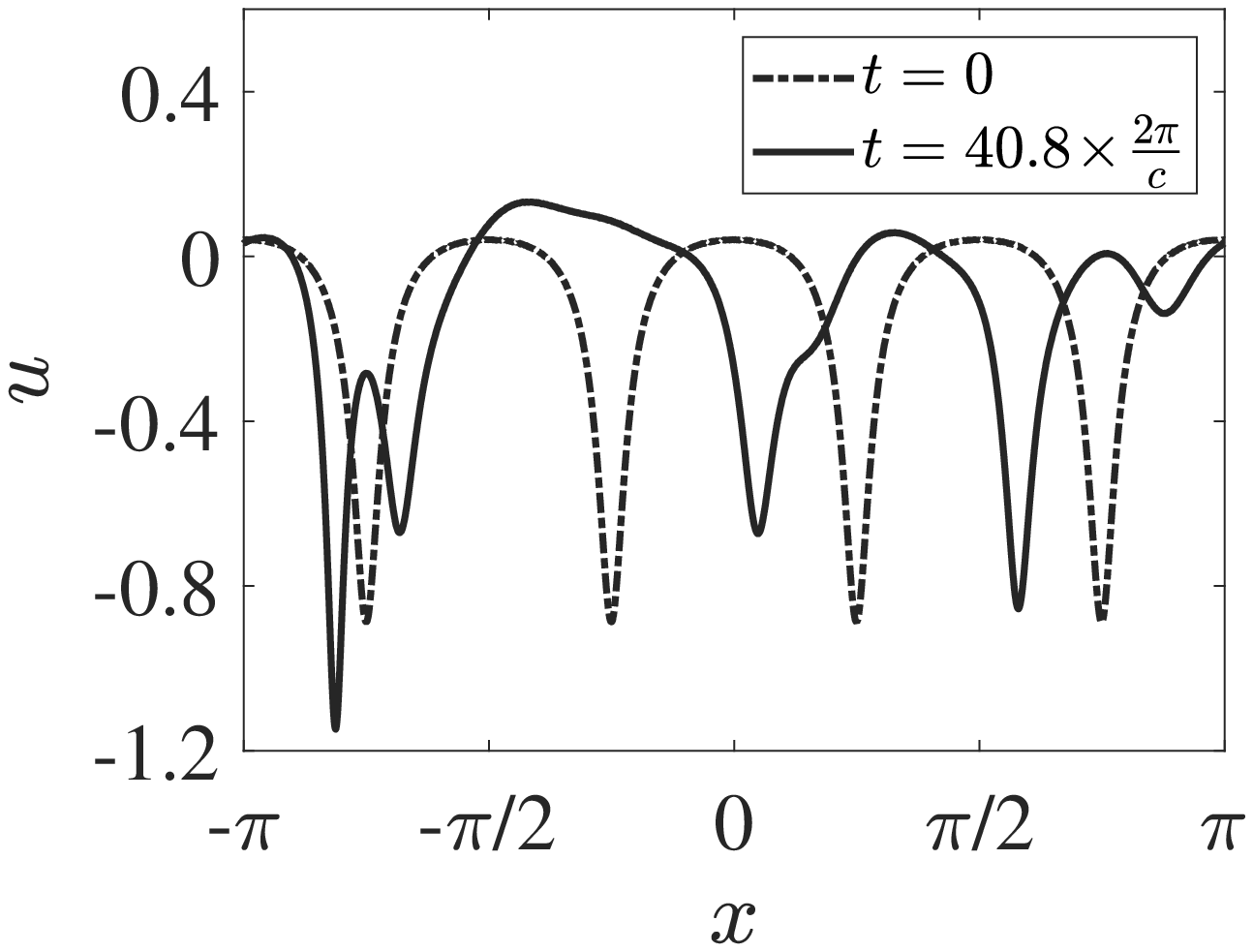}
\end{center}
\vspace*{-11pt}
\caption{
$T=T(1,4)+0.0001$. Similar to Figures~\ref{fig15} and \ref{fig16}, but $k=4$, almost connecting the $k=1$ branch (f) and almost the end of the numerical continuation (g), for which $c\approx0.93897389482$ (f) and $c=0.4$ (g).}
\label{fig17}
\end{figure}

\begin{figure}[htp]
\begin{center}
\includegraphics[height=.26\textheight, angle =0]{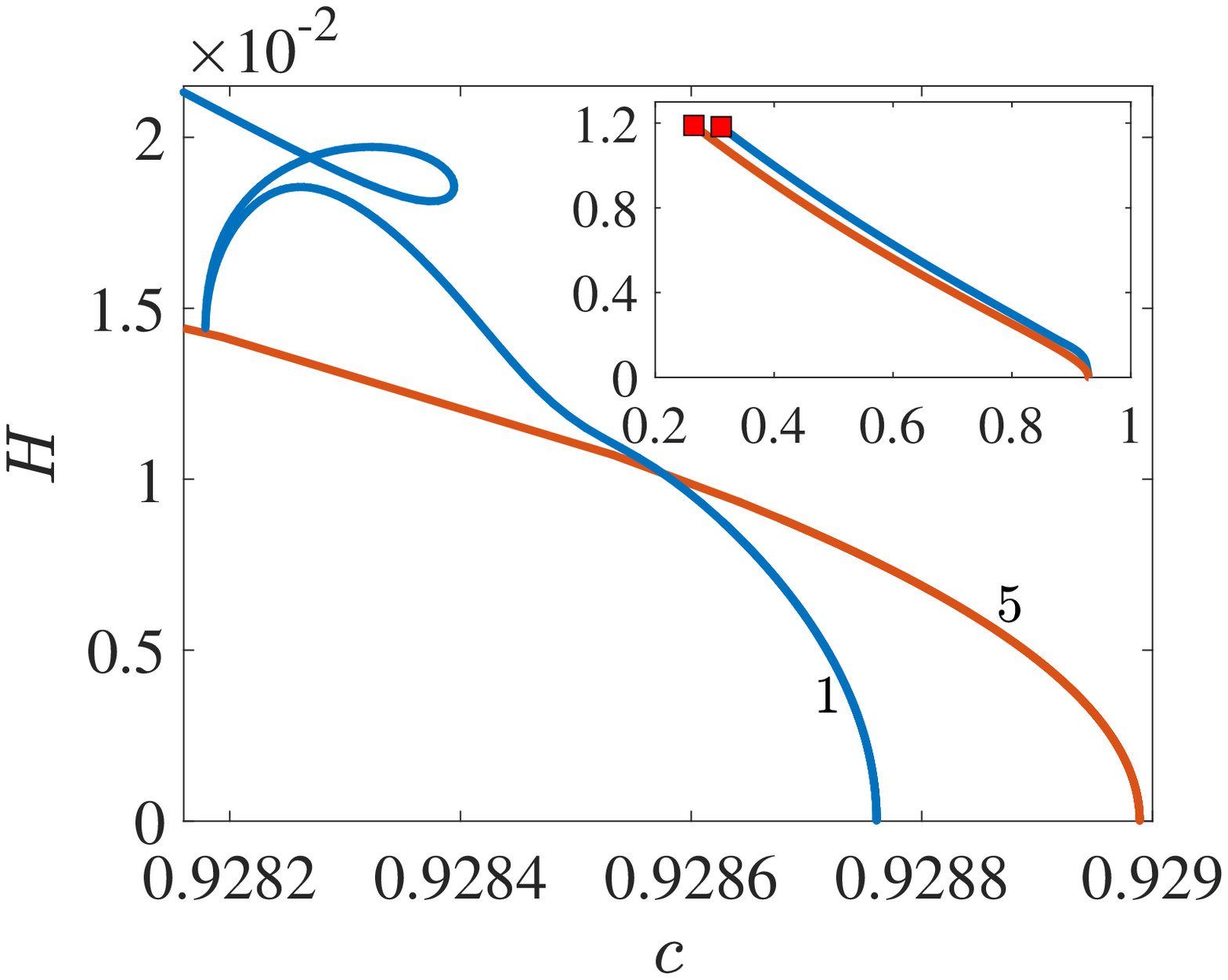}
\includegraphics[height=.26\textheight, angle =0]{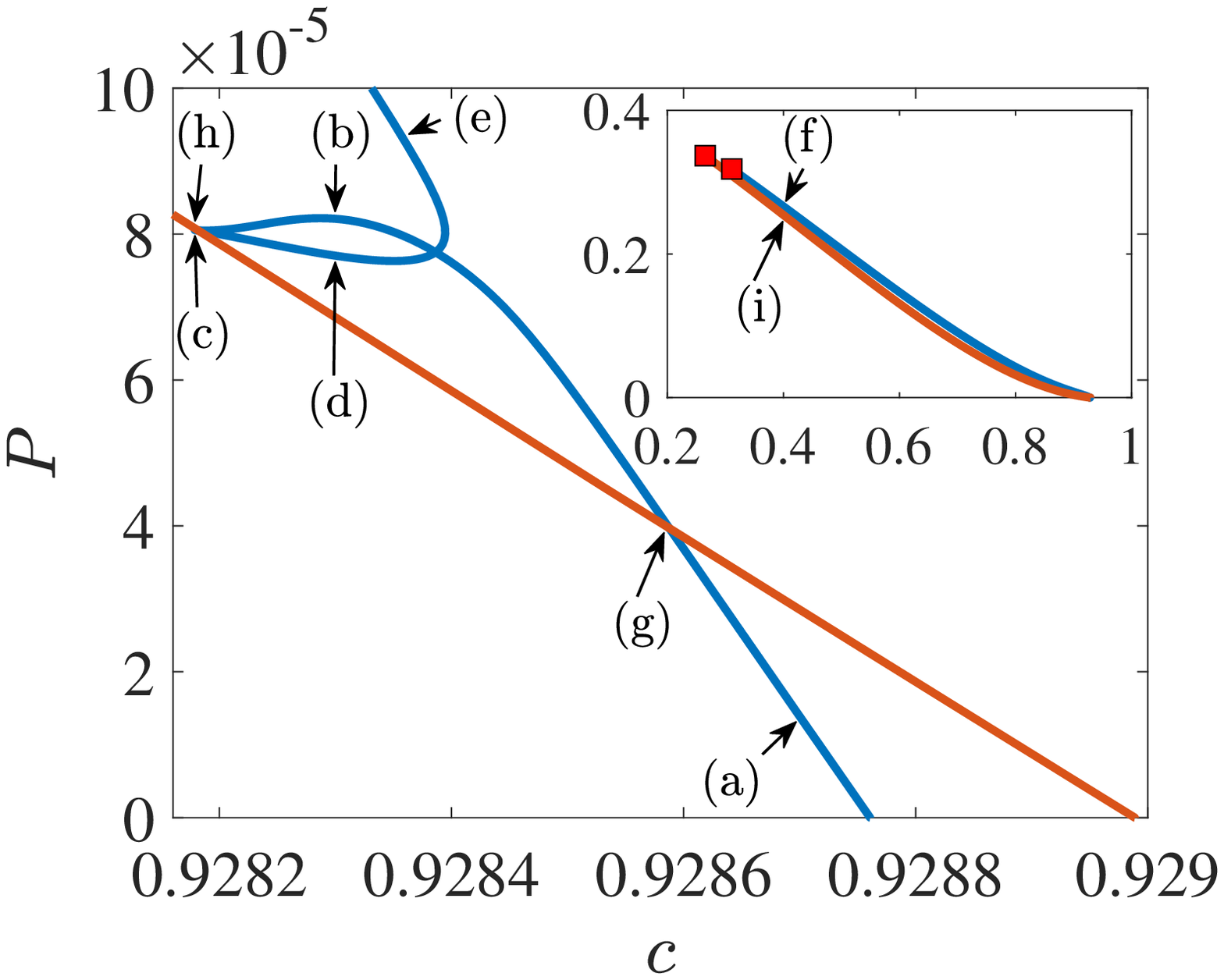}
\end{center}
\vspace*{-11pt}
\caption{
%
$T=T(1,5)+0.0001$. $H$ and $P$ vs. $c$ for $k=1$ (blue) and $k=5$ (orange). The red squares in the insets correspond to the limiting admissible solutions, for which $c \approx 0.3109760939$ ($k=1$) and  $c \approx 0.26500852187$ ($k=5$). See Figures~\ref{fig19}, \ref{fig20}, \ref{fig21} for the profiles at the points labelled with (a) through (i).}
\label{fig18}
\end{figure}

\begin{figure}[htp]
\begin{center}
\includegraphics[height=.21\textheight, angle =0]{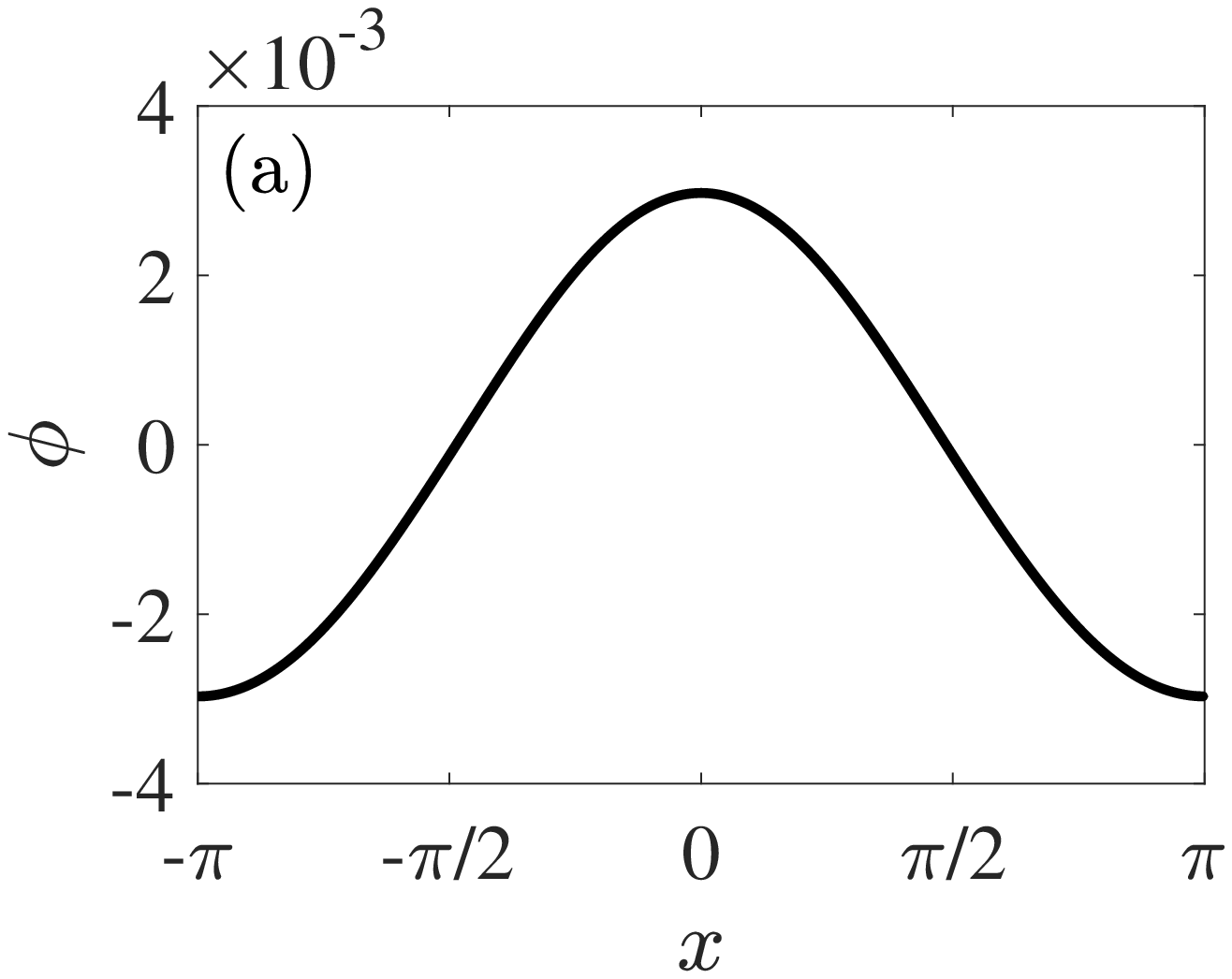}
\includegraphics[height=.21\textheight, angle =0]{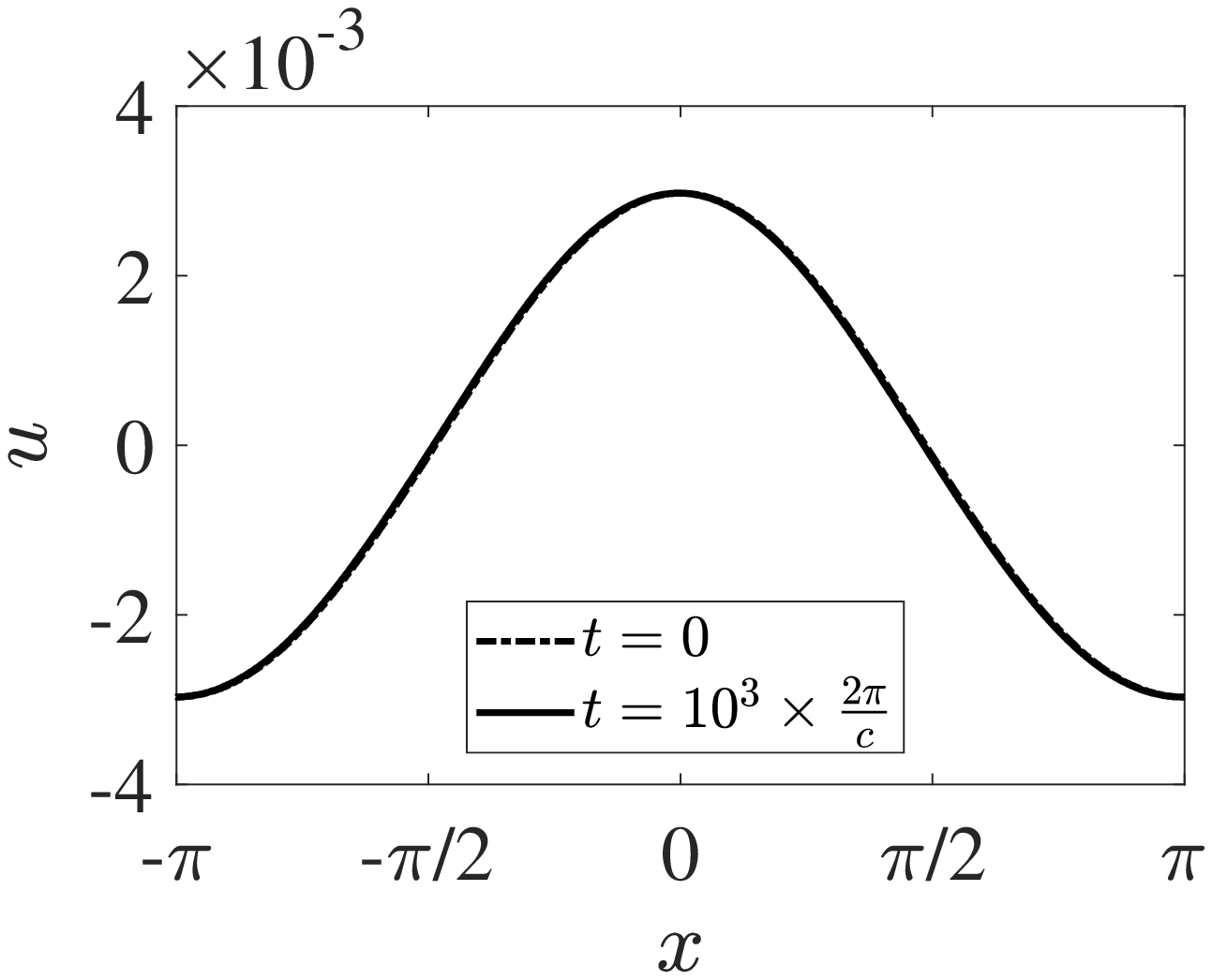}\\
\includegraphics[height=.21\textheight, angle =0]{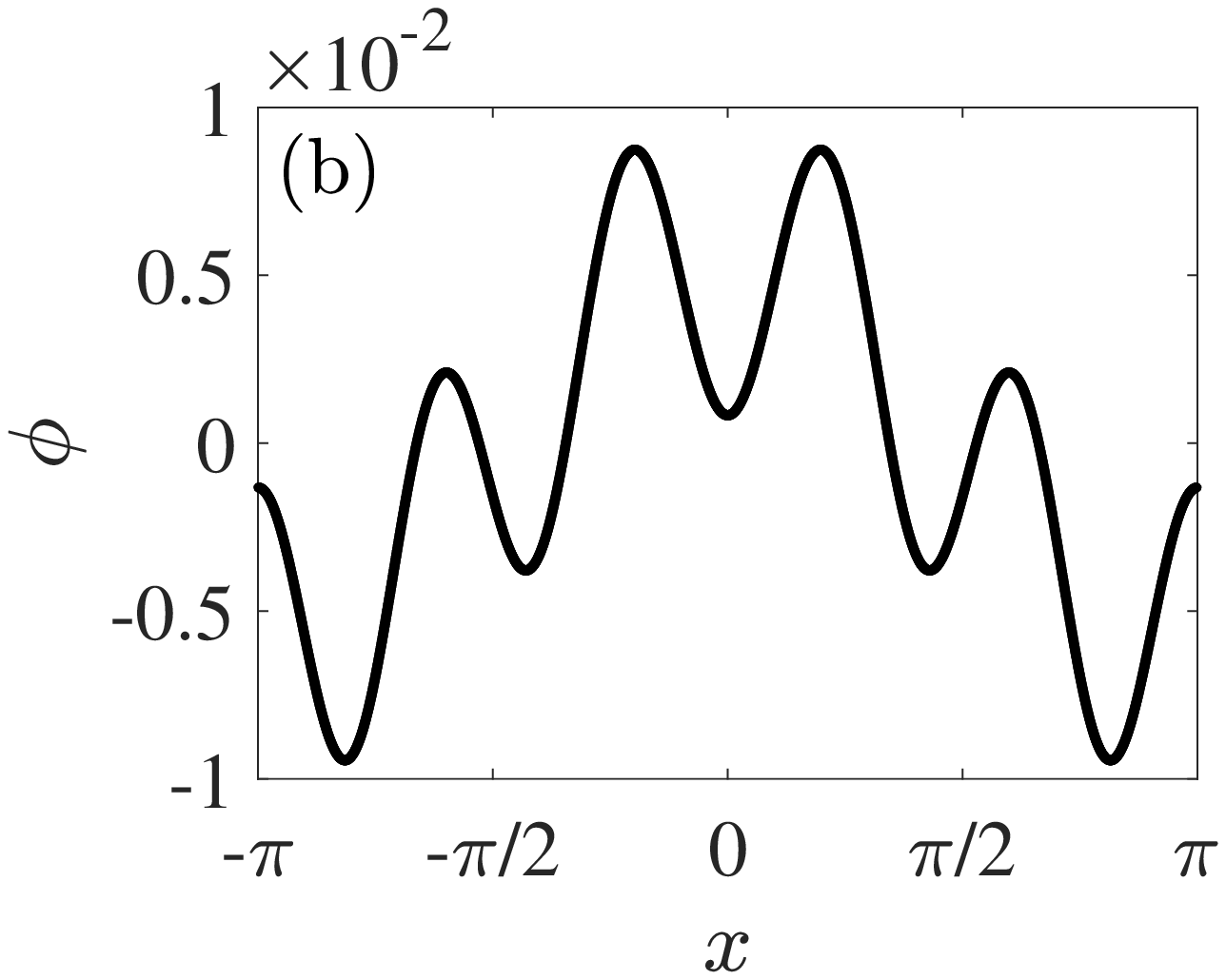}
\includegraphics[height=.21\textheight, angle =0]{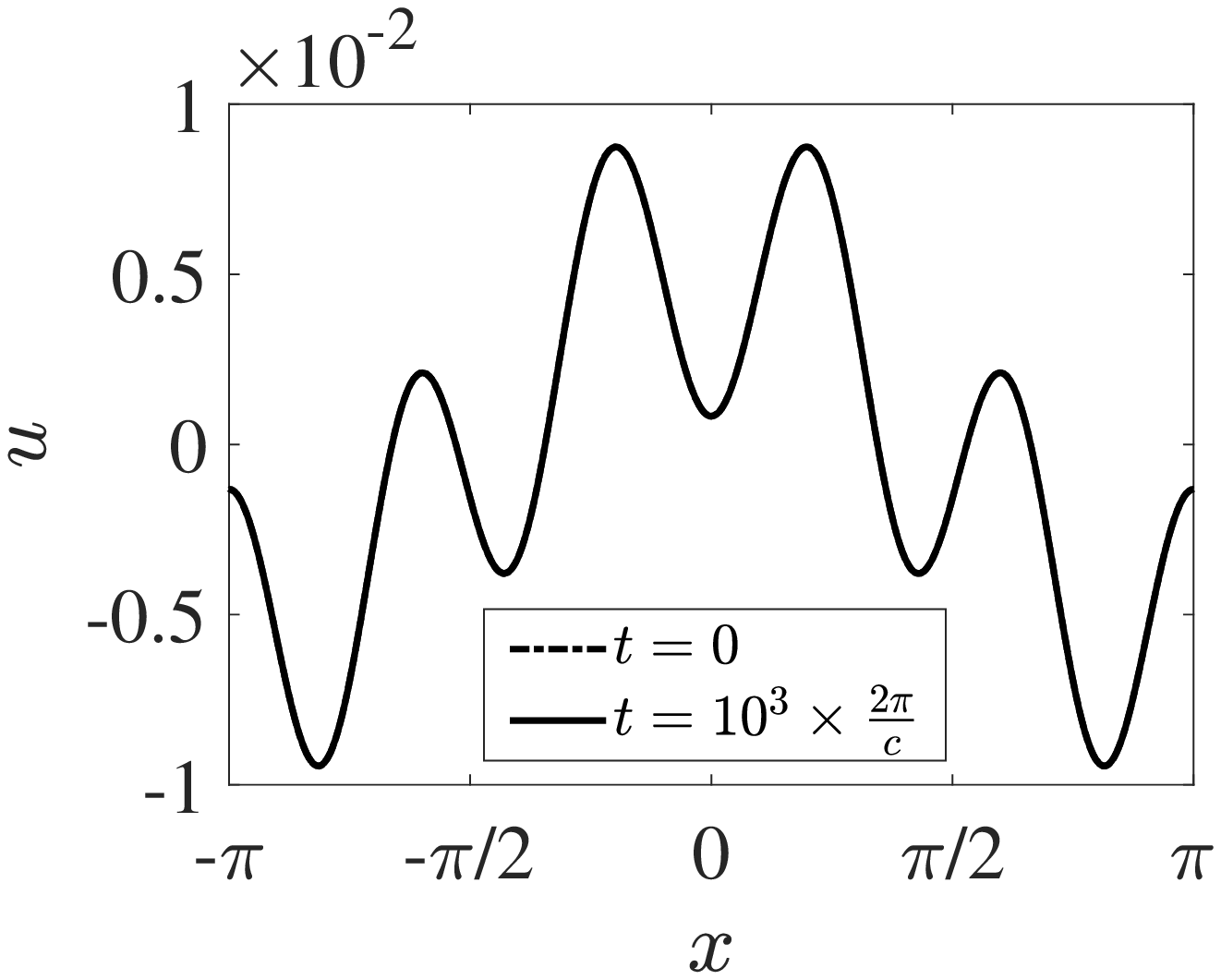}
\end{center}
\vspace*{-11pt}
\caption{
$T=T(1,5)+0.0001$, $k=1$. Left column: profiles at the points labelled with (a) and (b) in the right panel of Figure~\ref{fig18}, prior to (a) and past (b) crossing the $k=5$ branch, for which $c=0.9287$ (a) and $0.9283$ (b). Right column: profiles perturbed by small random noise at $t=0$ (dash-dotted) and of the solutions after $10^3$ periods (solid), after translation of the $x$ axis (a).}
\label{fig19}
\end{figure}

\begin{figure}[htp]
\begin{center}
\includegraphics[height=.21\textheight, angle =0]{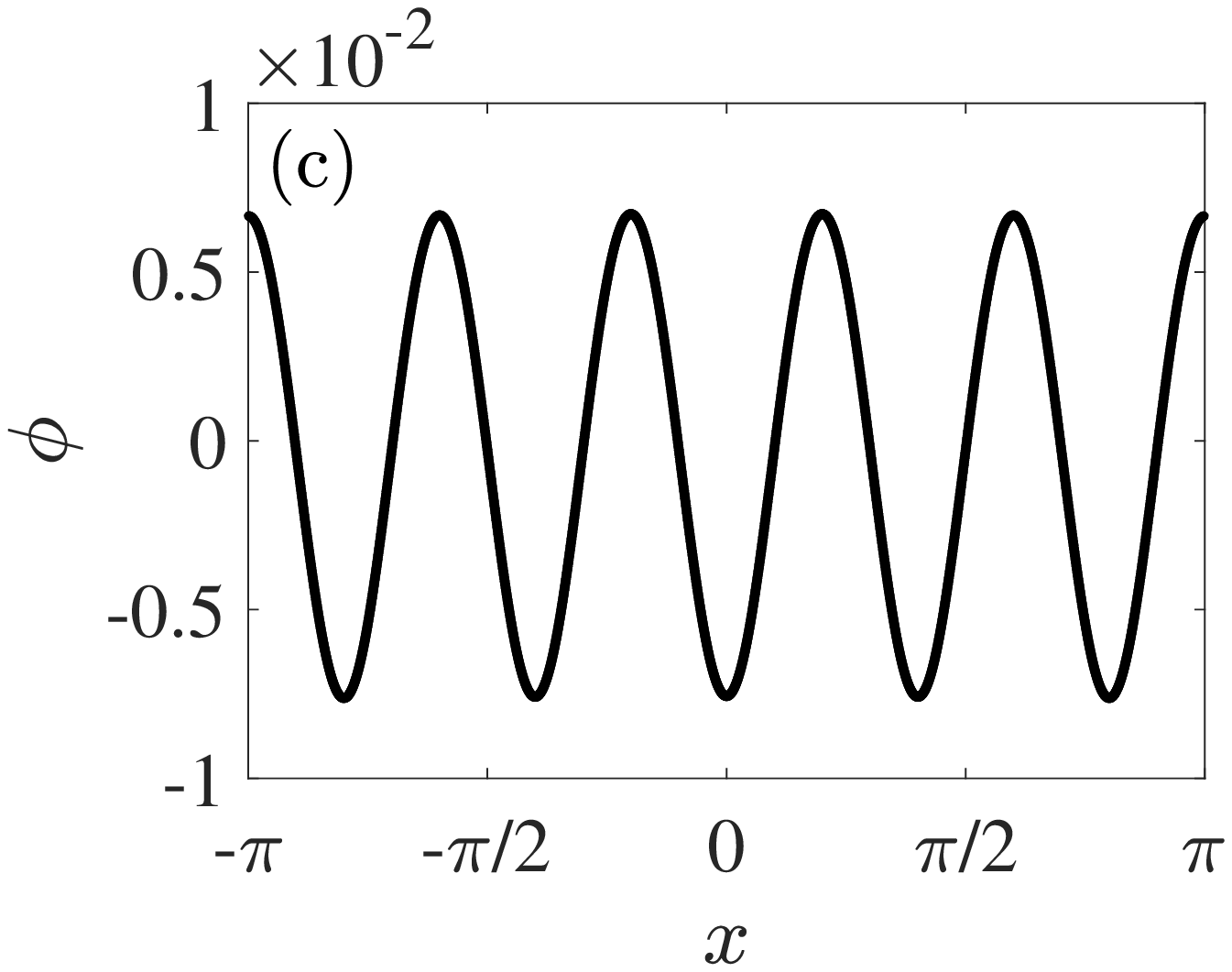}
\includegraphics[height=.21\textheight, angle =0]{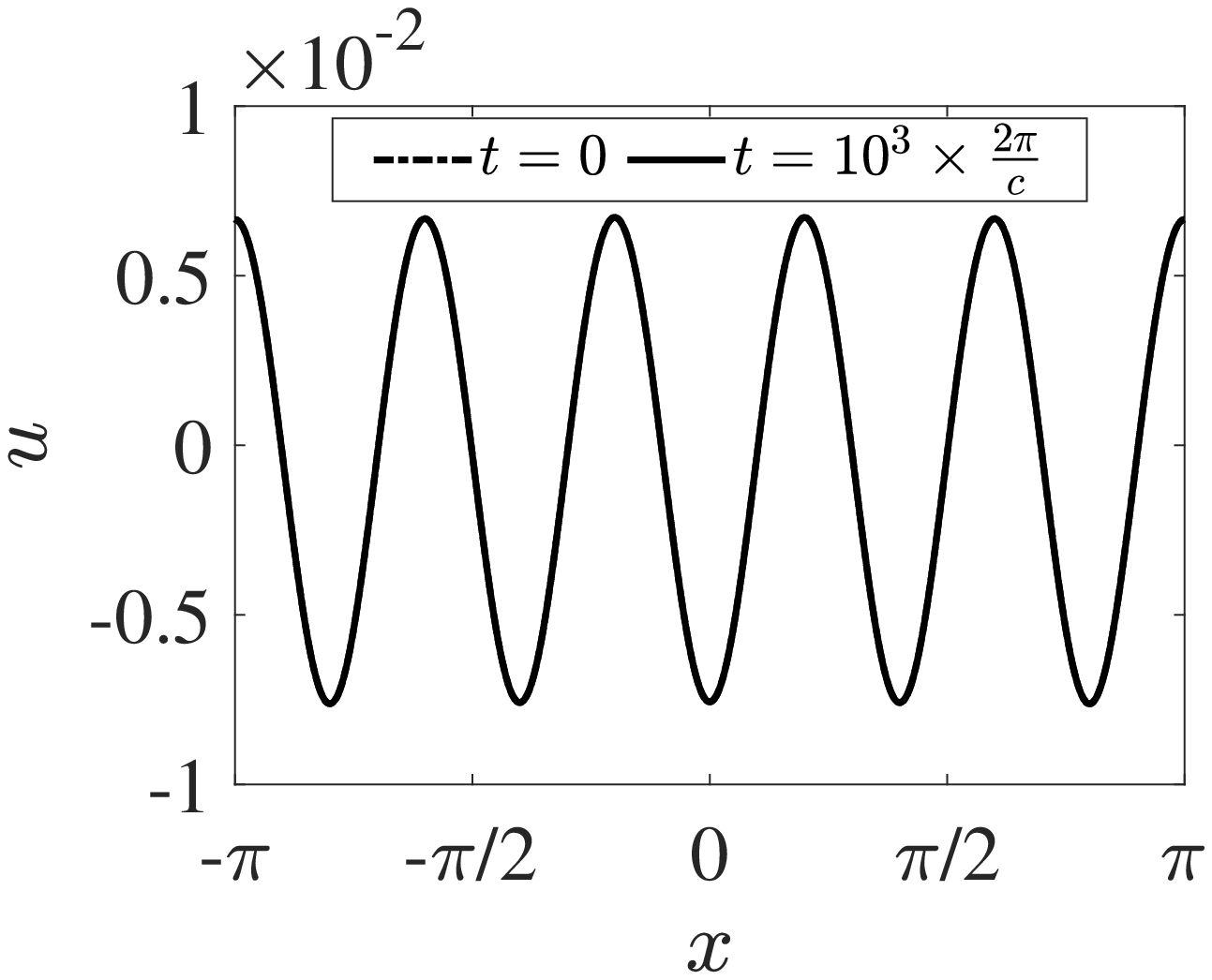}\\
\includegraphics[height=.21\textheight, angle =0]{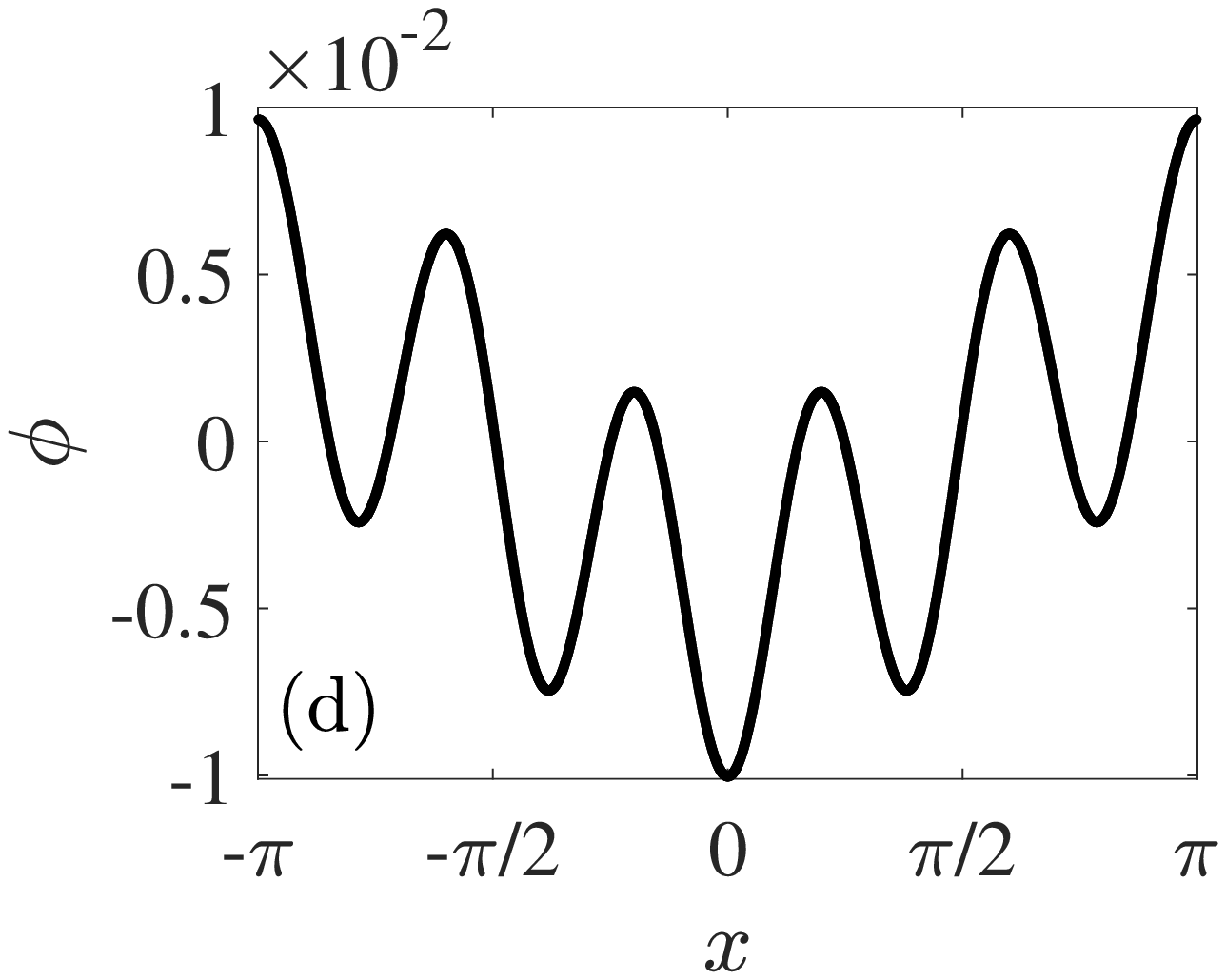}
\includegraphics[height=.21\textheight, angle =0]{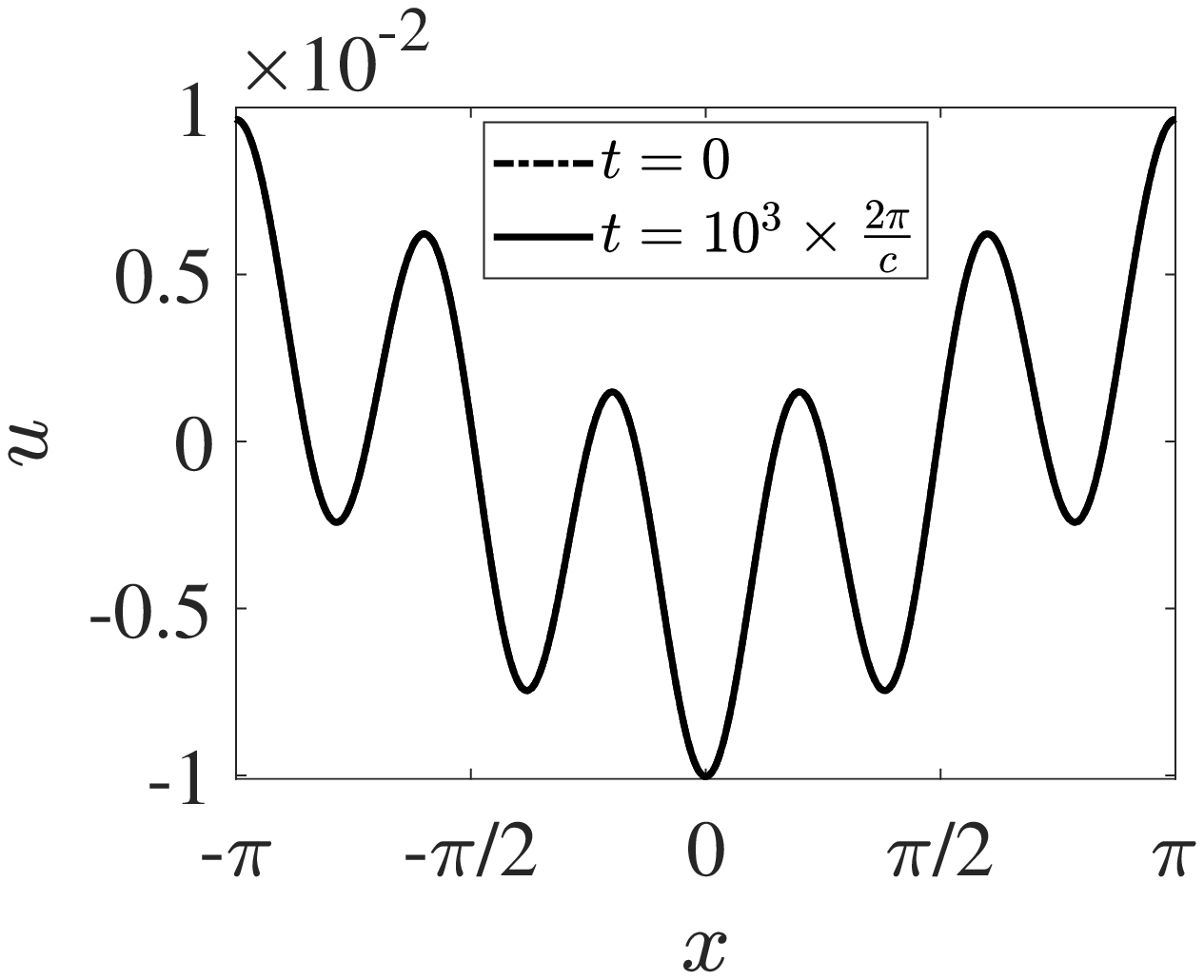}\\
\includegraphics[height=.21\textheight, angle =0]{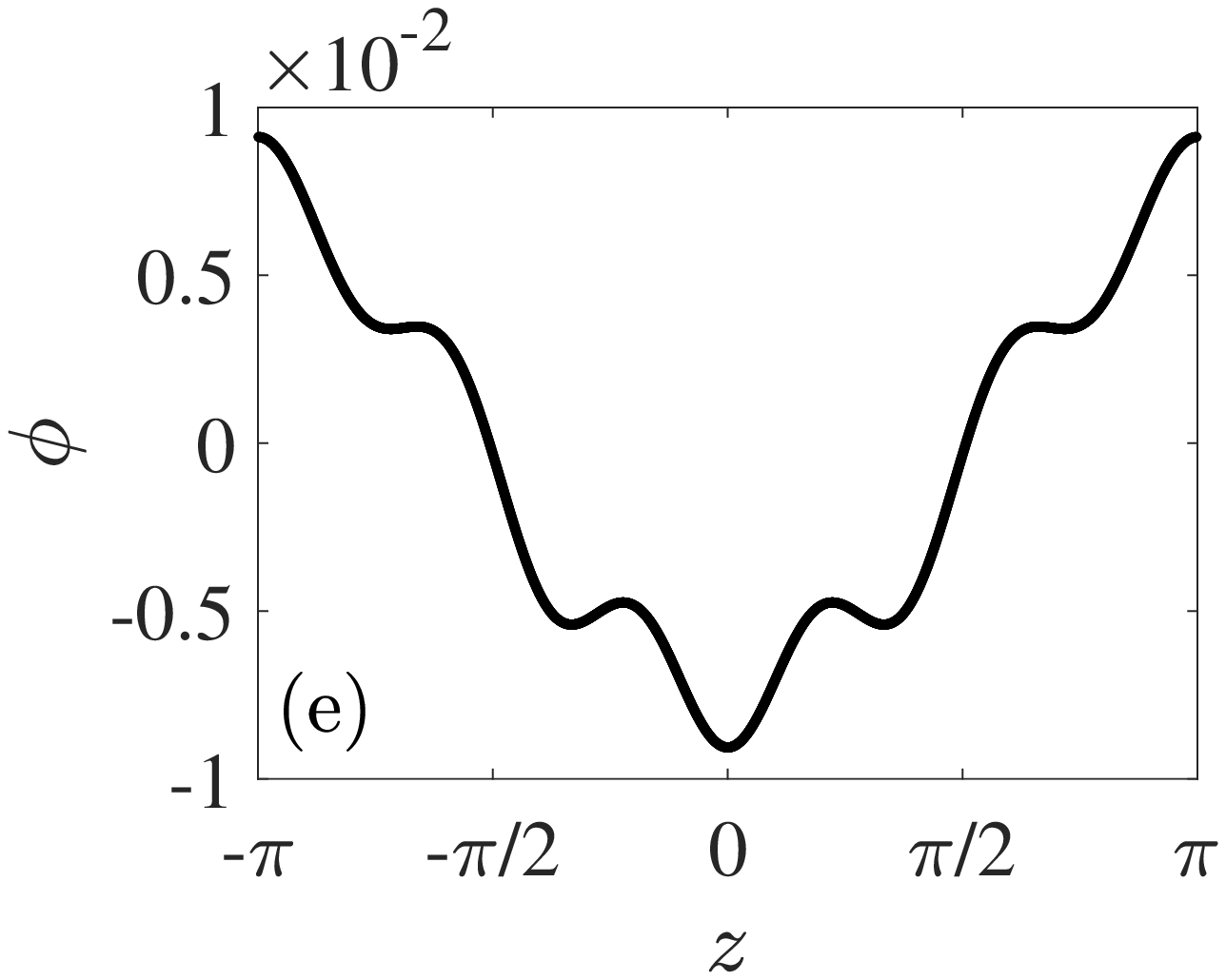}
\includegraphics[height=.21\textheight, angle =0]{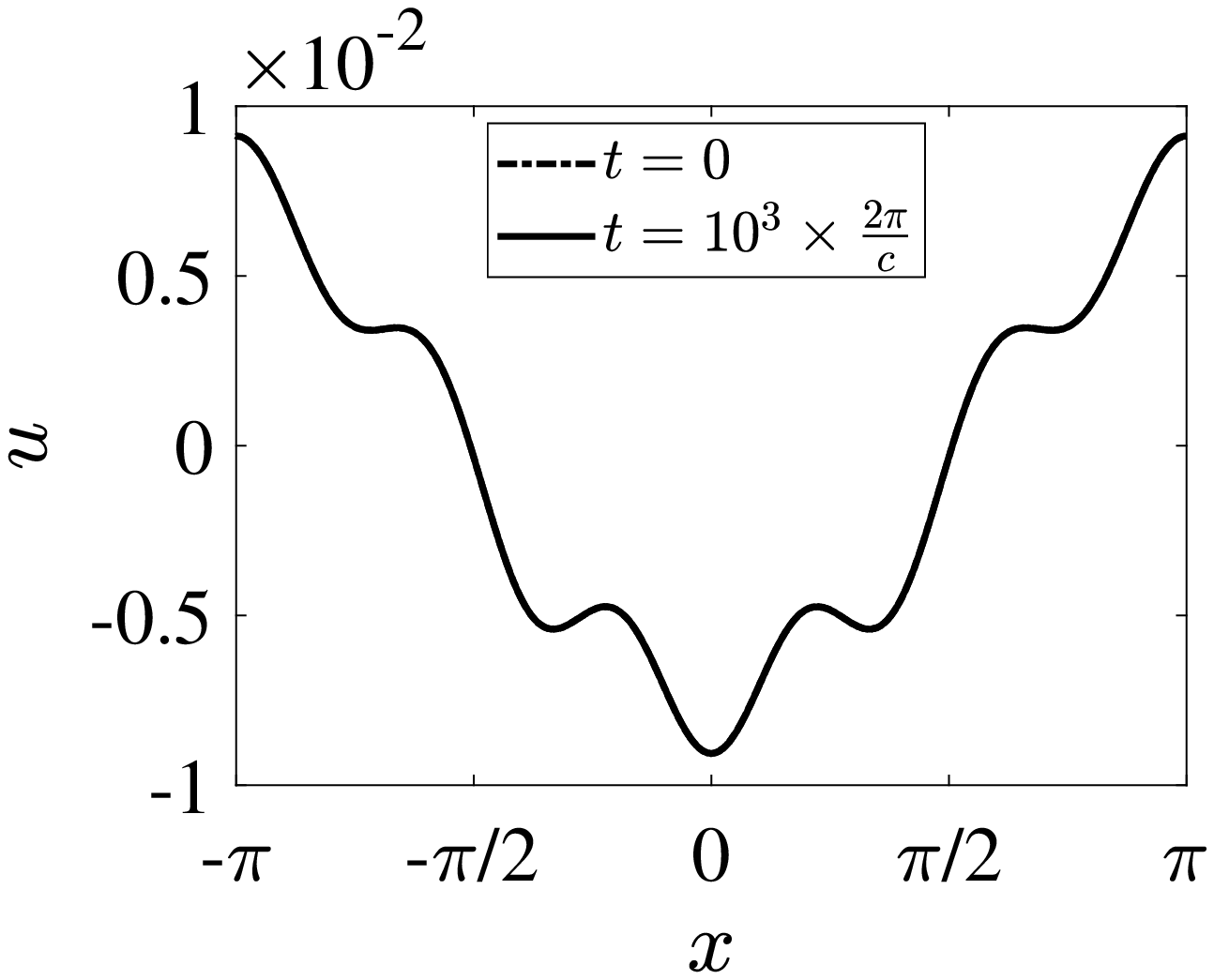}\\
\includegraphics[height=.21\textheight, angle =0]{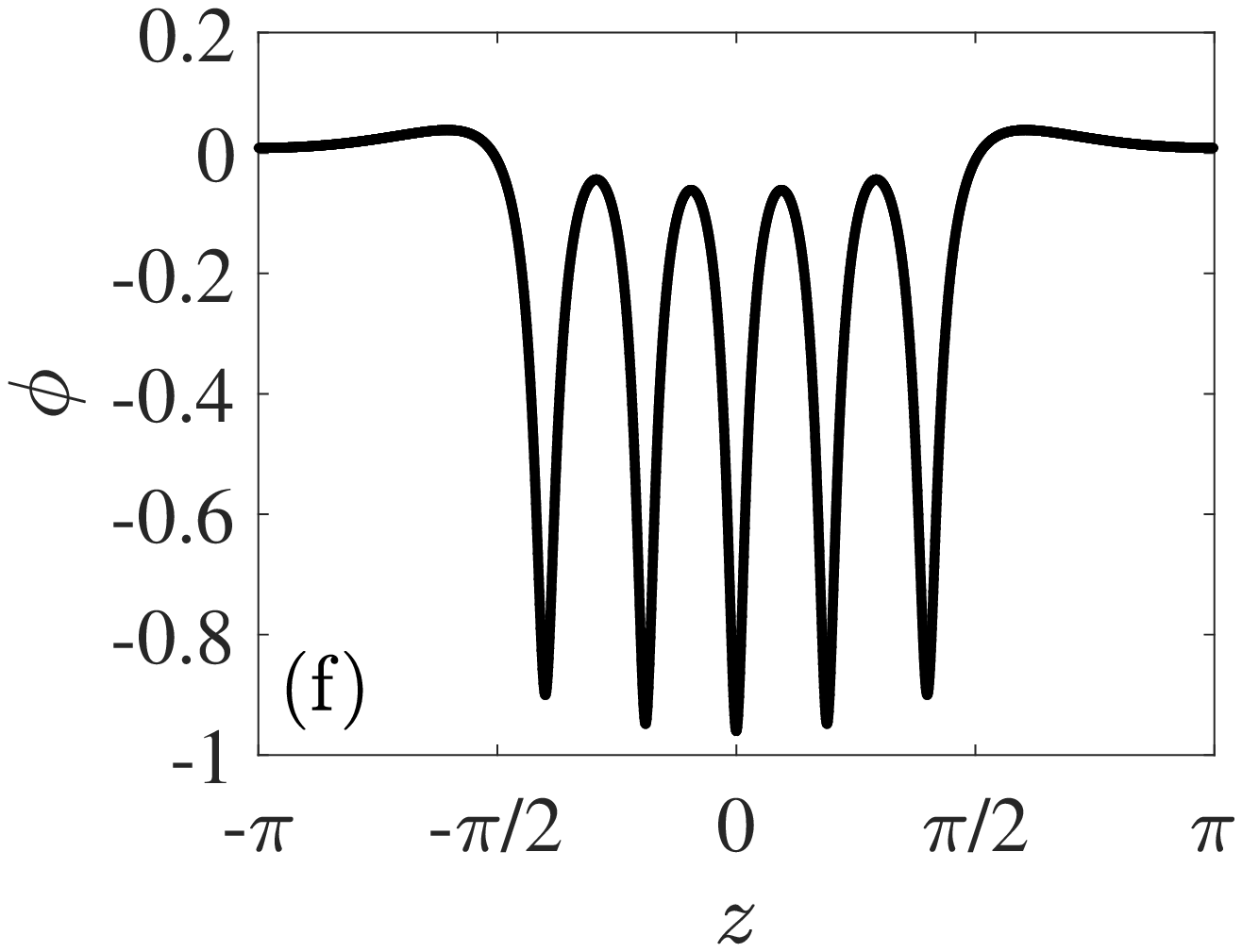}
\includegraphics[height=.21\textheight, angle =0]{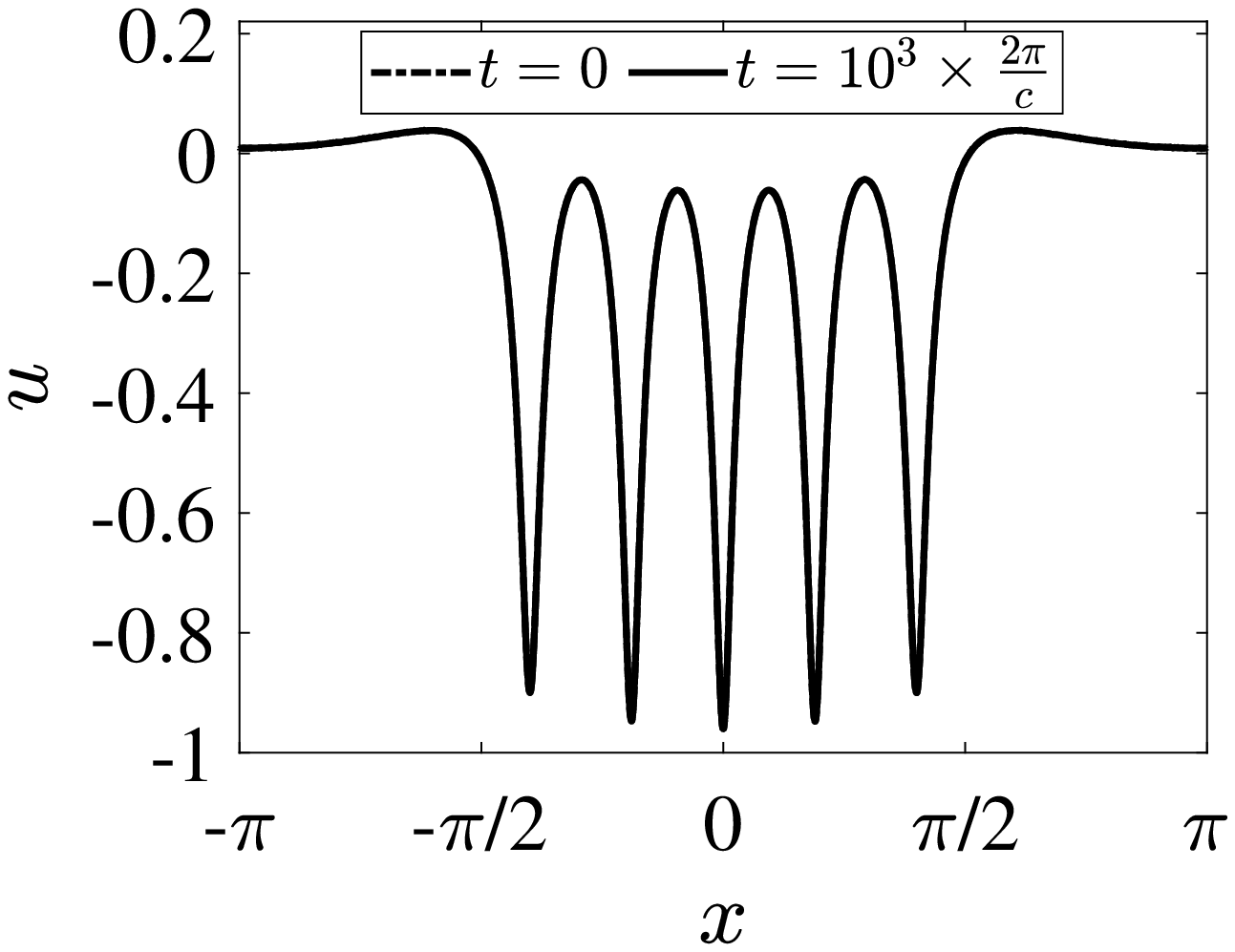}
\end{center}
\vspace*{-11pt}
\caption{
$T=T(1,5)+0.0001$, $k=1$. Similar to Figure~\ref{fig19}, almost connecting the $k=5$ branch (c), prior to (d) and past (e) crossing the $k=1$ branch itself, and almost the end of the numerical continuation (f), for which $c=0.92817880141$ (c), $0.9283$ (d), $0.92836$ (e), and $0.4$ (f).}
\label{fig20}
\end{figure}

\begin{figure}[htp]
\begin{center}
\includegraphics[height=.21\textheight, angle =0]{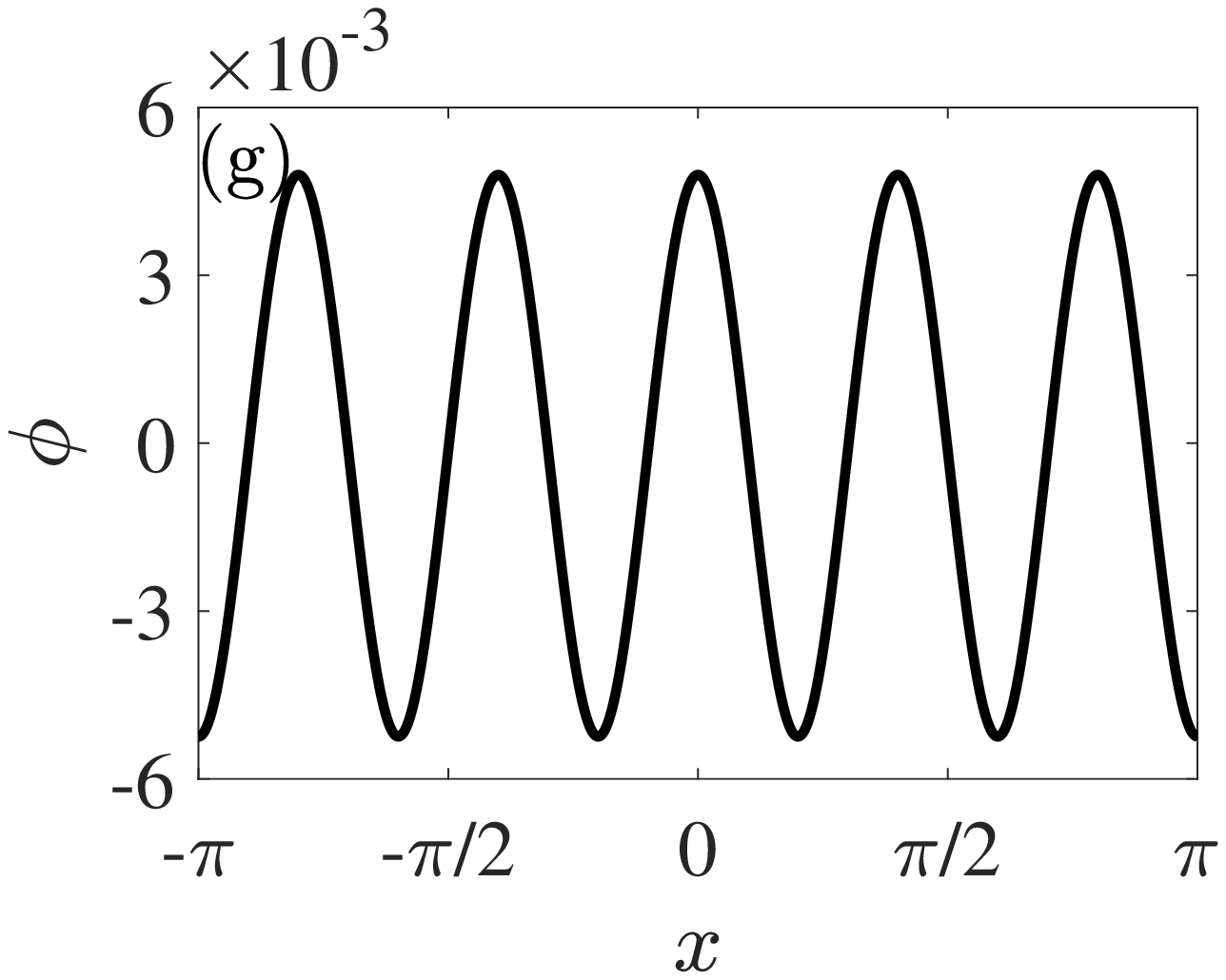}
\includegraphics[height=.21\textheight, angle =0]{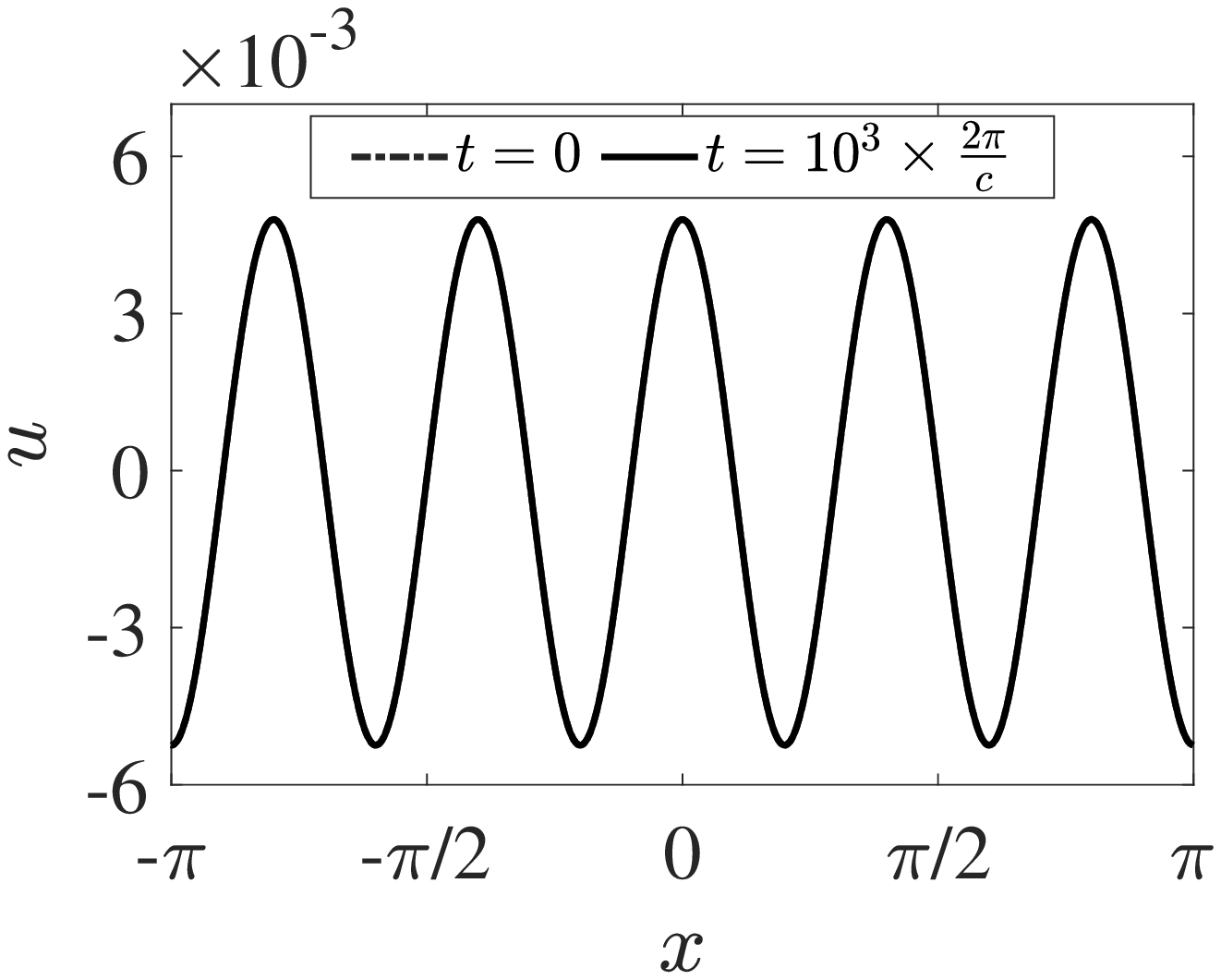}\\
\includegraphics[height=.21\textheight, angle =0]{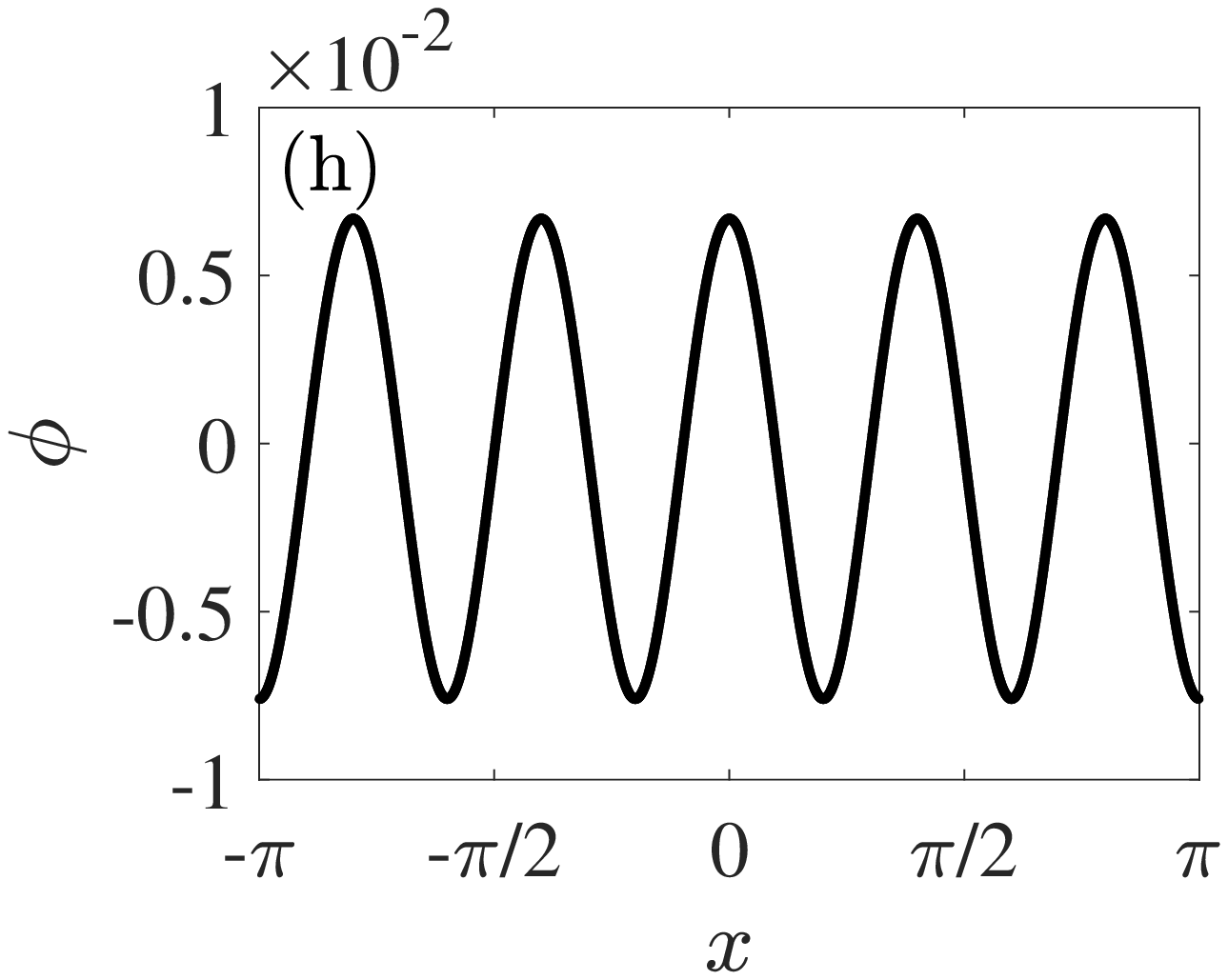}
\includegraphics[height=.21\textheight, angle =0]{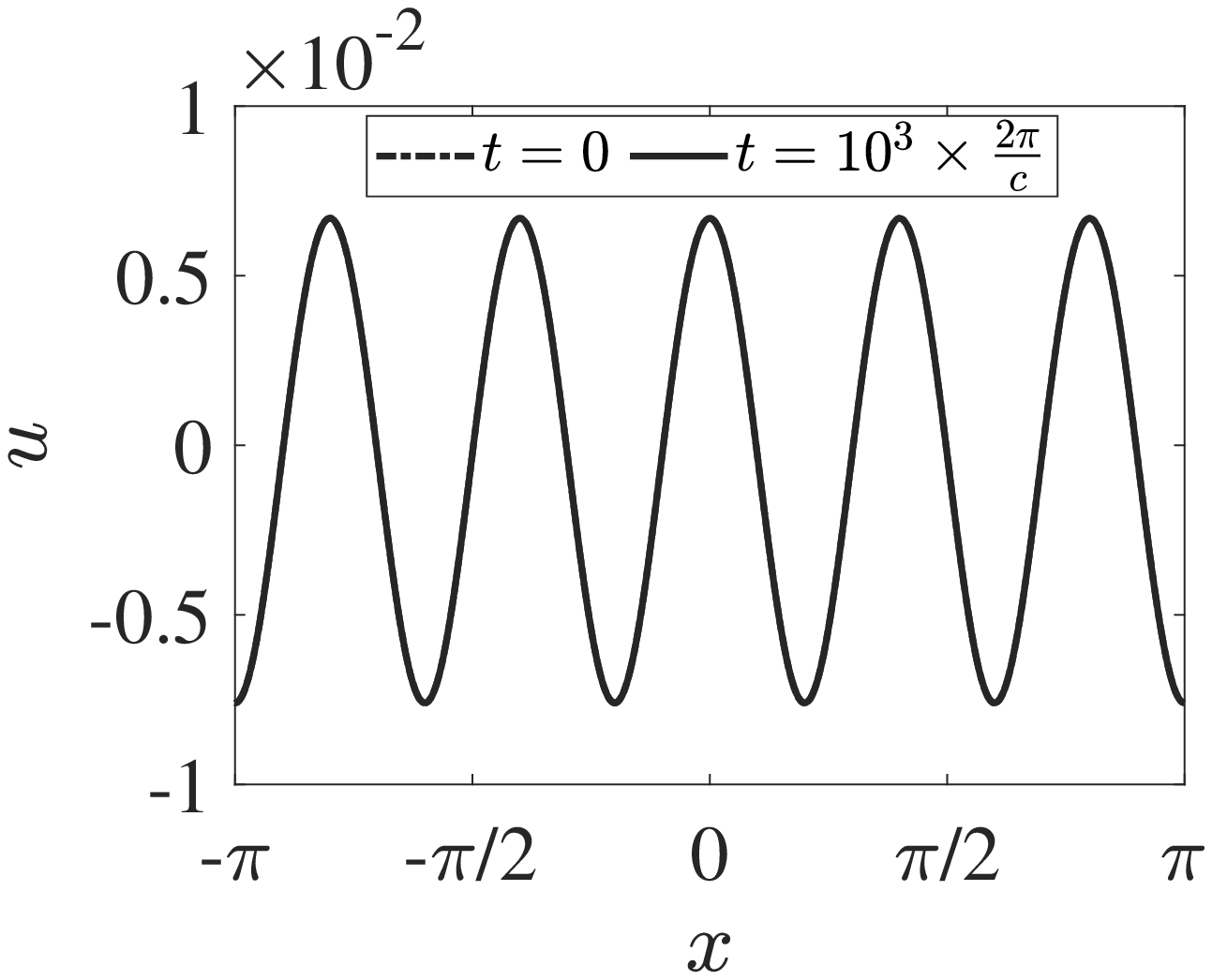}\\
\includegraphics[height=.21\textheight, angle =0]{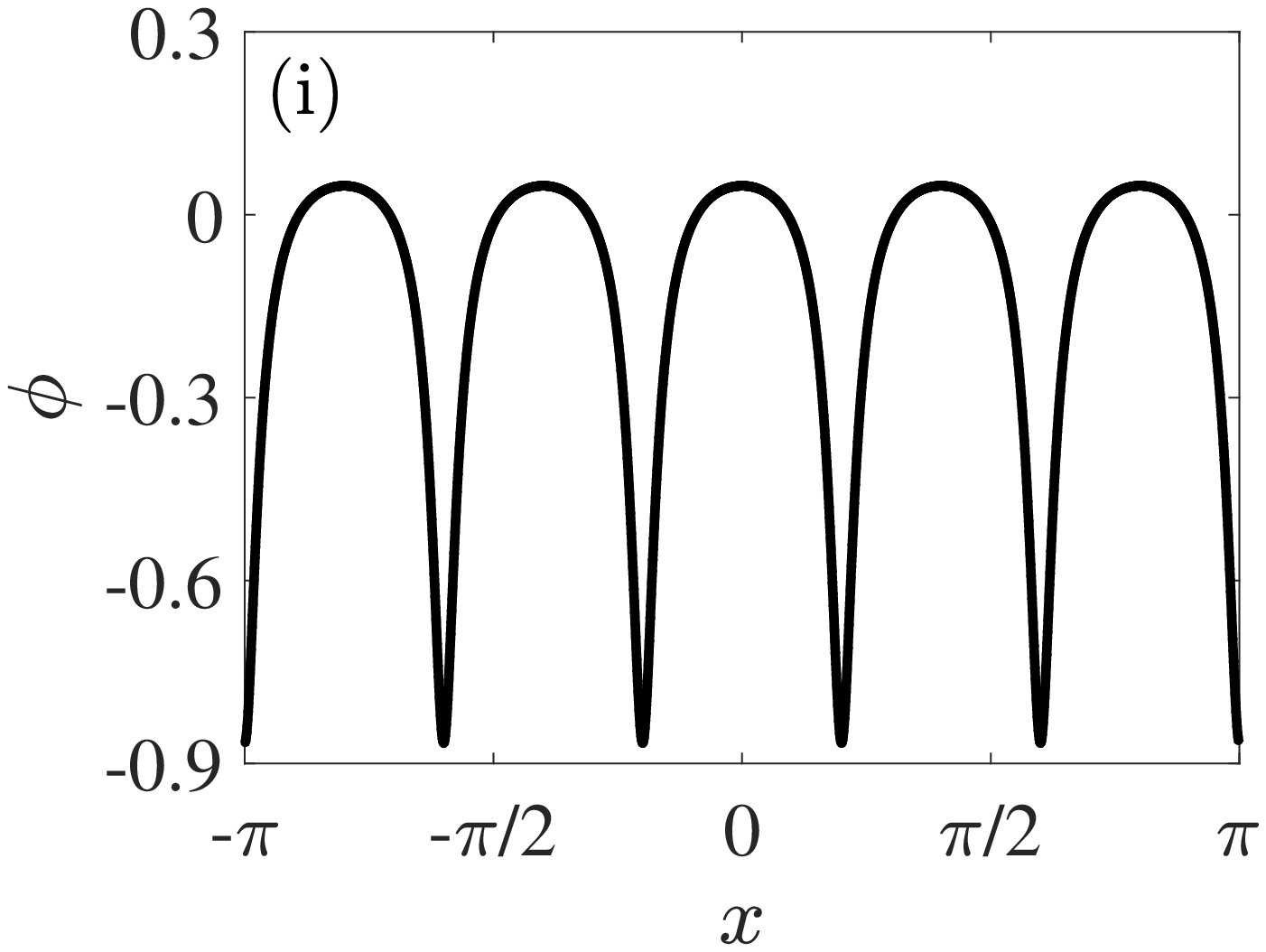}
\includegraphics[height=.21\textheight, angle =0]{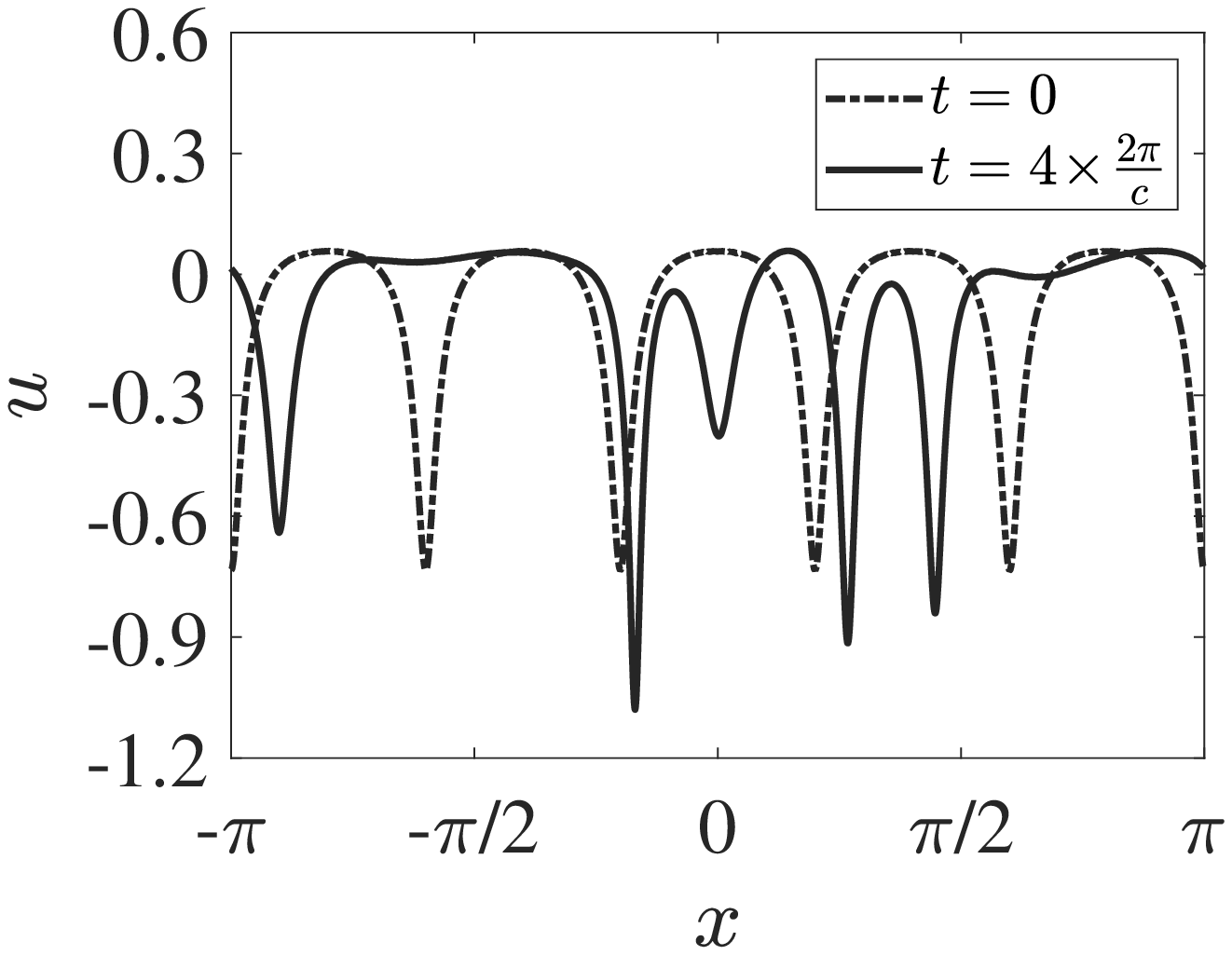}
\end{center}
\vspace*{-11pt}
\caption{
$T=T(1,5)+0.0001$, $k=5$. Left column: profiles at the points labelled with (g)-(i) in the right panel of Figure~\ref{fig19}, almost crossing the $k=1$ branch (g), almost connecting the $k=1$ branch (h), and almost the end of the numerical continuation (i), for which $c=0.9285878$~(g), $0.92817880141$~(h), and $0.4$~(i). Right column: profiles perturbed by small random noise at $t=0$ (dash-dotted) and of the solutions at lager times (solid).}
\label{fig21}
\end{figure}

\end{document}